%% file: CBBE_Rev.tex
\documentclass[12pt]{article}

\usepackage{amsmath,amssymb}
\usepackage{graphicx}
\usepackage{caption}
\usepackage[dvipsnames]{xcolor}
\usepackage{colortbl}
\usepackage{dcolumn}
\usepackage{bm}
\usepackage{float}
\usepackage{epigraph}

\usepackage[semicolon,round,nonamebreak]{natbib}

\usepackage{hyperref}
\hypersetup{ colorlinks = true, citecolor = black, urlcolor = blue}

\input{definitions}
\include{definitions_ext}

\input{def.tex}

\title{Data Assimilation in the Geosciences\\ {\large An overview on methods, issues and perspectives}}

\author{
Alberto Carrassi\thanks{Nansen Environmental and Remote Sensing Center, Bergen, Norway},
Marc Bocquet\thanks{CEREA, joint laboratory \'Ecole des Ponts ParisTech and EDF R\&D, Universit\'e Paris-Est, Champs-sur-Marne, France},
Laurent Bertino\footnotemark[1] and
Geir Evensen\thanks{IRIS, Bergen, Norway} \footnotemark[1]
}

\date{}
\begin{document}
\maketitle

\begin{center}

\subsubsection*{\small Article Type: Overview}


\hfill \break
\thanks

\subsubsection*{Abstract}
\begin{quote}
We commonly refer to state-estimation theory in geosciences as {\em data assimilation}. This term encompasses the entire sequence of operations that,
starting from the observations of a system, and from additional statistical and dynamical information (such as  a dynamical evolution model), provides an 
estimate of its state. Data assimilation is standard practice in numerical weather prediction, but its application is becoming widespread in many other areas of
climate, atmosphere, ocean and environment modeling; in all circumstances where one intends to estimate the state of a large dynamical system based on
limited information.
While the complexity of data assimilation, and of the methods thereof, stands on its interdisciplinary nature across statistics, dynamical systems and numerical
optimization, when applied to geosciences an additional difficulty arises by the continually increasing sophistication of the environmental models. Thus, in
spite of data assimilation being nowadays ubiquitous in geosciences, it has so far remained a topic mostly reserved to experts. We aim this overview article at
geoscientists with a background in mathematical and physical modeling, who are interested in the rapid development of data assimilation and its growing domains of
application in environmental science, but so far have not delved into its conceptual and methodological complexities.
\end{quote}

\end{center}

\clearpage





\renewcommand{\baselinestretch}{1.0}
\normalsize

\clearpage

\section{\sffamily \Large Introduction \label{sec:INTRO}} 
The purpose of this article is to provide a comprehensive, state-of-the-art overview of the methods and challenges of data assimilation (DA) in the
geosciences. We aim the article at geoscientists confronted with the problem of combining data with models and need to learn DA, but who are intimidated by the
vast, and technical, literature. This work may guide them through a first journey into the topic while being at the same time as complete and precise as
possible. 

The finest mathematical details, at the crossing between different areas such as numerical methods, algebra, statistics or dynamical systems, are essential to
grasp the sense of the DA problem, to reveal its interdisciplinary nature and, for many, its beauty.  We have nevertheless avoided all technicalities that were
not, in our opinion, strictly necessary and have worked to make the narrative intelligible for geoscientists, or climate scientists, who do not usually possess
a strong background in mathematics.

We provide the readers with an extensive bibliography, but also recognize that it is beyond the scope of this work to make it entirely
exhaustive. In the last decade or so, DA has attracted much attention. The topic has recently reached the rank of a discipline per se, as testified by
the appearance of books such as \citet{van2015nonlinear}, \citet{Reich-Cotter-2015}, \citet{law2015data}, \citet{asch2016data} or \cite{fletcher2017data}.
These works have mainly addressed the mathematical dimension and formulation of the DA problem and complement seminal
books on DA \citep[e.g.,][]{daley1993atmospheric, Kalnay2002, evensen2009}, that had spread the knowledge about DA and shaped it as an
independent discipline. The present article places itself somehow in between these two classes of textbooks and aims at bridging the mathematical
side of the DA methods with the practicalities and the physical intuitions and know-how that have guided its dramatic development in
climate science and geosciences in general.

The paper is structured to provide a first general formulation of the state estimation problem, from a Bayesian perspective, in Sect.~\ref{sec:form}. Having defined
the issue, and illustrated the intrinsic difficulties to adopt the Bayesian approach in the geosciences, the two most popular families of DA algorithms, both
based on a Gaussian approximation, namely the Kalman filter and the variational methods, are described in Sect.~\ref{sec:GaussMeth} and in the complementary Appendices {\color{red} A--D}. Section~\ref{sec:EnsMeth}
is entirely devoted to the ensemble methods, from their origin, through the most relevant successful variants and up to the new frontier of
hybrid ensemble-variational methods.  Four selected topics, characterizing the DA problems in the geosciences,
are the subjects of Sect.~\ref{sec:SpecTop}, and are examples of the specific type of approximations and compromises, as well as the level of
innovations DA has achieved. Finally Sect.~\ref{sec:Persp} presents a prospect of the recent challenges and current directions of research and developments,
with special attention to two, the particle filter and coupled DA.

\section{\sffamily \Large State estimation: formulation of the problem \label{sec:form}}

\subsection{Premises \label{sec:form1}}

The problem we intend to solve is the estimation of the state of a system, say the atmosphere, the ocean or any component of the Earth system or its whole, at any arbitrary past, present and future time. We possess two complementary, but both incomplete and inaccurate, sources of information: the {\it observations} and the {\it model}. 
Data assimilation provides the conceptual and methodological tools to tackle the problem by extracting synergies between model and observations and by exploiting their respective informational content.
Given the nature of the modeling and observation infrastructure in the geosciences, DA is conveniently formalized as a discrete-model/discrete-observation estimation problem, as we will do in this overview. 
We remark however that, the {\it nudging method} \citep[e.g.,][]{hoke76,nudging}, one of the simplest and most straightforward approach to assimilate data in a
dynamical model, is better formulated as continuous-model/continuous-data problem. Nudging will not be addressed in this overview, but it is worth mentioning that it has recently awoken new attention that has brought the introduction of new advanced formulations \citep[e.g.,][]{auroux08,PCL16}, and to the study of its connection with the synchronization problem \citep{duane06}. 
Interested readers can find a complete treatment of the continuous-continuous and discrete-continuous cases in many textbooks on estimation theory \citep[see, e.g.,][]{jazwinski1970,bain2009fundamentals}.

Throughout the text the following notation convention is used: $\x\in{\mathbb R}^m$ means that $\x$ is a $m$-dimensional vector whose components are real numbers; $f:{\mathbb R}^l \rightarrow {\mathbb R}^n $ signifies that the function $f$ transforms an $l$-dimensional vector into an $n$-dimensional one; model values at discrete times are indicated as $\x(t_k)=\x_k$.  
We can now formalize the two ingredients entering the estimation problem. 

\vspace{\baselineskip}
{\bf The dynamical model} - Let us assume that a model of the natural processes of interest is available as a discrete stochastic-dynamical system, 
\be
\label{eq:model}
\x_{k} = \mathcal{M}_{k:k-1}(\x_{k-1},\blam) + \eeta_{k}.
\ee
Here $\x_{k}\in{\mathbb R}^m$ and $\blam\in{\mathbb R}^p$ are the model state and parameter vectors respectively, $\mathcal{M}_{k:k-1}: {\mathbb R}^m \rightarrow {\mathbb R}^m$ is usually 
a nonlinear, possibly chaotic, function from time $t_{k-1}$ to $t_k$, and $\eeta_k\in{\mathbb R}^m$ is the model error, represented here as a stochastic
additive term, although it could be included into the parenthesis without loss of generality. The model parameters may include the external forcings or the
boundary conditions. 

The model error, $\eeta_k$, is intended to represent the error of a model prediction of the true unknown process at time $t_k$, initialized from the true state (perfect initial condition) at time $t_{k-1}$. It accounts for the cumulative effect, over the interval $t_{k}-t_{k-1}$, of errors in the parameters, $\blam$, errors in the numerical schemes used to integrate Eq.~\eqref{eq:model} as well as the effect of the unresolved scales.
The two latter arise from the spatio-temporal discretization from physical laws (e.g., the Navier Stokes equations) expressed as partial differential equations on a continuous media, into difference equations on a discrete grid or finite spectral modes. 
The appropriateness of the stochastic formulation of the model error is questionable in many cases \citep[see, e.g.,][]{nicolis2003dynamics}. Nevertheless it has the advantage of fitting very well
to the probabilistic Bayesian approach to the DA problem, as we will clarify later. Alternative forms of model error treatment in DA have been proposed
recently, including a deterministic one \citep[see][and references therein]{carrassi2016deterministic}, but they will not be developed in this overview.         

The dynamical model can also include an explicit dependence on time (i.e., be non-autonomous), as can be the case if the system is subject to climate change
driven by a time-dependent forcing, such as radiative forcing, anthropogenic changes in greenhouse gases and aerosol concentrations. 
In such case the system may not have an attractor nor an invariant measure on it (in practice it does not possess statistical equilibrium), 
a situation that again would hamper the development of a consistent statistical framework for DA. 
Recent studies on the pullback or random attractor \citep{Chekroun_et_al_2011,Dijkstra_2013} may open the path to suitable formulations of the estimation problem for non-autonomous systems.  

\vspace{\baselineskip}
{\bf The observation model} - Noisy observations of $\x_k$ are available at discrete times and are represented as components of the observation vector $\y_k\in{\mathbb R}^d$. 
Assuming the noise is additive, they are related to the model state vector through 
\be
\label{eq:obs}
\y_k={\mathcal H}_k(\x_k)+\epsi_k.
\ee
Equation~\eqref{eq:obs} defines the, generally nonlinear, observation operator, ${\mathcal H}:{\mathbb R}^m \rightarrow {\mathbb R}^d$, from model to observational space, which often involve
spatial interpolations, convolutions or spectral-to-physical space transformation in spectral models. 
Transformations based on physical laws for indirect measurements, such as radiative fluxes used to measure temperatures, can also be represented in this way \citep[e.g.,][]{Kalnay2002}.
To simplify the notation, we have assumed the observation dimension is constant, so that $d_k=d$.  

Similarly to model error, the observational error, ${\bm\epsilon}_k$, is also represented as a stochastic (i.e., random) additive term, and accounts for the instrumental error of the observing devices, deficiencies in the formulation of the observation operator itself, and the {\it error of representation (or representativeness)} \citep{lorenc1986analysis,janjic2017}. 
The latter arises from the presence of unresolved scales and represents their effect on the scales explicitly resolved by the model. The error of
representativeness is difficult to estimate due to lack of information at small scales, but ubiquitous in Earth science, because the description of a continuum fluid is made by an inevitably limited
(albeit always growing) number of discrete grid points (or spectral bands); see \citet{cohn1997introduction} for a discussion on this issue and the related ``change of
support'' techniques in geostatistics \citep{chilesdelfiner}. Note that the additivity of noise is also a simplification since
more general noisy observations $\y_k={\mathcal H}_k(\x_k,\epsi_k)$ could be considered.

\vspace{\baselineskip}
Confronting model with data is inherent to the scientific methods since Galileo's era. Nevertheless, the DA problem in Earth science has some characteristic criticalities that makes it unique. In geosciences we usually have $d \ll m$, i.e., the amount of available data is insufficient to fully describe the system and one cannot strongly rely on a data-driven approach: the model is paramount.  
It is the model that fills the spatial and temporal gaps in the observational network: it propagates information from observed-to-unobserved areas and from the
observing times to any other causally related. This capability has been pivotal in DA for numerical weather prediction (NWP) with notable examples being the ocean areas and the Southern
Hemisphere, where the lack of routine observations platforms has been compensated by the dynamical model \citep[see, e.g.,][]{daley1993atmospheric,
Kalnay2002}. The other peculiarity is the use of massive dataset ($d\approx\mathcal{O}(10^7)$) and huge models states ($m\approx\mathcal{O}(10^9)$). 
Thus, DA methods are designed to achieve the best possible use of a never sufficient (albeit constantly growing) amount of data, and to attain an efficient data-model fusion, in a short period of time (typically $3-6$ hours for NWP).
This poses a formidable computational challenge, and makes DA an example of big-data problems.

\subsection{Bayesian formulation of the state estimation problem \label{sec:form2}}

With the two complementary pieces of information in hand, model and data, we can move forward and formalize their fusion. Because of the assumed random nature
of both the model and observational error, they can be described in terms of {\it probability density functions} (pdfs), and the Bayesian approach offers a
natural framework to understand the DA problem. Our main focus here is on state estimation, i.e., the issue of estimating $\x$, based on the model and the
observation. Nevertheless, in many physical applications, one is often interested in the joint estimate of the system's state and of its parameters, $\blam$
\citep[see, e.g.,][and references therein]{eve09a}. We shall allude to parameter estimation later, but its extensive exposition is beyond
the scope of this overview.

In the Bayesian formulation, the output of the estimation process is the {\it posterior distribution} $p(\x\vert\y)$ of the unknown process $\x$ conditioned on the data $\y$, which can be obtained using {\it Bayes' rule}
\be
\label{eq:Bayes-Th}
p(\x\vert\y)=\frac{p(\y\vert\x)p(\x)}{p(\y)}\, .
\ee
In Eq.~\eqref{eq:Bayes-Th}, the three components of the Bayesian inference appear: $p(\x)$ is the {\it prior pdf} that gathers all the knowledge before
assimilating the new observations, $p(\y\vert\x)$ is the {\it likelihood of the data} conditioned on the state $\x$ (i.e., what would be the observation if the true state
were known?), and $p(\y)$ is the marginal distribution of the observation, $p(\y)=\int \! \mathrm{d}\x \, p(\y\vert\x)p(\x)$, i.e., the distribution of $\y$ whichever the value of the state. The distribution $p(\y)$is independent of $\x$ and is treated as a normalization coefficient. The postulate of a prior is a distinctive feature of the Bayesian approach,
which allows the introduction of arbitrary information about the system before data are included.
Its choice is subjective and one can in principle use any distribution that suits a study's specific purposes, either based on climatology (i.e., from
historical knowledge about the system), on theoretical physical principles or even on subjective expert's opinion. However, in many practical cases, the search
for a good informative prior is not straightforward, although its choice may strongly affect the results, sometimes adversely. 

So far, we have left the times of $\x$ and $\y$ undefined. 
Both the model and the observational error sequences, $\{\eeta_k: k=1, \ldots ,K\}$ and $\{\epsi_k: k=1, \ldots, K\}$ are assumed to be independent in time, mutually
independent, and distributed according to the pdfs $p_{\eeta}$ and $p_{\epsi}$, respectively. These pdfs are related to the prior and likelihood terms of 
Bayes' rule as follows: 
\begin{align}\label{eq:trans}
p(\x_{k}\vert\x_{k-1})& = p_{\eeta}[\x_{k} - \mathcal{M}_{k:k-1}(\x_{k-1})] , \\
p(\y_{k}\vert\x_{k})& = p_{\epsi}[\y_{k} - \mathcal{H}_k(\x_k)],
\end{align} 
where the model dependency on the parameters, $\blam$, has been dropped to simplify the notation. 
Let us define the sequences of system states and observations within the interval $[t_0,t_K]$ as $\x_{K:0}=\{\x_K,\x_{K-1},...,\x_0\}$ and
$\y_{K:1}=\{\y_K,\y_{K-1},...,\y_1\}$ respectively. Since the observational errors are assumed to be independent in time we can split the products of the probabilities 
\be
\label{eq:ObsInd}
p(\y_{K:1}\vert\x_{K:0}) = \prod_{k=1}^{K}p(\y_k\vert\x_k) = \prod_{k=1}^{K}p_{\epsi}[\y_{k} - \mathcal{H}_k(\x_k)],
\ee
meaning that the mutual likelihood of all the observations in the interval $t_K-t_0$ is the product of the individual likelihoods at each time.
 We will further assume that the process is Markovian, which means that the state $\x$ at time $t_k$, conditioned on all past states, only depends on the most
 recent state at time $t_{k-1}$ and split the prior pdf accordingly 
\be
\label{eq:Markov}
p(\x_{K:0}) = p(\x_0)\prod_{k=1}^{K}p(\x_k\vert\x_{k-1}) = p(\x_0)\prod_{k=1}^{K} p_{\eeta}[\x_{k} - \mathcal{M}_{k:k-1}(\x_{k-1})].
\ee
By combining Eq.~\eqref{eq:ObsInd} and \eqref{eq:Markov} using Bayes' rule, Eq.~\eqref{eq:Bayes-Th}, we get the posterior distribution as a product 
\be
\label{eq:Posterior}
p(\x_{K:0}\vert\y_{K:1}) \propto p(\x_0)\prod_{k=1}^{K}p(\y_k\vert\x_k)p(\x_k\vert\x_{k-1}) = p(\x_0)\prod_{k=1}^{K}p_{\epsi}[\y_{k} - \mathcal{H}_k(\x_k)]p_{\eeta}[\x_{k} - \mathcal{M}_{k:k-1}(\x_{k-1})].
\ee 
Equation~\eqref{eq:Posterior} is of central importance: it states that a new update can be obtained as soon as new data is available; it is called a hidden
Markov chain in statistics. It is worth to mention that, although the hypothesis of uncorrelated-in-time model and observational errors provides a notable mathematical advantage and it has been key to obtain Eq.~\eqref{eq:Posterior}, that is in fact almost never a valid one in realistic geosciences applications. Model error between two successive updates will be in most cases very naturally time-correlated. Similarly, observational error will also be time-correlated when, for instance, measurements are taken by the same instruments (e.g. from a satellite at two successive passages) and are processed using a physical models. 


Depending on which time period is needed for state estimation, it is possible to define {\it three estimation problems} \citep{Wiener49}:
\begin{enumerate}
\item {\bf Prediction}: estimate $p(\x_l\vert\y_{k:1})$ with $l>k$.
\item {\bf Filtering}: estimate $p(\x_k\vert\y_{k:1})$.
\item {\bf Smoothing}: estimate $p(\x_{K:0}\vert\y_{K:1})$, or selected marginals of this pdf, such as $p(\x_l\vert\y_{K:1})$, with $0 \le l < K$. 
\end{enumerate}
A schematic illustration of the three problems is given in Fig.~\ref{fig:Fig1}.

\begin{figure}[b!]
\includegraphics[height=21cm, width=16cm]{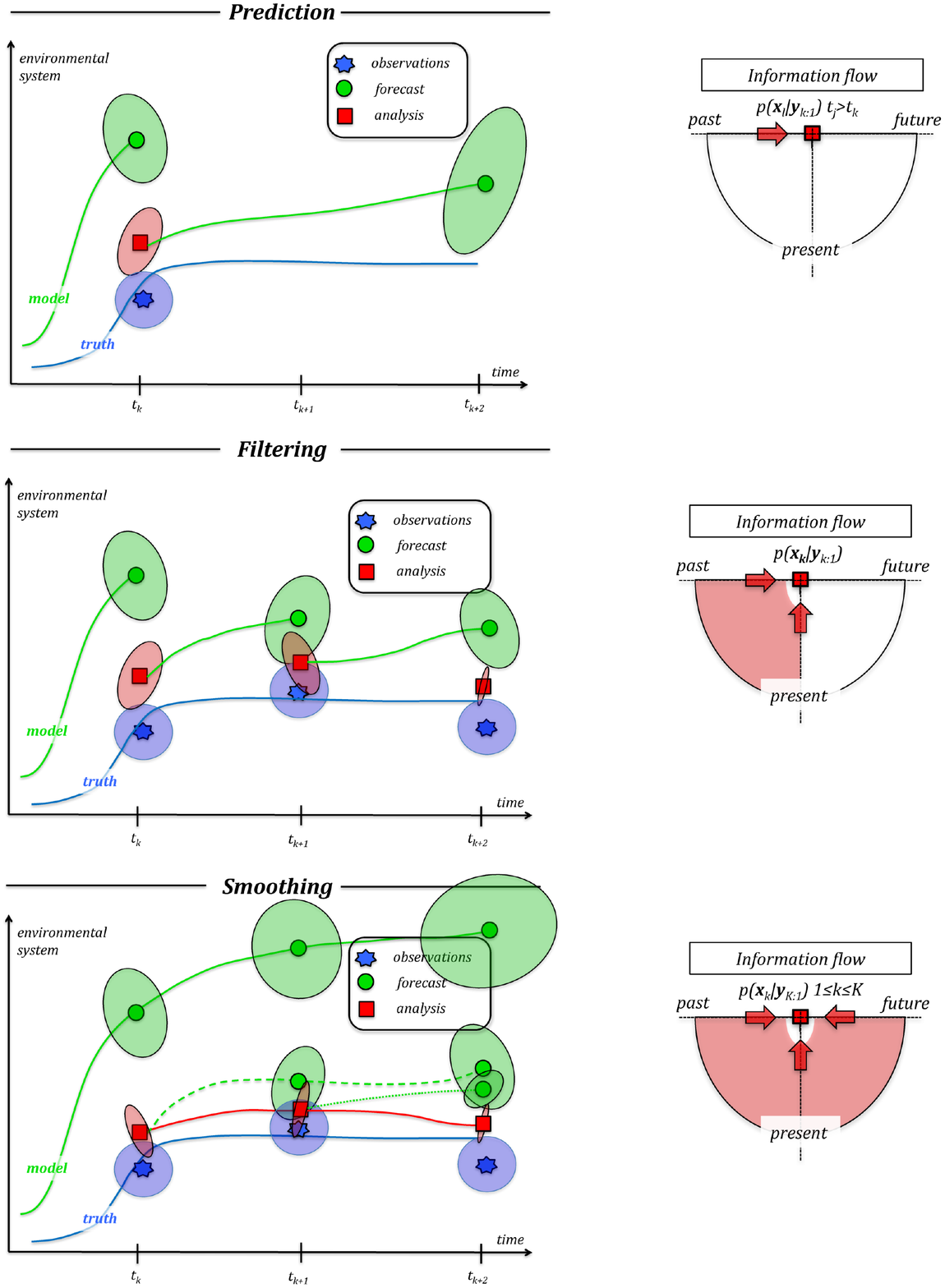}
\caption{Caption next page}
\end{figure}
\addtocounter{figure}{-1}
\begin{figure}[t!]
\caption{\label{fig:Fig1} 
Illustration of the three estimation problems: {\bf prediction} (top), {\bf filtering} (middle) and {\bf smoothing} (bottom). The true unknown signal is represented by the blue line.
Observation (blue), forecast (green) and analysis (red) pdfs  are displayed as ellipsoids of proportional size, i.e. the smaller the size the smaller the estimated uncertainty, so the larger the confidence; observational error is assumed constant. The associated blue stars for the observations, green squares for the forecast and red square for the analysis have to be intended as point-estimators based on the corresponding pdfs; one, not unique, choice can be the mean of the pdfs (cf. Sect.~\ref{sec:GaussMeth}, for a discussion on the choice of the estimator).  
{\bf Prediction} (top panel): an analysis is produced at $t_k$ using the forecast and the observation at $t_k$: the analysis uncertainty is smaller than the forecast uncertainty. From the analysis at $t_k$ a prediction is issued until $t_{k+2}$. The prediction error grows in time (as exemplar of a chaotic behavior typical in geophysical systems; cf.\ Sect.~\ref{sec:AUS}) and the forecast uncertainty at $t_{k+2}$ (the green ellipsoid) is larger than the analysis uncertainty at $t_k$ (red ellipsoid). Information is propagated only forward from $t_k$ as depicted in the information flow diagram (top right). 
{\bf Filter} (mid panels): a prediction is issued from the analysis at $t_k$ until the next observations at $t_{k+1}$; the forecast uncertainty at $t_{k+1}$ is larger than that in the analysis at $t_k$. At time $t_{k+1}$, a new analysis is performed by combining the forecast and the observations at $t_{k+1}$. The analysis uncertainty (red ellipsoid) is smaller than both the forecast and the observational uncertainties (green and blue ellipsoid/circle). From the analysis at $t_{k+1}$ the process is repeated and a new forecast until $t_{k+2}$ is issued, and a new analysis is performed using the observation at $t_{k+2}$. 
The information flow diagram (mid right) depicts how the information is carried from both the past (as in the prediction problem) and from the present using current data.
{\bf Smoother} (bottom panels): all observations between $t_k$ and $t_{k+2}$ contribute simultaneously to the analysis, which is now the entire trajectory within the smoothing interval $[t_k,t_{k+2}]$. 
At the final time, $t_{k+2}$, the smoother and filter have assimilated the same amount of observations, so their analyses at $t_{k+2}$, and their associated estimated uncertainties, are approximately the same (compare the red ellipsoids at $t_{k+2}$ for smoother and filter), but the smoother is more accurate at any other time in the window. The smoother solutions at $t_k$ and $t_{k+1}$ provide initial conditions for predictions until $t_{k+2}$ (dotted and solid green lines respectively). At final time, $t_{k+2}$, there are three forecasts initialized respectively by the analyses at $t_{k-1}$ (not shown), $t_{k}$ and $t_{k+1}$, and the associated uncertainties (green ellipsoids) are inverse proportional to the length of the prediction, with the forecast initialized at $t_{k+1}$ being the most accurate. 
}

\end{figure}

The {\it prediction problem} (Fig.~\ref{fig:Fig1} top panel) is formally addressed by integrating Eq.~\ref{eq:trans}, i.e.~by solving the Chapman-Kolmogorov equation for the propagation of a pdf by the model 
\be 
\label{eq:Chap-Kol}
p(\x_l\vert\y_{k:1}) = \int \! \mathrm{d}\x_k \, p_{\eeta}[\x_l - \mathcal{M}_{l:k}(\x_{k})]p(\x_k\vert\y_{k:1}),
\ee 
given the conditional pdf at time $t_k$, $p(\x_k\vert\y_{k:1})$. 

The {\it filtering problem} (Fig.~\ref{fig:Fig1} mid panel) is the most common in geophysical applications, and is characterized by sequential processing, in which
measurements are utilized as they become available \citep{jazwinski1970,Bengtsson81}. An {\it analysis step}, in which the conditional pdf $p(\x_k\vert\y_{k:1})$ is updated
using the latest observation, $\y_k$, alternates with a forecast step which propagates this pdf forward until the time of a new observation. The analysis is based on the application of Bayes' Eq.~\eqref{eq:Bayes-Th}, which becomes
\be
\label{eq:filt-Bay}
p(\x_k\vert\y_{k:1}) = \frac{p_{\epsi}[\y_{k} - \mathcal{H}_k(\x_k)]p(\x_k\vert\y_{k-1:1})}{\int \! \mathrm{d}\x_k \, p_{\epsi}[\y_{k} - \mathcal{H}_k(\x_k)]p(\x_k\vert\y_{k-1:1})}\, ,
\ee
while in the {\it prediction step} one integrates the Chapman-Kolmogorov equation \eqref{eq:Chap-Kol} from $k-1$ to $k$.  The process is then repeated, sequentially, with the outcome of the Chapman-Kolmogorov equation providing the prior distribution for the next analysis step.  

Estimating the conditional pdf of the state at any time $t_k$, $0\le k\le K$, based on all observations (past, present and future), is known as the {\it
smoothing problem} (Fig.~\ref{fig:Fig1} bottom panel). This is relevant when one is interested in a retrospective analysis after the observations have been
collected, and it is very useful in generating {\it reanalysis} as well as in the estimation of the parameters $\blam$. Reanalysis programs 
offer an homogeneous reconstructed estimate of the system based on observations covering an extended period of time (up to a century),
using state-of-the-art model and DA methods of the time when the reanalysis is produced. Reanalyses are paramount in climate diagnosis and several NWP, ocean and climate
centers issue their own reanalysis products \citep[e.g., among many others, ][]{dee2011era,xie2017quality}.

The joint smoothing pdf is given in Eq.~\eqref{eq:Posterior}, and is based on the recursive use of Bayes' formula for the update at each $t_k$, and the Chapman-Kolmogorov equation for the pdf propagation from $t_{k-1}$ to $t_k$.
To see this, the smoothing distribution at time $t_k$ can be written using the subsequent state at time $t_{k+1}$
\be
\label{eq:SmoothMarg}
p(\x_k\vert\y_{K:1}) = \int \! \mathrm{d}\x_{k+1} \, p(\x_k\vert\x_{k+1},\y_{K:1})p(\x_{k+1}\vert\y_{K:1}).
\ee
Furthermore, we note the following
\be
\label{eq:backPass}
p(\x_k\vert\x_{k+1},\y_{K:1}) = p(\x_k\vert\x_{k+1},\y_{k:1}) = \frac{p(\x_{k+1}\vert\x_{k})p(\x_k\vert\y_{k:1})}{p(\x_{k+1}\vert\y_{k:1})},
\ee
where we used again the fact that observations $\{\y_{k+1},\ldots,\y_{K}\}$ are independent of $\x_k$ when $\x_{k+1}$ is known (see Sect.~\ref{sec:form2}), and $p(\x_k\vert\y_{k:1})$ is the filter solution (the analysis) at time $t_k$. 
Together, Eq.~\eqref{eq:SmoothMarg} and \eqref{eq:backPass} suggest a forward-backward recursive algorithm, in which a forward filtering is followed by a backward smoothing.
Starting from $p(\x_0)$, the forward phase consists of a sequential filter from $t_1$ to $t_K$ which allows one to estimate, and to store, the sequence of filter pdfs
$p(\x_k\vert\y_{k:1})$. In the backward phase, from $t_{K-1}$ to $t_1$, one evaluates Eq.~\eqref{eq:backPass} using the previously stored filter solutions, and finally
estimates the smoothing pdf recursively via Eq.~\eqref{eq:SmoothMarg} using the smoothing pdf, $p(\x_{k+1}\vert\y_{K:1})$, from the previous iteration. We will see in
Sect.~\ref{sec:KS} that the above recursion possesses an analytical solution when the dynamical and observational model are linear and all error pdfs are Gaussian. 

We conclude the section with some comparative considerations between the filter and smoother. The filtering solution at an arbitrary $t_k$ ($0\le k \le K$) is obtained by
sequential updating until $t_k$ and it thus accounts for all observations before $t_k$. In contrast, the smoothing solution at the same time $t_k$  
also accounts for future observations until $t_K$, and it is thus generally more accurate than the filtering one. At the final time $t_K$ both solutions have incorporated the same amount of data, so that in the absence of approximations, they will coincide \citep[e.g.,][]{kalnay2010}. 

\section{\sffamily \Large A route to solutions: the Gaussian approximation \label{sec:GaussMeth}}

With the pdfs for smoother, filter and prediction in hand, it still remains to be decided which estimator suits best a specific
problem. Those pdfs, in fact, describe fully the probability of all possible states of the system but hardly make good targets for an estimation algorithm. 
Two ``natural'' options, the {\it mean} of the distribution or its {\it mode}, characterize two traditional approaches to state estimation. 
Methods designed targeting the mean are named {\it minimum squared error} estimators because, irrespective of the properties of the pdf (i.e., whether or not it is symmetric
and unimodal), the mean is always the minimum squared error estimate \citep{jazwinski1970}. The mode is the peak of the distribution, the most probable state, and methods
targeting it are referred to as {\em maximum a posteriori} estimators. Other possible options are viable, such as the median, but they are generally more difficult to compute
do not have the same wide range of applicability as the mean or the mode. 

Despite its appealing and clear logic, the huge dimensions of typical models and datasets used in environmental science hamper the use of a fully Bayesian approach.
The problem dimension renders it extremely difficult, or practically impossible, to define and to evolve the pdfs. 
To overcome this computational, albeit fundamental, issue it is usually assumed that the uncertainties of each piece of information (observations, model and prior) are
Gaussian distributed. This hypothesis leads to a substantial simplification: the pdfs can be completely described by their first and second moments: the mean and the covariance matrix.

The Gaussian approximation is at the core of most DA procedures successfully used in the geosciences, and this overview will largely discuss methods that rely upon it to different extents.  
This section presents two main approaches to the estimation problem in which DA in the geosciences is rooted: the minimum variance Kalman-like (Sect.~\ref{sec:KFS}) and the maximum a posteriori variational methods (Sect.~\ref{sec:Var}). Section~\ref{sec:GaussMeth} is complemented by four Appendices ({\color{red} A-D}) where interested readers can find additional details on the methods.  

\subsection{The linear and Gaussian case - The Kalman filter and smoother \label{sec:KFS}}

The dynamical and observational models, Eqs.~\eqref{eq:model} and \eqref{eq:obs}, are both assumed to be linear
\begin{align}
\label{eq:modellin}
\x_k & = \bM_{k:k-1}\x_{k-1}+\eeta_k, & \eeta_k\sim\mathcal{N}(\bzero,\bQ_k), \\ 
\label{eq:obslin}
\y_k & = \bH_k\x_k+\epsi_k, & \epsi_k\sim\mathcal{N}(\bzero,\bR_k), 
\end{align}
with $\bM_{k:k-1}$ and $\bH_k$ being $m\times m$ and $d\times m$ matrices respectively. 
The observational and model noises are assumed to be mutually uncorrelated and in time, unbiased, and Gaussian distributed with covariance matrices $\bR_{k}\in {\mathbb R}^{d\times d}$ and $\bQ_{k}\in {\mathbb R}^{m\times m}$ respectively. 

\subsubsection{Kalman filter \label{sec:KF}} 

Under the aforementioned hypotheses of all sources of errors being Gaussian and both the observational and dynamical models being linear, the Kalman filter is the exact (optimal) solution to estimate recursively the conditional mean and the associated uncertainty, the conditional covariance, of the filtering pdf $p(\x_k\vert\y_{k:0})$.  
The process is sequential, meaning that observations are assimilated in chronological order, and the KF alternates a {\it forecast step}, when the mean and covariance of the prediction conditional pdf, $p(\x_k\vert\y_{k-1:0})$, are evolved, with an {\it analysis step} in which the mean and covariance of the filtering conditional pdf, $p(\x_k\vert\y_{k:0})$, are updated.      
By virtue of the Gaussian assumption, the mean and covariance suffice to fully describe the pdfs and the KF provides an exact set of closed equations.
The KF formulae are given below without their derivation, which goes beyond the scope of this overview, but the reader can find numerous alternative ways in classical books or reviews on filtering \citep{jazwinski1970, Bengtsson81, cohn1997introduction, talagrand1997assimilation, wikle2007bayesian, bain2009fundamentals}. 

Let us indicate the forecast/analysis mean state and covariance as $\x^{\mathrm{f/a}}$ and $\bP^{\mathrm{f/a}}$ respectively. The KF equations then read 
\begin{align}
\label{eq:KF-fcst1}
\mathrm{Forecast~Step} \qquad\qquad\qquad & \x^\mathrm{f}_k=\bM_{k:k-1}\x^\mathrm{a}_{k-1}, \\ 
\label{eq:KF-fcst2}
                                 & \bP^\mathrm{f}_{k}=\bM_{k:k-1}\bP^\mathrm{a}_{k-1}\bM^{\rm T}_{k:k-1}+\bQ_{k}. 
\end{align}
\begin{align}
\label{eq:KF-anl1}
\mathrm{Analysis~step} \qquad\qquad\qquad & \bK_k=\bP^\mathrm{f}_k\bH_k^{\rm T}(\bH_k\bP_k^\mathrm{f}\bH_k^{\rm T}+\bR_k)^{-1}, \\
\label{eq:KF-anl2}
				 & \x^\mathrm{a}_k=\x^\mathrm{f}_k+\bK_k(\y_k-\bH_k\x^\mathrm{f}_k), \\
\label{eq:KF-anl3}
				 & \bP^\mathrm{a}_k= (\bI_k-\bK_k\bH_k)\bP^\mathrm{f}_k.
\end{align}
Given as inputs the matrices $\bQ_k$, $\bR_k$, $\bH_k$ and $\bM_k$, for $k\ge 1$, and the initial condition for the mean, $\x_0^\mathrm{a}=\x_0$, and initial error covariance,
$\bP_0^\mathrm{a}=\bP_0$, Eqs.~\eqref{eq:KF-fcst1}--\eqref{eq:KF-anl3} estimate sequentially the state, $\x^{\mathrm{f/a}}_k$, and the associated error covariance,
$\bP^{\mathrm{f/a}}_k$, at any time $k>1$. The matrix $\bK_k\in{\mathbb R}^{m\times d}$ is the Kalman gain and it contains the regression coefficients of the optimal linear
combination between the prior (given here by the forecast $\x^\mathrm{f}_k$) and the observations. The resulting state estimate, the analysis $\x^\mathrm{a}_k$, has minimum
error variance and is unbiased. In fact, the KF Eqs.~\eqref{eq:KF-fcst1}--\eqref{eq:KF-anl3} can be obtained without the need of the Gaussian hypothesis and the KF solution is also said the {\it Best Linear Unbiased Estimator}, BLUE \citep[see, e.g.,][their Sect.~3.4.3]{asch2016data}. Similarly the KF equations can also be seen and derived using a maximum entropy approach \citep[see, e.g.,][]{mitter2005information,giffin2014kalman}.

\subsubsection{Kalman smoother \label{sec:KS}}

We discussed in Sect.~\ref{sec:form2} that a recursive estimate of the pdf of the system state at any arbitrary time in the observing window, conditioned on all observations in the window, i.e. the pdf $p(\x_k\vert\y_{K:1})$, can be obtained using a forward and backward recursions, in which a forward-in-time filter is followed by a backward-in-time smoother.
Under the hypotheses of linear dynamical and observational models and Gaussian errors, this bi-directional recursion can be solved analytically and it is referred to as the {\it Kalman smoother} \citep[KS,][]{jazwinski1970}.
Like for the KF, we describe the KS formulae in the following, but we do not provide their derivation which can found, for instance, in \citet{jazwinski1970}. 

Assume that a forward in time KF has already been implemented using Eqs.~(\ref{eq:KF-fcst1}--\ref{eq:KF-anl3}) for $k=1,\ldots,K$, and the forecast and analysis means and covariances, $\x_k^{\mathrm{f/a}}$ and $\bP_k^\mathrm{f/a}$, have been computed and stored. With the filtering solutions at disposal we can run the KS recursion backward in time, for $k=K-1,\ldots,1$, to compute the smoothing mean and covariance, $\x_k^\mathrm{sm}$ and $\bP^\mathrm{sm}_k$, according to 
\begin{align}
\label{eq:KS-backrec0}
& & \bS_k & = \bP^\mathrm{a}_k\bM^{\rm T}_{k+1:k}(\bM_{k+1:k}\bP^\mathrm{a}_{k}\bM^{\rm T}_{k+1:k}+\bQ_{k+1})^{-1} & = \bP^\mathrm{a}_k\bM^{\rm T}_{k+1:k}\bP^{-f}_{k+1}, \\
\label{eq:KS-backrec1}
& \mathrm{KS~Mean} & \x^\mathrm{sm}_k & = \x^\mathrm{a}_k+\bS_k(\x^\mathrm{sm}_{k+1}-\x^\mathrm{f}_{k+1}) & , \\ 
\label{eq:KS-backrec2}
& \mathrm{KS~Covariance} & \bP^\mathrm{sm}_{k} & = \bP^\mathrm{a}_k + \bS_k(\bP^\mathrm{sm}_{k+1}-\bP^\mathrm{f}_{k+1})\bS_k^{{\rm T}}  &
\end{align}
with initial conditions, at time $t_K$, $\x^\mathrm{sm}_K = \x^\mathrm{a}_K$ and $\bP^\mathrm{sm}_K = \bP^\mathrm{a}_K$. As anticipated in Sect.~\ref{sec:form2}, the KF and KS
solutions are equivalent at final time $t_K$, but not at earlier times. The KS formulation given above is also known as {\it Rauch-Tung-Striebel} smoother; more details on the smoothing techniques can be found in \citet{cosme2012smoothing}. 

Appendix {\color{red} A} discusses in more details some key properties of the Kalman filter and smoother. 

\subsection{\sffamily \large Variational approach \label{sec:Var}}

Variational methods are designed to estimate the model trajectory that ``best fits'' all the observations within a prescribed observing window $[t_0, t_K]$. Given that the model trajectory is adjusted globally to fit all observations simultaneously, the state estimate at an arbitrary time $t_k$, $t_0\le t_k\le t_K$, is influenced by all observations within the window. 
The problem is formalized as the one of minimizing an appropriate scalar {\it cost function} that quantifies the model-to-data misfit \citep{thompson1969reduction,sasaki1970}. 

Assuming data are available at each time $t_k$ in the window (this assumption can easily be relaxed without any major changes in the exposition that follows), the variational problem reads
\be  
\label{eq:var-gen}
\x^{\rm a}_{K:0} = \mathrm{argmin}({\mathcal J}(\x_{K:0})) 
\qquad k=1,\ldots,K. 
\ee  
where the notation $argmin$ is used to signify that $\x^{\rm a}_{K:0}$ is the solution for which the cost function, ${\mathcal J}(\x_{K:0}) $, attains its minimum. 
Note that ${\mathcal J}(\x_{K:0})$ is a function of the entire model trajectory within the window 
and the variational analysis solution, the trajectory $\x^{\rm a}_{K:0}$, provides the closest possible fit to the data, while being consistent with the model dynamics over the entire window. Note furthermore that, in formulating the variational problem in Eq.~\eqref{eq:var-gen}, we did not require the model being linear and we have not yet made any hypothesis on the statistical nature of the errors. The most common forms of the variational problem in the geosciences are discussed and derived in Sect.~\ref{sec:var-stat}. For example, when 
both observational and model errors are assumed uncorrelated in time, unbiased, and Gaussian distributed with covariances $\bR_k$ and $\bQ_k$ respectively, these errors are both
penalized in the cost function as follows: 
\begin{equation}
\label{eq:w4dvar}
{\mathcal J}(\x_{K:0}) =
\frac{1}{2}\sum_{k=0}^K\left\| \y_k - {\mathcal H}_k(\x_k)\right\|_{\bR^{-1}_k}^2 +\frac{1}{2}\sum_{k=1}^K\left\| \x_k - {\mathcal M}_{k:k-1}(\x_{k-1})\right\|_{\bQ^{-1}_k}^2 +\frac{1}{2}\left\| \x_0 - \x^\rmb \right\|_{\bB^{-1}}^2 , 
\end{equation}
where the notation, $\left\| \x \right\|^2_\bA = \x^\T \bA\x$, is used to indicate the weighted Euclidean norm. In Eq.~\eqref{eq:w4dvar} we furthermore assumed to have a
prior, commonly referred to as {\it background} in variational methods literature, for the initial state at the start of the window, in the form of a Gaussian pdf with mean, $\x^{{\mathrm b}}\in{\mathbb R}^m$, and covariance, $\bB\in{\mathbb R}^{m\times m}$, respectively. We will recall and derive Eq.~\eqref{eq:w4dvar} in Sect.~\ref{sec:var-stat}.

The calculus of variations can be used to find the extremum of ${\mathcal J}(\x_{K:0})$ and leads to the corresponding Euler-Lagrange equations \citep[see,
e.g.,][]{le1986variational,bennett1992inverse}. These arise by searching for the solution of a constrained minimization problem in which the solution is required 
to follow the model dynamics by appending the model equations such that
\be
\label{eq:lagrang}
{\mathcal L}(\x_{K:0},\bGamma) = {\mathcal J}(\x_{K:0}) + \sum_{k=1}^{K} \bGamma_k^{{\mathrm T}}(\x_k-\mathcal{M}_{k:k-1}(\x_{k-1})-\bpsi_k),
\ee
with $\bGamma_k^{{\mathrm T}}\in{\mathbb R}^m$ the Lagrange multiplier vector at time $t_k$. The model error term can also be part of the control variable and it is indicated here as $\bpsi_k$ to differentiate it from the stochastic model error, $\eeta_k$, of Eq.~\eqref{eq:model}.   
Although rigorous and formally appealing, solving the Euler-Lagrange equations for problems with the size and complexity of typical geophysical systems is practically impossible, unless drastic approximations are introduced. When the dynamics are linear and the amount of observations is not very large, the Euler-Lagrange equations can be solved with the {\it method of representers} which has been very successful in the early days of DA for the ocean \citep{bennett1992inverse}, and later extended to nonlinear dynamics by \citet{uboldi2000time}. 
A nice property of the representer method is that it reduces the dimension of the problem to the number of measurements (smaller than the whole model state at all times or the model state dimension; see \citet{bennett1992inverse}). 

Nevertheless, the representers method is still far from being applicable for realistic high-dimensional problems such as NWP. An attractive alternative is represented by the {\it descent methods}, which use the gradient vector of the cost function in an iterative minimization procedure \citep[][see also Appendix {\color{red} B}]{talagrand1987variational}. This latter approach, in its many variants, is adopted in most of the operational NWP centers that employ variational assimilation \citep{fisher2001developments}. At the cost function minimum, its gradient must vanish and its Hessian matrix be positive definite (this condition is the extension to vector functions of the classical condition for a minimum of a scalar function for which its first derivative, the gradient, must vanish and its second derivative, here the Hessian, must be positive, here positive definite):
\be
\label{eq:min-cond}
\nabla_{\x_{K:0}}{\mathcal J}(\x_{K:0})=\bzero, \qquad \nabla^2_{\x_{K:0}}{\mathcal J}(\x_{K:0}) > 0,
\ee
where $\bzero\in{\mathbb R}^{m\times K}$ is a matrix with all entries equal to zero, and the second inequality stands for the positive definiteness of the Hessian matrix. The {\it control variable} with respect to which the  minimization is performed is the entire model trajectory over the observing window. The variational problem defined by Eq.~\eqref{eq:var-gen} is usually referred to as {\it weak-constraint} given that the model dynamics is affected by errors and it thus constitutes a weak constraint during the minimization \citep{sasaki1970}. 

An important particular case is given by the {\it strong-constraint} variational assimilation in which the model is assumed to be perfect \citep{penenko1976variational,lewis1985use,le1986variational} (i.e., $\bpsi_k=0$). In this case, the dynamical model is (assumed) purely deterministic and its solutions are uniquely determined by specifying the initial conditions. This transforms the problem from a constrained into an unconstrained one and the cost function into a function of the initial condition alone, ${\mathcal J}(\x_{K:0})\Rightarrow{\mathcal J}(\x_{0})$; its gradient with respect to the state at $t_0$ reads
\be
\label{eq:grad-strong}
\nabla_{\x_{0}}{\mathcal J}(\x_{K:0}) = \sum_{k=1}^{K}\bM^{{\mathrm T}}_{k:0}\nabla_{\x_{k}}{\mathcal J}(\x_{k}) + \nabla_{\x_{0}}{\mathcal J}(\x_{0}).
\ee
where we have used the fact that $\x_{k} = \mathcal{M}_{k:0}(\x_{0}) $, and applied the chain rule for differentiating compositions of functions.
The matrix $\bM^{{\mathrm T}}_{k:0}$ is the {\it adjoint operator}, the adjoint of the tangent linear model, and implies a backward integration from $t_k$ to $t_0$. The gradient, Eq.~\eqref{eq:grad-strong}, can be obtained through a set of operations involving both forward nonlinear and backward adjoint integration; see Appendix {\color{red} B} for more details on the minimization procedures. 

The drastic reduction of the number of control variables, from the entire trajectory $\x_{K:0}$ for the weak-constraint into the initial
state alone $\x_0$ for the strong-constraint case, together with the use of the adjoint methods \citep{lions1971optimal}, have made possible the successful use and diffusion of strong-constraint variational methods, or some suitable approximation of them, in atmospheric and oceanic DA \citep[see, e.g.,][]{Kalnay2002,talagrand2010variational}.

\subsubsection{Statistical interpretation and 4DVar \label{sec:var-stat}}

The form of the cost function reflects how we
intend to balance the data and the model relative to each other and, again, the statistical approach offers a natural framework to accomplish this task.   
A natural choice 
would be to set the cost function proportional to the conditional posterior pdf, $p(\x_{K:0}\vert\y_{K:0})$, for which a suitable functional
form, desirably smooth and derivable, must be given.

This can be done, for instances, under the hypothesis of Gaussian errors, and it leads to a significant simplification.
Recall the Gaussian assumptions made in formulating the Kalman
filter (KF) and Kalman smoother (KS) in Sect.~\ref{sec:KFS}: the prior, model and observational errors are all Gaussian distributed. We can
now substitute these Gaussian pdfs into Bayes's Eq.~\eqref{eq:Posterior}, to obtain the desired posterior distribution; it is the product of
three Gaussian pdfs and is thus itself Gaussian
\begin{align}
\label{eq:logp}
\ln p(\x_{K:0}\vert\y_{K:0}) & \propto -\left[ \frac{1}{2}\sum_{k=0}^K\left\| \y_k - {\mathcal H}_k(\x_k)\right\|_{\bR^{-1}_k}^2 +\frac{1}{2}\sum_{k=1}^K\left\| \x_k - {\mathcal M}_{k:k-1}(\x_{k-1})\right\|_{\bQ^{-1}_k}^2 +\frac{1}{2}\left\| \x_0 - \x^\rmb \right\|_{\bB^{-1}}^2\right] , \\
\label{eq:Jw4DVar}
 & = -{\mathcal J}^{\mathrm{w4DVar}}(\x_{K:0})
\end{align} 
which provides the cost function already given in Eq.~\eqref{eq:w4dvar}.
Maximizing $\ln p(\x_{K:0}\vert\y_{K:0}) $, i.e., finding the most likely trajectory (the analysis $\x^{\rm a}_{K:0}$) is equivalent to minimizing the term between the squared brackets. Equation~\eqref{eq:Jw4DVar} defines the cost function of the {\it weak-constraint 4DVar} \citep[e.g.][]{zupanski1997general,vidard2004variational,tr2006accounting,tremolet2007model}. When both the dynamical and observational models are linear (i.e., ${\mathcal M}_{k:k-1}={\mathbf M}_{k:k-1}$ and ${\mathcal H}_{k}={\mathbf H}_{k}$) the solution, $\x^{\rm a}_{K:0}$ coincides with the solution of the Kalman smoother Eqs.~\eqref{eq:KS-backrec0}--\eqref{eq:KS-backrec2}. In the general nonlinear case, the minimization can be carried out iteratively using the gradient of the cost function as described in Appendix {\color{red} B}. 
The control variable for the minimization is the entire trajectory, $\x_{K:0}$, and the estimate at each time $t_k$ is influenced by all observations in the window, before and after $t_k$. The gradient \citep[not shown; see][]{tr2006accounting, CV2010} with respect to $\x_k$ includes a term proportional to the innovation, $\y_k - {\mathcal H}_k(\x_k)$, because data are assumed to be available at each time step. In the general case, when the model time discretization is finer than the observational temporal density, the corresponding innovation term disappears. 
Since the weak-constraint 4DVar solves for the full trajectory $\x_{K:0}$ within the window, it does not need the assumption of uncorrelated model error. Nevertheless, if the model error is assumed uncorrelated in time, the cost function requires only the model error spatial covariances at times $t_k$ and not their temporal correlation, and the model error cost function term (the second one in the rhs of Eq.~\eqref{eq:logp}) reduces to a single summation over the time steps weighted by the
inverse of the model error covariances \citep{tr2006accounting}. Even in this case however, the size of the control variable is huge, $\x_{K:0}\approx\mathcal{O}(m\times K)$ (i.e. the
state vector dimension times the temporal steps in the assimilation window), and it is difficult to estimate reliably the model error covariances, $\bQ_k$ (cf.\ Appendix
{\color{red} A}). The development of specific solutions have thus been necessary to make the weak-constraint 4DVar feasible in the geosciences \citep[see,
e.g.,][]{griffith2000adjoint,vidard2004variational,tremolet2007model}, and its use in operational settings has only been very recent \citep{lindskog2009weak,fisher2011weak,ngodock20144dvar}. 
The weak-constraint 4DVar in the presence of time correlated model error is described in \citet{CV2010}, while its hybrid ensemble-variational formulation (cf.\ Sect.~\ref{sec:EnVar}) has been studied recently by \citet{amezcua2017weak}. An interesting alternative conceptualization of the weak-constraint 4DVar is described in \citet{ye2015improved} that uses path-integral and annealing methods to solve the problem.

In the more widely used {\it strong-constraint 4DVar} it is assumed that the model is perfect, which reduces the cost function to
\be
\label{eq:J-strong}
{\mathcal J}^{\mathrm{s4DVar}}(\x_{0}) = \frac{1}{2}\sum_{k=0}^K\left\| \y_k - {\mathcal H}_k\circ{\mathcal M}_{k:0}(\x_0)\right\|_{\bR^{-1}_k}^2 +\frac{1}{2}\left\| \x_0 - \x^{{\mathrm b}} \right\|_{\bB^{-1}}^2, 
\ee
with the $\circ$ symbol representing the composition of operators and the control variable is now ``only'' the system's state at the beginning of the window, $\x_0$. 
Similarly to the equivalence between the solutions of the weak-constraint 4DVar and of the Kalman smoother, in the linear case the solution of the strong-constraint 4DVar at the end of the window, $\x_{K}$, will be equivalent to that of a Kalman filter, Eqs.~\eqref{eq:KF-fcst1}--\eqref{eq:KF-anl3}, that had started from the same initial condition and had assimilated the same observations. 
The gradient of Eq.~\eqref{eq:J-strong} can be obtained by applying Eq.~\eqref{eq:grad-strong} and reads
\be
\label{eq:grad-s4DVar}
\nabla_{\x_0}{\mathcal J}^{\mathrm{s4DVar}}(\x_{0}) = -\sum_{k=0}^K \bM^{{\mathrm T}}_{k:0} \bH_k^{{\mathrm T}}\bR^{-1} \left[\y_k - {\mathcal H}_k\circ{\mathcal M}_{k:0}(\x_0)\right] + \bB^{-1} (\x_0 - \x^{{\mathrm b}}),
\ee
which reveals the effects of the adjoint operator, $\bM^{{\mathrm T}}_{k:0} $, to project backward the influence of the observation at any $t_k$. 
The final analysis is obtained using an iterative process starting by providing a first guess, $\x_0^{i=0}$, and evaluating ${\mathcal J}^{\mathrm{s4DVar}}(\x^i_{0})$ and the
gradient at each iteration. As explained in Appendix {\color{red} B}, the process is then repeated until the convergence criteria are met. Then using the final minimum state
at $\x^{\rm a}_0$ the analyzed trajectory is obtained over the entire window, $\x^{\rm a}_{K:0}$, by integrating the deterministic model from $t_0$ to $t_K$. 
This trajectory represents the most likely model solution fitting the observations within the window and given the prescribed initial condition and data uncertainty. An extension of the strong-constraint 4DVar to account for time-correlated observational error in an operational weather forecast context has been introduced by \cite{jarvinen1999variational}.

Appendix {\color{red} C} discusses in more details some key properties of the variational methods, while three popular approximations of the Kalman filter and variational methods, namely the {\it extended Kalman filter}, the {\it incremental 4DVar} and the {\it 3DVar} are described in Appendix {\color{red} D}.
Finally, a schematic illustration of the different problems tackled by the weak-constraint 4DVar, strong-constraint 4DVar, and 3DVar (cf.\ Appendix {\color{red} D}) is given in Fig.~\ref{fig:Fig2}.
\begin{figure}[b!]
\includegraphics[width=16cm, height=21cm]{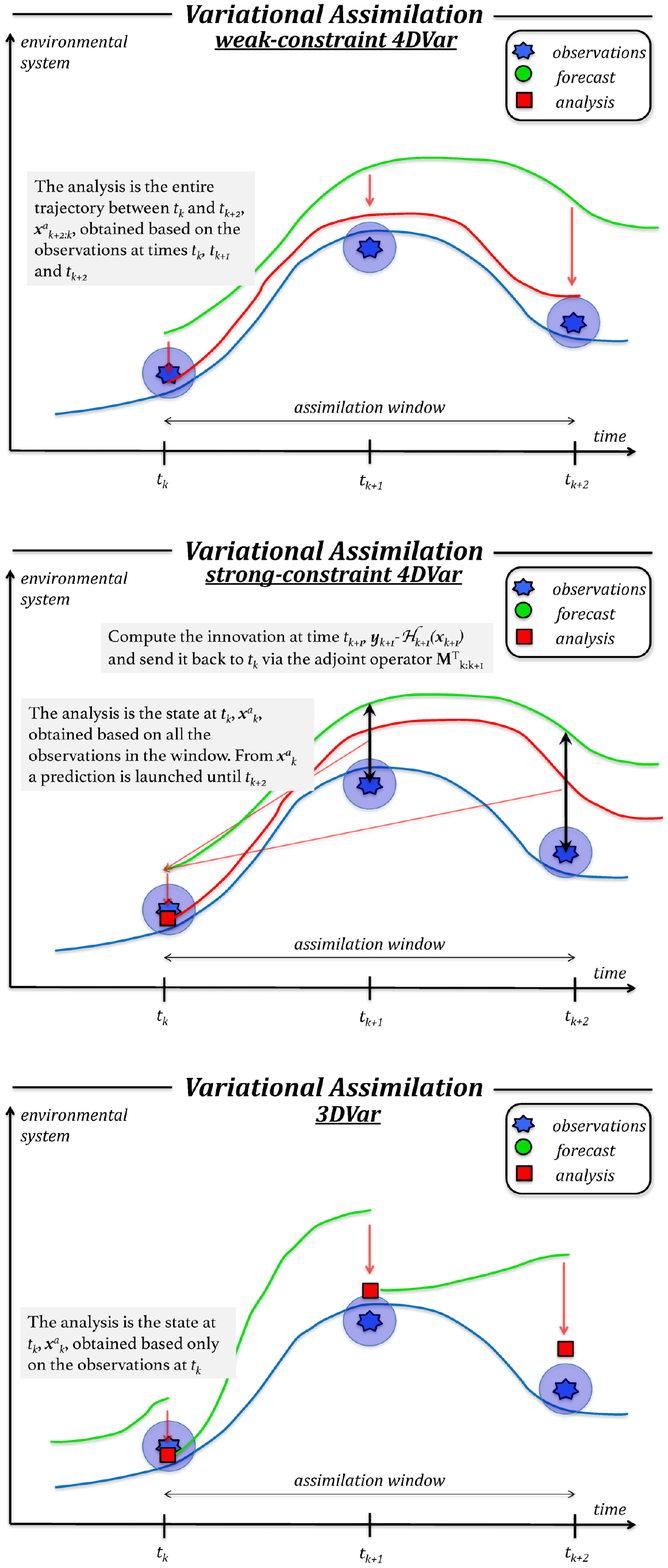}
\caption{Caption next page.}
\end{figure}
\addtocounter{figure}{-1}
\begin{figure}[t!]
\caption{\label{fig:Fig2}Illustration of the three variational problems: {\bf weak-constraint 4DVar} (w4DVar, top panel), {\bf strong-constraint 4DVar} (mid panel) and {\bf 3DVar} (bottom panel). In the w4DVar the control variable is the entire trajectory within the window (from $t_k$ to $t_{k+2}$ in Fig.~2), $\x_{k+2:k}$, so the corrections (continuous red arrow) are computed and utilized at the observation times. The analysis trajectory (red line) is moved toward the observations. In the s4DVar, the control variable is the state at the beginning of the window, $\x_k$, and the corrections are computed at observation times (black arrows) but then propagated back to $t_k$ (red arrows) using the adjoint model (see Eq.~\eqref{eq:grad-s4DVar}): once the analysis at the initial time is computed, it is then used as initial condition to run a forecast until $t_{k+2}$. In the 3DVar the corrections are computed and utilized at each observation times $t_k$ sequentially, so that the analysis at $t_k$, ${\bf x}^{\rm a}_k$, takes into account only the observations at $t_k$ and is then used as initial condition for a forecast until the next observations at $t_{k+1}$ and so on. Variational methods do not explicitly compute an estimate of the uncertainty (cf. Appendix {\color{red} B}) so the colorful ellipsoids of Fig.~1 are not present here.}
\end{figure}

\section{\sffamily \Large Ensemble methods}
\label{sec:EnsMeth}

A Monte Carlo approach is at the basis of a class of algorithms referred to as {\it ensemble-based methods} \citep{evensen2009} of which the {\it ensemble Kalman filter}, EnKF, and the {\it ensemble Kalman smoother}, EnKS, are the most celebrated examples.  
The EnKF and EnKS are still based on the Gaussian hypothesis: their analysis update steps are only based on the mean and covariance. The transition density is, however, approximated using an ensemble of realizations of the prior $p(\x_{k}\vert\x_{k-1})$, in practice by running an ensemble of trajectories of the model dynamics. The EnKF and its variants \citep[][and  cf.\ Sect.~\ref{sec:EnKF}--\ref{sec:EnKFSQRT}]{evensen2009,asch2016data} are nowadays among the most popular approaches for DA in high-dimensional systems and have been successfully applied in atmospheric and oceanic contexts. Evidence has emerged that a small number of particles, typically $100$, is sufficient in many applications, hence making EnKF feasible in situations where the forward step of the DA is computationally expensive. 
A version of the EnKF is operational for the atmospheric model at the Canadian Meteorological Centre, CMC \citep{houtekamer2005atmospheric}
Also, the EnKF is used in the ocean system TOPAZ (see Sect.~\ref{sec:TOPAZ}) developed at the Nansen Environmental and Remote Sensing Centre \citep[NERSC,][]{sakov2012topaz4}. The meteorological office of Norway is running an operational
version of TOPAZ and produces operational forecasts for the Arctic Ocean, sea ice, and ecosystem,
for the European Copernicus Marine Services (\href{www.marine.copernicus.eu}{www.marine.copernicus.eu}).

There have been numerous alternative EnKF-like methods, and we will not attempt here to discuss them all; a recent review of the EnKF for atmospheric DA is given by \citet{houtekamer2016review}. However the many different formulations fall almost all into two main categories, namely the {\it stochastic} and the {\it square-root} (also called {\it deterministic}) ones, and this is the subject of the next sections. We shall then discuss the EnKS in Sect.~\ref{sec:EnKS} and devote Sect.~\ref{sec:Infl-Loc} to the description of the important countermeasures that must be put in place to make the EnKF and EnKS work in realistic high-dimensional nonlinear cases. This part of the article closes by presenting an outlook of the current frontier of the hybrid ensemble-based and variational methods in Sect.~\ref{sec:EnVar}.  

\subsection{\sffamily \large Stochastic ensemble Kalman filter \label{sec:EnKF}}
The EnKF as introduced by \citet{eve94a,bur98a,eve03a} is based on the use of an 
ensemble representation for error statistics.
Thus, instead of storing a full covariance matrix,  we can represent the same error statistics using
an appropriate ensemble of model states.   Given an error covariance matrix,
an ensemble of finite size provides an approximation of the error covariance matrix, and,   
as the size $N$ of the ensemble increases, the errors in the Monte Carlo sampling
decrease proportionally to $1/\sqrt{N}$.  

The derivations that follow will assume a linear observation operator for the sake of simplicity. 
However ensemble-based methods can straightforwardly incorporate a nonlinear observation operator \citep{evensen2009}.

\subsubsection{Representation of error statistics}

Given an ensemble of model realizations at a certain time, it is convenient to introduce the ensemble matrix 
\be
 \label{eq:Efa}
 \bE^{{\rm f,a}} = [\x^{{\rm f,a}}_1, \ldots, \x^{{\rm f,a}}_N] \in \Re^{m\times N} , 
\ee
which contains the $N$ ensemble members in its columns.

The  error covariance matrices $\bP_k^\rmf$ and $\bP_k^\rma$ for the predicted and analyzed estimate
in the KF are defined in terms of the true state in the KF Eq.~(\ref{eq:KF-fcst2}--\ref{eq:KF-anl3}). 
However, since the true state is not known, we will instead define the ensemble-anomaly covariance matrices around the ensemble mean.
Given the ensemble mean
\be
\barx^{{\rm f,a}} = \frac{1}{N}\sum_{n=1}^N \x_n^{{\rm f,a}},
 \label{eq:Xfamean}
\ee
we define the ensemble-anomaly matrix as
\be
\X^{{\rm f,a}} = \frac{1}{\sqrt{N-1}}[\x^{{\rm f,a}}_1-\barx^{{\rm f,a}}, \ldots, \x^{{\rm f,a}}_N-\barx^{{\rm f,a}}] \in \Re^{m\times N}.
 \label{eq:Xfa}
\ee
We can now write the ensemble-based error covariance matrices for the forecast and analysis ensembles as
\begin{align}
\lp\bP^\rme\rp^\rmf &= (\bX^{\rmf})(\bX^{\rmf})^{{\rm T}} , \label{eq:Pf2}\\
\lp\bP^\rme\rp^\rma &= (\bX^{\rma})(\bX^{\rma})^{{\rm T}} ,  \label{eq:Pa2}
\end{align}
where the superscript ``$\rme$'' denotes that the quantities (e.g., matrices in Eq.~\eqref{eq:Pf2} and 
\eqref{eq:Pa2}) are estimated based on the ensemble. 

Equations~\eqref{eq:Pf2} and \eqref{eq:Pa2} embed an interpretation where
the ensemble mean is the best estimate, and the spread of the ensemble around the mean
is a natural definition of the error in the ensemble mean.
We can interpret an ensemble of model states as a Monte Carlo representation of the pdf $p(\x_0)$ in Eq.~(\ref{eq:Markov}).

\subsubsection{Prediction of error statistics}
Given a pdf $p(\x_0)$ at time $t_0$, the joint pdf from time $t_0$ to $t_k$ is
given by Eq.~(\ref{eq:Markov}), which involves a multiplication by the transition density defined in Eq.~(\ref{eq:trans}). 
It was shown in \citet{eve94a} that if the pdf $p(\x_0)$ at time $t_0$ is represented by an ensemble of model states,
then the multiplication with the transition density is equivalent to integrating each realization
of $\x$ according to the model equations as defined in Eq.~(\ref{eq:model}).
Thus, from a finite ensemble of model states, we can
compute an ensemble approximation of the joint pdf $p(\x_{k:0})$  for the time interval from $t_0$ to $t_k$. 

\subsubsection{Analysis scheme}
The novel idea of \citet{eve94a} was to design an alternative update scheme that could work directly on the ensemble and where
it was not necessary to compute the full covariances as defined by Eq.~(\ref{eq:Pf2}) and (\ref{eq:Pa2}). 
\citet{eve94a} showed that, by updating each individual ensemble member according to the standard KF
Eq.~(\ref{eq:KF-anl1}) and (\ref{eq:KF-anl2}), the updated ensemble will have the correct mean and covariance matrix for the update in agreement with
Eq.~\eqref{eq:KF-anl3} in the standard KF.  \citet{bur98a} further proved that, in order for the EnKF analysis error covariance, $\lp\bP^\rme\rp^\rma$, to be consistent with that of the KF, it is essential to treat the observations as
random variables having a distribution with mean equal to the observed value and covariance equal to $\bR$.
Thus, given a vector of observations, $\y \in \Re^d$, we define an ensemble of perturbed observations
\begin{equation}\label{eq:dj}
   \y_n  = \y + \epsi_n, \qquad 1\le n\le N ,
\end{equation}
which we store in the observation matrix
\begin{align}
 \bY_\rmo&=[\y_1, \ldots , \y_N] \in \Re^{d\times N} ,\label{eq:Eps} 
\end{align}
which columns are the perturbed measurements $\y_n \in \Re^d$.  Then we define the corresponding 
matrix of the normalized anomaly ensemble of the observations $\dY_\rmo \in \Re^{d\times N}$ as
\begin{equation}
\begin{split}
 \dY_\rmo&=\frac{1}{\sqrt{N-1}}[\y_1-\y, \ldots , \y_N-\y] \\ 
         &= \frac{1}{\sqrt{N-1}}[\epsi_1, \ldots , \epsi_N]  . 
\end{split}
\label{eq:Eps-norm} 
\end{equation}

By subtracting any nonzero mean from the $N$ random draws $\epsi_n$, we ensure that the simulated random measurement errors have
zero ensemble-mean and thus the random perturbations do not introduce any bias in the update.
Next we define the ensemble covariance matrix of the measurement errors as
\begin{equation}
\bR^\rme= \dY_\rmo(\dY_\rmo)^\rmT.
\label{eq:obsenscov}
\end{equation}
In the limit, $N\rightarrow\infty$, of infinite ensemble size, this matrix converges
to the prescribed error covariance matrix $\bR$ used in the KF.

The analysis step in the EnKF consists of updates performed on each of the ensemble members, as given by
\begin{equation}
\x^\rma_{n} = \x^\rmf_{n} + (\bP^\rme)^\rmf \bH^\rmT \Bigl[ \bH (\bP^\rme)^\rmf \bH^\rmT + \bR^\rme \Bigr] ^{-1}
      \Bigl[ \y_{n} -\bH \x_{n} \Bigr], \qquad 1\le n\le N .
\label{eq:ekfps}
\end{equation}
With a finite ensemble size, the use of the ensemble covariances introduces an approximation of the true covariances.  Furthermore,
if the number of measurements is larger than the number of ensemble members,
then the matrices $\bH (\bP^\rme)^\rmf \bH^\rmT$ and $\bR^\rme$ are singular, and we must use a pseudo-inversion.  

Equation~(\ref{eq:ekfps}) implies the update of the ensemble mean 
\begin{equation}
\ol{\x^\rma}  = \ol{\x^\rmf} 
                  +  (\bP^\rme)^\rmf \bH^\rmT \Bigl[ \bH (\bP^\rme)^\rmf \bH^\rmT + \bR^\rme \Bigr]^{-1}
                     \Bigl[\ol{\y} -\bH \ol{\x^\rmf}  \Bigr]  ,
\label{eq:ekfpsa}
\end{equation}
where $\ol{\y}=\y$ since the measurement perturbations have ensemble mean equal to zero.
Thus, the relation between the analyzed and predicted ensemble mean
is identical to the relation between the analyzed and predicted state
in the standard KF, apart from the use
of $(\bP^\rme)^\rmf$ and $\bR^\rme$ instead of $\bP^{\rmf}$ and $\bR$.

If the ensemble updates $\x^\rma_{n}$ and $\ol{\x^\rma}$ from Eqs.~(\ref{eq:ekfps}) and (\ref{eq:ekfpsa}) 
are inserted back into the analyzed covariance Eq.~(\ref{eq:Pa2}), it was shown by \citet{eve94a} and \citet{bur98a} that
\begin{equation}
\lp\bP^\rme\rp^\rma = [ \bI - \bK^\rme \bH ] (\bP^\rme)^\rmf  ,
\label{eq:anacov}
\end{equation}
with $\bK^\rme$ being the ensemble-based Kalman gain matrix (cf.\ Sect.~\ref{sec:KF} and  Eq.~\eqref{eq:KF-anl1}). 
The result in Eq.~\eqref{eq:anacov} proves a consistency between the original KF and the EnKF (cf.\ Eq.~\eqref{eq:KF-anl3}).

Note that the EnKF analysis scheme is approximate since it does not correctly account for non-Gaussian contributions in the predicted ensemble. 
In other words, the EnKF analysis scheme does not solve the Bayesian update equation for a non-Gaussian pdf.  
On the other hand, the EnKF analysis scheme is not just a re-sampling of a Gaussian posterior distribution.  
Only the updates defined by the right-hand side of Eq.~(\ref{eq:ekfps}), which
we add to the prior non-Gaussian ensemble, are linear.
Thus, the updated ensemble inherits many of the non-Gaussian
properties from the forecast ensemble.   In summary, we have a computationally efficient analysis scheme
where we avoid re-sampling of the posterior.

\subsubsection{Formulation in terms of the ensemble}
\citet{eve03a} reformulated the EnKF analysis scheme in terms of the ensemble without reference to the ensemble covariance matrix,
which allows for an efficient numerical implementation and alternative interpretation of the method. 
We follow the slightly updated formulation from \citet{evensen2009}.

The analysis equation (\ref{eq:ekfps}), expressed in terms of the ensemble matrices, is
\begin{equation}
\bE^\rma=\bE^\rmf + (\bX^\rmf) (\bX^\rmf)^\rmT \bH^\rmT \left[ \bH (\bX^\rmf) (\bX^\rmf)^\rmT \bH^\rmT +\dY_\rmo\lp\dY_\rmo\rp^\rmT \right]^{-1} [\bY_\rmo - \bH  \bE^\rmf ],
\label{eq:ensanab}
\end{equation}
where we replace all the error covariance matrices by their ensemble representations.
\citet{eve03a} showed that the analysis equation (\ref{eq:ensanab}) could be written as
\begin{equation}
\label{eq:X5ana}
\bE^\rma = \bE^\rmf \bT,
\end{equation}
with the transformation matrix $\bT \in \Re^{N\times N}$ defined as
\begin{equation}
\bT= \bI_N + \bY^\rmT \bC^{-1} (\bY_\rmo - \bH  \bE^\rmf ).
\label{eq:Xdef}
\end{equation}
Here we have defined the observed ensemble-anomaly matrix 
\begin{equation}
\bY =   \bH \bX^\rmf \in \Re^{d\times N},
\label{eq:Sdef}
\end{equation}
and the matrix 
\begin{equation}
\bC =   \bY \bY^\rmT + \dY_\rmo \lp\dY_\rmo\rp^\rmT    \in \Re^{d\times d}, 
\label{eq:errcov1}
\end{equation}
while $\bI_N \in \Re^{N\times N}$ is the identity matrix. 
In practical implementations, it is common also to use the full-rank exact measurement error covariance matrix $\bR$ 
as an alternative to the product of measurement perturbations $\bR^\rme = \dY_\rmo \lp\dY_\rmo\rp^\rmT$,
although that comes at an additional computational cost unless one assumes $\bR$ to be diagonal \citep[see also][]{hoteit2015mitigating}.

\begin{figure*}[ph]
\begin{center}
\begin{tabular}{c}
\includegraphics[height=5.7cm, width=0.85\textwidth]{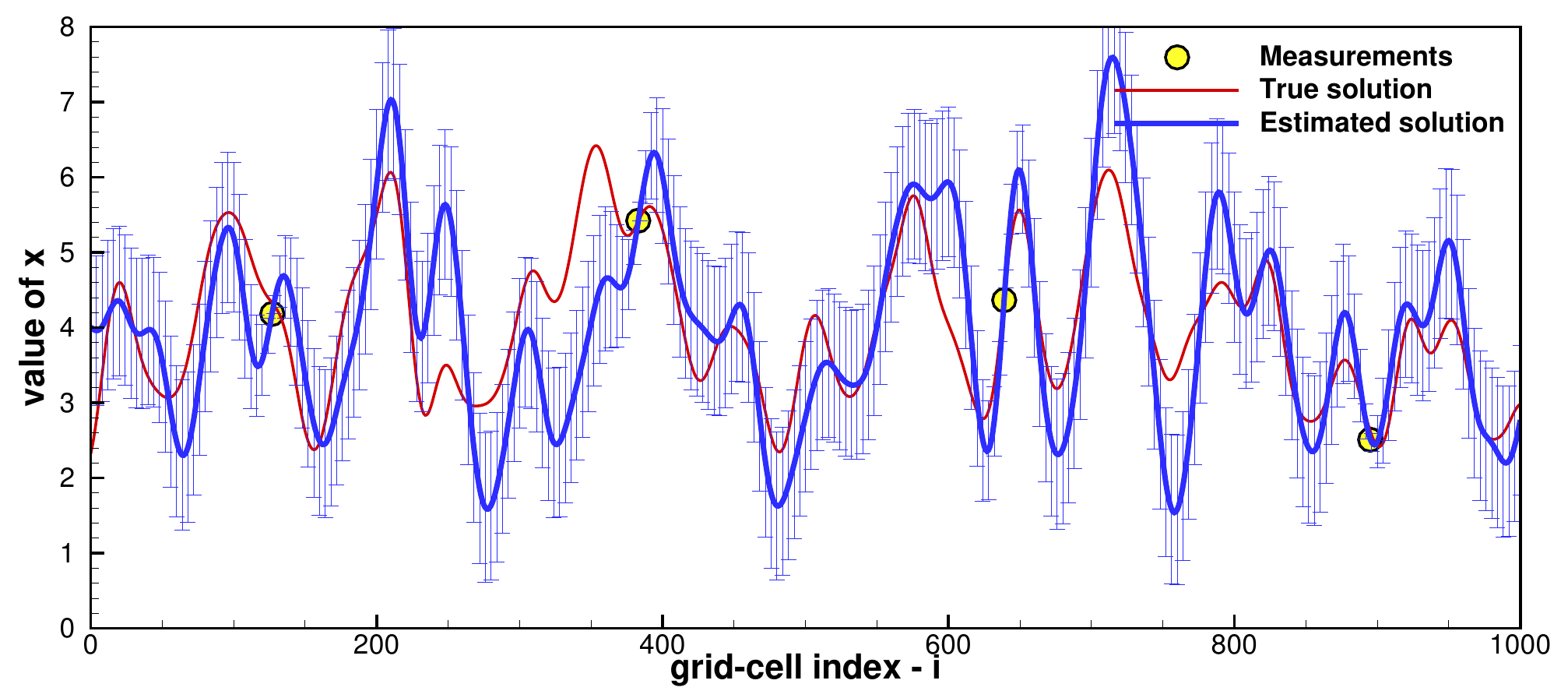}\\
\includegraphics[height=5.7cm, width=0.85\textwidth]{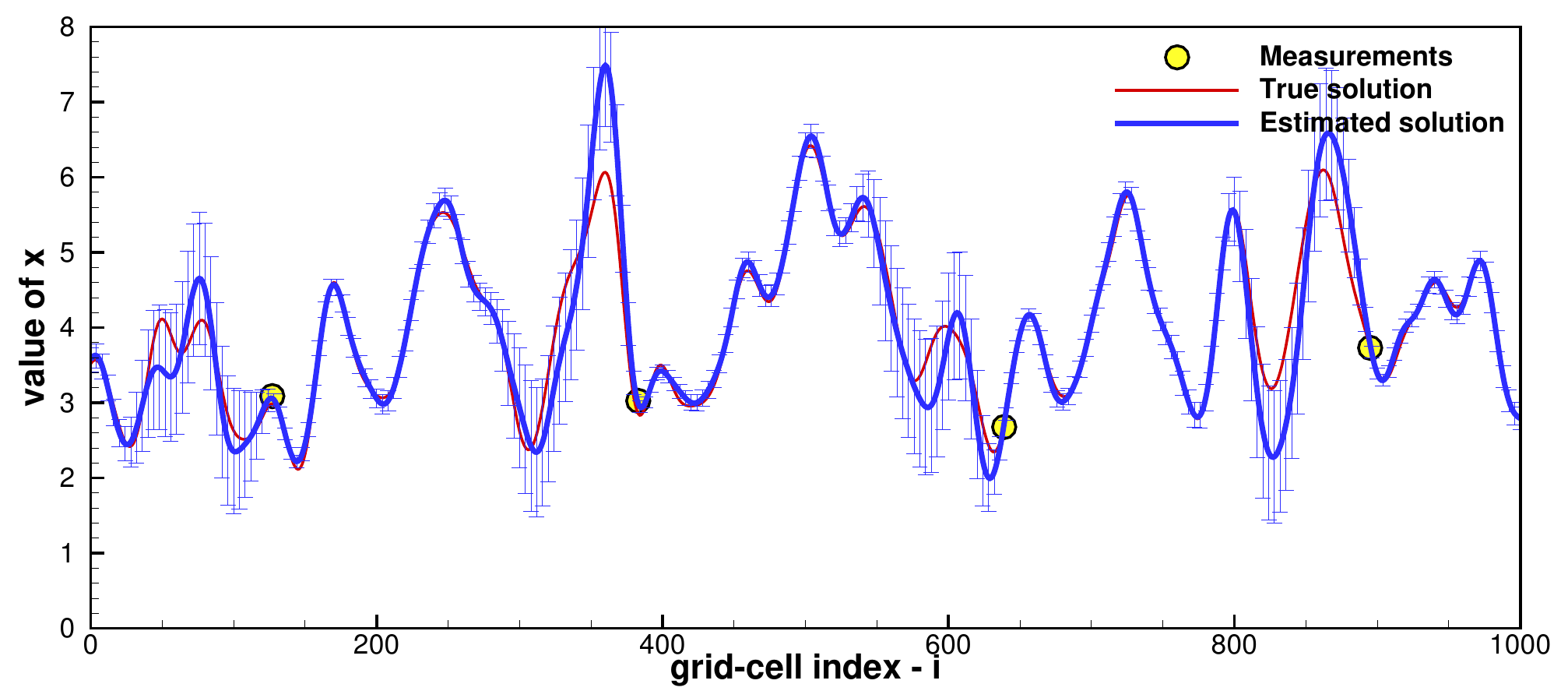}\\
\includegraphics[height=5.7cm, width=0.85\textwidth]{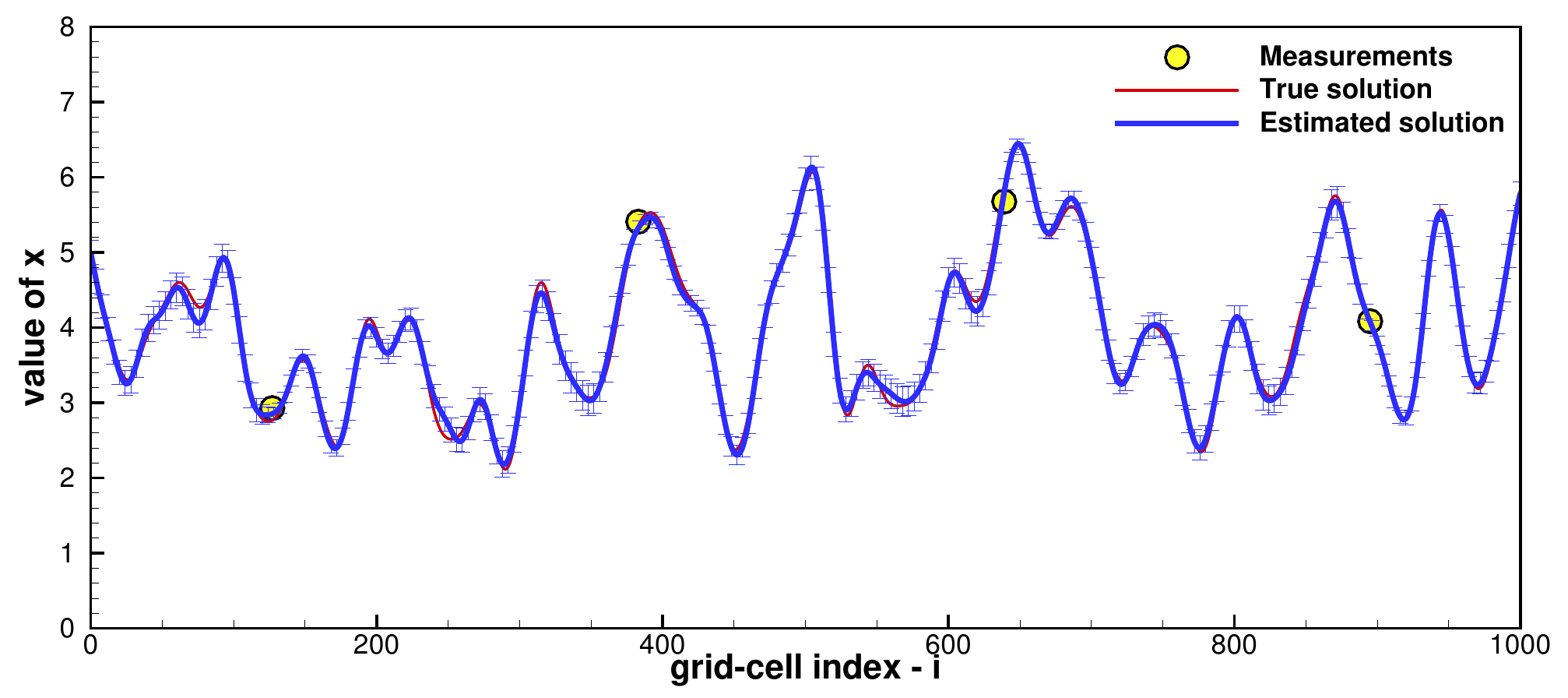}
\end{tabular}
\end{center}
\caption{A linear advection equation on a periodic domain (traveling wave from left to right) illustrates how
the KF and/or EnKF estimates the state at three different times, namely, (top) $t=5$, (middle) $t=150$, and (bottom) $t=300$.
The plots show the reference/true solution (red line), measurements (yellow circles), and error estimate (error-bars equivalent to one standard deviation of forecast error variance). The initial condition, observation and model error variance is $1.0$, $0.01$ and $0.0001$ respectively, and the ensemble size is $N=500$. In each of the panels we see that the error increases downstream, but also that downstream of an observation its effect is propagated by the wave and the errors are smaller on the right than on the left of the observation point. 
Similarly, by comparing the three panels for increasing time, we see how the errors decreases as a function of time, indicating a good performance of the filter. 
\label{fig:advsolB}}
\end{figure*}

The significant result from Eq.~(\ref{eq:X5ana}) is that the EnKF update ensemble becomes a combination of the forecast ensemble members,
and is searched within the space spanned by the forecast ensemble.  
It is clear that the formulation in Eq.~(\ref{eq:X5ana}) is a stochastic scheme due to the use of randomly perturbed measurements. 
The use of perturbed measurements allows for a natural interpretation of the EnKF as a Monte Carlo algorithm while making it easy to
understand and implement. 
An illustration is provided in Fig.~\ref{fig:advsolB}, which shows the EnKF mean solution and estimated error, as a function of the model grid and for three successive analysis times. 

An efficient and stable numerical implementation of the analysis scheme is discussed in \citet{evensen2009}, including the case in which
$\bC$ is singular or of low rank, e.g., due to the number of measurements being larger than the number of realizations or if measurements are linearly dependent on each other. 

In practice, the ensemble size is critical since the computational cost scales linearly with the number of
realizations.  That is, each realization needs to be integrated forward in time.  The cost associated with the ensemble
integration motivates the use of an ensemble with the minimum number of realizations that can provide acceptable accuracy.
We will see in Sect.~\ref{sec:AUS} how the optimal ensemble size to achieve satisfactory performance is also related to the 
dynamical properties, notably the degree of instabilities, of the model.  

The convergence of the solution of the stochastic EnKF for $N\rightarrow\infty$ has been studied by \cite{le2009large}: the rate of such convergence was proven to be $1/\sqrt{N}$, and in the linear dynamics and observation models case, the solution approaches that of the KF. In the general nonlinear case, and even when at least the observation operator is linear, the asymptotic solution differs from the fully Bayesian filter, and for any finite value of $N$, the individual ensemble members are not independent. 

There are two significant sources of sampling errors in the EnKF, namely (i) the use of a finite
ensemble of model realizations and (ii) the introduction of stochastic measurement perturbations \citep{eve04a,evensen2009}.
Besides, stochastic model errors influence the predicted error statistics, which are approximated by the ensemble. 
The stochastic perturbation of measurements used in EnKF can be avoided using a
square-root implementation of the analysis scheme, which is the topic of the next Section.

\subsection{\sffamily \large Deterministic square-root schemes \label{sec:EnKFSQRT}}
The perturbation of measurements used in the EnKF standard analysis Eq.~(\ref{eq:ensanab}) is an additional source of sampling error. 
However, some papers have introduced methods alternative to the traditional EnKF, i.e., \textit{square-root schemes},
which compute the analysis without perturbing the measurements and preserve precisely the posterior variance in the update.
Examples include the Ensemble Transform Kalman Filter (ETKF) by \citet{bis01a,whi02a}, 
the Ensemble Adjustment Kalman Filter (EAKF) by \citet{and01a},
and the Maximum Likelihood Ensemble Filter (MLEF) by \citet{zupanski2005}. 

\subsubsection{Updating the mean}
In a square-root scheme, the analyzed ensemble mean is computed from the standard KF analysis equation,
\begin{equation}
 \ol{\x}^\rma = \ol{\x^\rmf} + \bX^\rmf \bY^\rmT \bC^{-1} \Bigl(\ol{\y} - \bH \ol{\x^\rmf}\Bigr),
\label{eq:meanupdate}
\end{equation}
with the matrix $\bC$ defined in Eq.~\eqref{eq:errcov1} but with $\dY_\rmo \lp\dY_\rmo\rp^\rmT$ replaced by $\bR$, i.e.,
\begin{equation}
\bC =   \bY \bY^\rmT + \bR .  
\label{eq:errcov2}
\end{equation}
The ensemble mean in the square-root formulation is thus computed without the use of perturbed observations.

\subsubsection{Updating the ensemble perturbations}
We derive the deterministic algorithm for updating the ensemble perturbations starting from the traditional analysis equation~(\ref{eq:KF-anl3}) for the covariance update in the KF but
written in its ensemble-based representation in Eq.~\eqref{eq:anacov}.  We then insert the expression for the Kalman gain matrix to obtain
\begin{equation}
 \bigl(\bP^\rme\bigr)^\rma =
 (\bP^\rme)^\rmf -  
(\bP^\rme)^\rmf \bH^\rmT
      \Bigl[\bH (\bP^\rme)^\rmf \bH^\rmT + \bR\Bigr]^{-1}
 \bH (\bP^\rme)^\rmf.
\label{eq:kfanalysis}
\end{equation}
Using the ensemble representation of the error covariance matrices, Eqs.~\eqref{eq:Pf2} and \eqref{eq:Pa2}, along with Eqs.~\eqref{eq:Sdef} and \eqref{eq:errcov2}, we can 
rewrite Eq.~\eqref{eq:kfanalysis} as
\begin{equation}
\lp\bX^{\rma}\rp \lp\bX^{\rma}\rp^{\rmT} = 
(\bX^\rmf)\Bigl[ \bI_N - \bY^\rmT\bC^{-1}\bY \Bigr](\bX^\rmf)^{\rmT} = (\bX^\rmf)\Bigl[ \bI_N + \bY^\rmT\bR^{-1}\bY \Bigr]^{-1}(\bX^\rmf)^\rmT.
\label{eq:kfanalysis2} 
\end{equation}
We obtain the second expression by making use of the Sherman-Morrison-Woodbury formula \citep[see][their Sect.~2.1.4]{golub2013matrix}.

We can derive the square-root filters by factorizing the symmetric expression in Eq.~(\ref{eq:kfanalysis2}). 
The simplest, but not a very efficient approach, is to compute the eigenvalue factorization of the matrix 
$\bI_N - \bY^\rmT\bC^{-1}\bY = \bZ \bLam \bZ^\rmT$, and then the update as
\begin{equation}
 \bX^{\rma} = \bX^\rmf \bZ \sqrt{\bLam},
\label{eq:sqrtsimple}
\end{equation}
which defines a symmetric factorization of Eq.~(\ref{eq:kfanalysis2}).
However, this update equation does not preserve the mean (it is biased).
It was shown \citep{wang2004better, sak08b, liv08a} that the \textit{symmetric square root} preserves the zero mean in the updated perturbations.
Accordingly, the update in Eq.~(\ref{eq:sqrtsimple}) must be replaced  by 
\begin{equation}
 \bX^{\rma} = \bX^\rmf \bZ \sqrt{\bLam} \bZ^\rmT,
\label{eq:sqrtsym}
\end{equation}
which is another symmetric factorization of Eq.~(\ref{eq:kfanalysis2}) \citep[see, e.g.,][]{hunt2007}.
Consequently, if the predicted ensemble members have a non-Gaussian distribution, then the updated distribution retains the shape,
although the variance is reduced \citep[see the review in][their Sect.~4]{raanes2015extending}. The importance of the symmetry in the square-root methods was first recognized by \citet{ott2004} and then adopted by \citet{hunt2007}.

The operational use of the square-root schemes requires more focus on numerical stability and efficiency. Here we assumed that $\bC^{-1}$ exists,
which is not always the case and particularly not so when the number of measurements is larger than the ensemble size nor when $\bR$ is not full rank. 
For a more elaborate derivation of numerically stable schemes, we refer to the initial publications by \citet{and01a,whi02a,bis01a} and the reviews by
\citet{tip03a} and \citet{nerger2012unification}. \citet{eve04a,eve09a,evensen2009} derived a numerically efficient square-root filter that computes the inversion in the
subspace spanned by the measured ensemble perturbations $\bY$. The scheme works equally well with a non-diagonal measurement error-covariance matrix and in the case when $\bC$ is of low rank. 

Equation~\eqref{eq:kfanalysis2} is also known as {\it right transform} as it applies to the right-hand side of the anomalies, in the $N$-dimensional 
ensemble space \citep{asch2016data}. Similarly, {\it left transform} expressions also exist, that apply to the $m$-dimensional state-space
\citep[see][their Sect.~6.4, for an extensive treatment of the subject]{asch2016data}.

A randomization of the analysis update can also be used to generate updated perturbations that better resemble a Gaussian distribution, see \citet{eve04a}.
Thus, we write the symmetric square root solution Eq.~(\ref{eq:sqrtsym}) as
\begin{equation}
\bX^{\rma} =  \bX^\rmf \bZ \sqrt{\bLam} \bZ^\rmT \bPhi^\rmT,
\label{eq:sqrtanaD}
\end{equation}
where $\bPhi$ is a mean-preserving random orthogonal matrix, which can be computed using the algorithms from \citet{pham2001stochastic} or \citet{sak08a}.
Note again that the random
rotation in the square-root filter, contrary to the measurement perturbation used in EnKF, eliminates all previous
non-Gaussian structures from the forecast ensemble.
Efficient means to account for stochastic model noise within square-root filters are discussed in \citet{raanes2015extending}.

\subsection{\sffamily \large Ensemble Kalman Smoother (EnKS)\label{sec:EnKS}}
The EnKS is a straightforward extension of the EnKF in which we use the output of the latter 
as a prior in the  EnKS. Like the EnKF uses the ensemble covariances
in space to spread the information from the measurements, the EnKS uses
the ensemble covariances to spread the information in both space and time (also backward).
We need to reintroduce here the time index. 

Assume we have measurements available at discrete times $t_k$, $k=1,\ldots,K$. We compute the EnKF solution recursively
by assimilating the observations when they are available and then propagate the ensemble until the next measurements (cf.\ Sec.~\ref{sec:EnKF}). 
The members of the EnKF can be stored at any time instant when we desire  a smoother update.

The analysis update at a time $t_k$ (which does not have to be an observation time)
from measurements available at a later time $t_j$, $1\le k<j\le K$, reads
\begin{equation}
\bE^\rma_k=\bE^\rmf_k+\bX^\rmf_k \bY_j^\rmT \bC_j^{-1} (\bY_{\rmo,j} - \bY_j) \quad 1\le k<j\le K,
\label{eq:enksana1}
\end{equation}
where $\bY_j$ from Eq.~(\ref{eq:Sdef}), and $\bC_j$ from Eq.~(\ref{eq:errcov1}) are
evaluated using the ensemble and measurements at $t_j$. 
From the right-hand side of Eq.~(\ref{eq:enksana1}) we recognize the 
product $\bX^\rmf_k \bY_j^\rmT$ as the covariance between the predicted measurements at $t_j$ and 
the model state at $t_ < t_j$.

The update at the time $t_k$ uses precisely the same combination of ensemble members
as was defined by $\bT_k$ in Eq.~(\ref{eq:Xdef}) for the EnKF analysis at the time $t_k$. 
Thus, we can compactly write the EnKS analysis as
\begin{equation}
 \bE^{\rma,{\rm EnKS}}_k = \bE^{\rma,{\rm EnKF}}_k \prod_{j=\hat{k}}^K   \bT_j,
\label{eq:EnKSana2}
\end{equation}
where $\hat{k}$ corresponds to the first data time after $t_k$, and $t_K$ is the last measurement time.
It is then a simple exercise to compute the EnKS analysis as soon as the EnKF solution has been found.
The computation requires only the storage of the transform matrices $\bT_k \in \Re^{N\times N}$, for $k=1,\ldots,K$, 
and the EnKF ensemble matrices for the times when we want to compute the EnKS analysis.
Note that the EnKF ensemble matrices are large, but it is possible to store only specific variables at selected locations where the EnKS solution is needed.
The equivalence between the formulation of the EnKS given above with the Rauch-Tung-Striebel smoother (cf.\ Sect.~\ref{sec:KS})
even in the nonlinear, non-Gaussian, case is discussed in \citet{raanes2016ensemble}.

An example of implementation of an ensemble smoother is represented by the "no cost smoother" (aka as {\it Running in place}, RIP) of \citet{kalnay2010} that provides
explicitly the weights of the ensemble members at the end of the assimilation window, and then applies those same weights throughout the window. The RIP/no-cost smoother has
been used to accelerate the spin up of the Local Ensemble Transform Kalman Filter \citep[LETKF,][]{hunt2007}, to increase its accuracy and to allow longer time windows. The no
cost smoother designed originally designed for the LETKF has been adapted to the Ensemble Square-Root Filter (EnSFR) by \citet{wang2013iterative} and is used operationally to
accelerate the spinup of the EnSRF forecast of tornadoes and severe storms at the {\it Center for Analysis and Prediction of Storms} of the University of Oklahoma
(\href{www.caps.ou.eu}{www.caps.ou.edu}). A recent study that analyses and compares ensemble smoothers for solving inverse problems can be found in \citet{eve18b}.

\subsection{\sffamily \large Making it work: Localization and Inflation}
\label{sec:Infl-Loc}

The reduction of dimensionality obtained through the Monte Carlo treatment of the KF that yielded the EnKF
comes at a price. Indeed, with a limited number, $N$, of anomalies, the sample covariance matrix is severely
rank-deficient. It would be surprising if such sample error covariance matrix could be a good substitute for the, possibly,
full-rank true one.

Assume that the true error covariance matrix is $\bB$. Further define $\bP^\rme$ as the sample covariance
matrix meant to approximate $\bB$ and obtained from $N$ samples, (as in Eq.~\eqref{eq:Pf2}) of the normal distribution with covariance matrix
$\bB$. Then it can be shown that for two distinct entry indices $i$ and $j$ corresponding to distinct locations:
\be
{\mathbb E}\([\bP^\rme-\bB]^2_{ij} \) = \frac{1}{N-1}\( [\bB]^2_{ij} + [\bB]_{ii}[\bB]_{jj} \) ,
\ee
with ${\mathbb E}$ indicating the expectation (average) operator.
In most spatial geophysical systems, $[\bB]_{ij}$ is expected to vanish fast (often exponentially) with the distance between sites
corresponding to $i$ and $j$.  By contrast, the $[\bB]_{ii}$ are the variances and remain finite, so that
\be
{\mathbb E}\([\bP^\rme-\bB]^2_{ij} \)  \underset{|i-j| \rightarrow \infty}{\sim}  \frac{1}{N-1} [\bB]_{ii}[\bB]_{jj} ,
\ee
meaning that both expressions coincide when the sites separation is sufficiently large.
Consequently, the right-hand-side arithmetically goes to $0$ with the ensemble size $N$.  Unfortunately, since
$[\bB]_{ij}$ is expected to vanish exponentially with the distance, we would have liked instead ${\mathbb
E}\([\bP^\rme-\bB]^2_{ij} \)$ to also vanish exponentially with the distance.  Hence, with $N$ finite (and usually 
$N\ll m$) the sample covariance $[\bP^\rme]_{ij}$ is
potentially a bad approximation of the vanishing true covariance, especially for large distances $|i-j|$. This
approximation generates spurious correlations between distant parts of the system, as a manifestation of the
rank-deficiency of the sample error covariance matrix.

The errors committed with such an approximation are usually referred to as \emph{sampling errors}.  When taking $\bP^\rme$ as
the forecast/background error covariance matrix, the accumulation of such errors over the EnKF data assimilation
cycles can be detrimental to the stability of the EnKF. Without counter-measures, the divergence of the EnKF in
high-dimensional geophysical systems is almost systematic.  Fortunately, more or less {\it ad hoc} fixes meant to
address the issue are known: \emph{localization} and \emph{inflation}. They are rather complementary and both of
them are often required.

\subsubsection{Localization}

Localization fundamentally relies on the exponential decrease of the correlations with the distance in geophysical
systems.  Under this condition, one can assume that the inter-dependence of distant parts of a physical system is
negligible. Hence, EnKF analyses could be made local \citep{houtekamer2001, hamill2001, hau02b, eve03a,
ott2004}. Localization comes in two flavors that we describe in the following.

\paragraph{Covariance localization}

The first approach, named \emph{covariance localization} (CL), seeks to regularize the sample error covariance matrix, with
the goal to mitigate the rank-deficiency of $\bP^\rme$ and the appearance of spurious error correlations. A mathematical means to
achieve this objective is to compute the Schur (or Hadamard) product of $\bP^\rme$ with a well chosen smooth correlation function
$\brho$.  We assume $\brho$ to have exponentially vanishing correlations for distant parts of the system and to be
representative of such dampening of real geophysical systems.  The Schur product of $\brho$ and $\bP^\rme$ is defined by
\be
\label{eq:schur}
\left[ \brho \circ \bP^\rme \right]_{ij} = [\brho]_{ij}[ \bP^\rme]_{ij} ,
\ee
i.e., a point-wise matrix multiplication. The Schur product theorem \citep{horn2012} ensures that this product is
positive semi-definite, and hence still represents a proper covariance matrix. For sufficiently regular
$\brho$, $\brho \circ \bP^\rme$ turns out to be full-rank (hence positive definite). Moreover, the spurious correlations
should be exponentially dampened as is obvious from Eq.~\eqref{eq:schur}.

\begin{figure}[H]
  \begin{tabular}{cc}
    \includegraphics[width=0.95\textwidth]{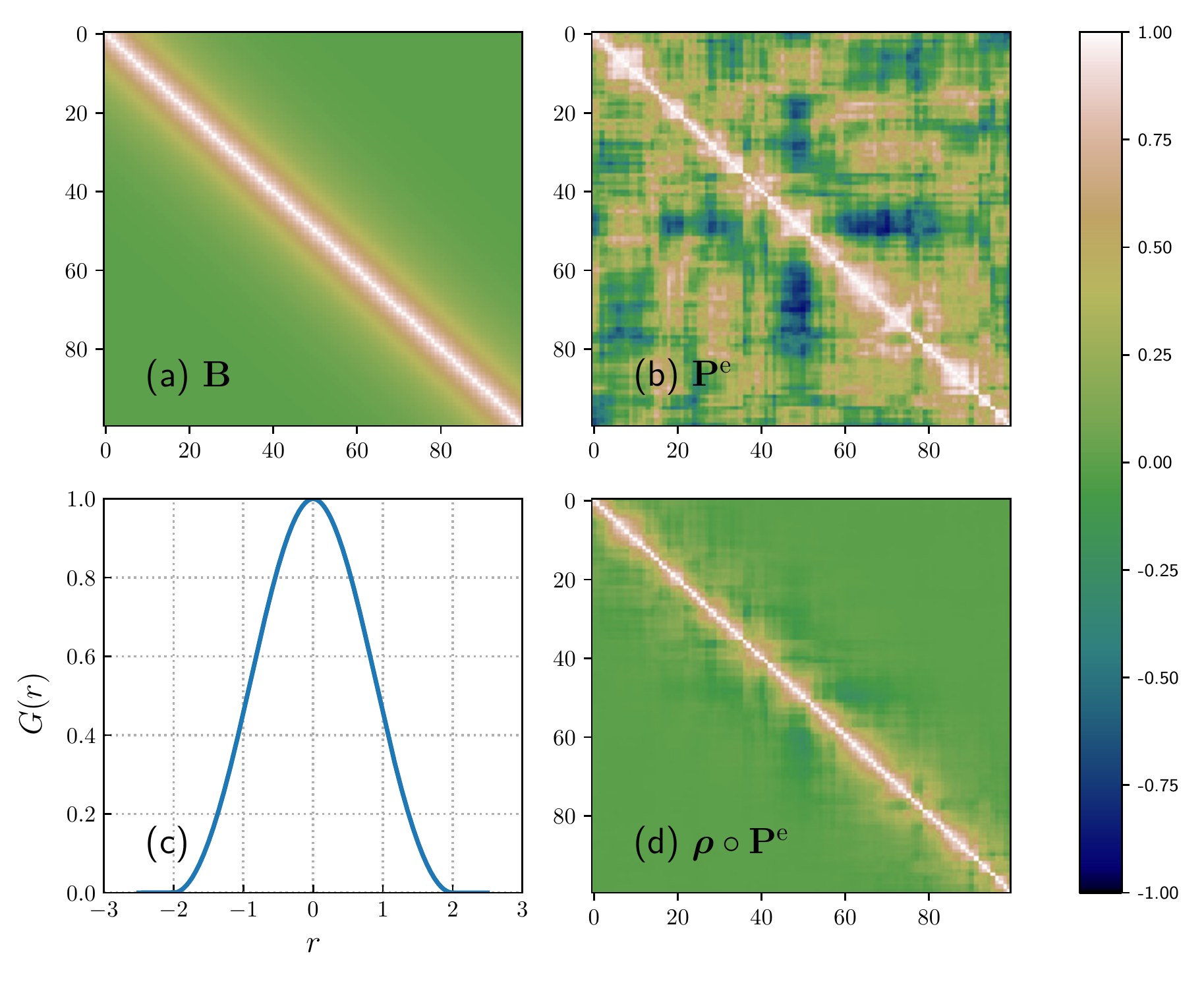} &
  \end{tabular}
  \caption{\label{fig:covloc} Panel a: True covariance matrix. Panel b: Sample covariance matrix. Panel c: Gaspari-Cohn
    correlation matrix used for covariance localization.  Panel d: Regularized covariance matrix obtained from a Schur product.}
\end{figure}

To illustrate the CL principle, we consider a multivariate ($m=100$) Gaussian distribution, with mean $\x^{\rm b}=0$.  Its
covariance matrix $\bB$ is given by $[\bB]_{ij} = e^{-\left|i-j\right|/L}$, where $i,j=1,\ldots,m$ and $L=10$ is the
correlation length.  An ensemble of $N=20$ members $\left\{ \x_{n} \right\}_{n=1,\ldots,N}$ is drawn from this
distribution from which we can compute the unbiased sample covariance matrix $\bP^\rme = \frac{1}{N-1}\sum_{n=1}^N
(\x_{n}-\barx)(\x_{n}-\barx)^\T$, with the mean $\barx = \frac{1}{N}\sum_{n=1}^N \x_{n}$, which are approximations
of $\bB$ and $\x^{\rm b}$, respectively.  The true and the sample covariance matrices are shown in
Fig.~\ref{fig:covloc}. The spurious correlations are obvious in panel Fig.~\ref{fig:covloc}b, where $\bP^\rme$ is displayed.  Let us define $\brho$
through the Gaspari-Cohn (GC) function \citep{gaspari1999}, 
\[
G(r) = \left\{
\begin{array}{ll}
{\rm if} \quad  0 \le r<1:& \quad  1 - \frac{5}{3}r^2 + \frac{5}{8}r^3 + \onehalf r^4 - \frac{1}{4}r^5, \\
{\rm if} \quad 1 \le r<2:& \quad    4 - 5r + \frac{5}{3}r^2 + \frac{5}{8}r^3 - \onehalf r^4 + \frac{1}{12}r^5 - \frac{2}{3r}, \\
{\rm if} \quad r \ge 2:& \quad  0   ,
\end{array} \right.
\]
which is a fifth-order piece-wise rational function.  The cutoff function is defined by
$r > 0 \mapsto G(r/c)$ where $c$ is a length scale which is called the localization radius. It mimics a
Gaussian distribution but vanishes beyond $r \ge 2c$, which is numerically efficient (in mathematical terms, $G$ is compactly supported).  The
Gaspari-Cohn cut-off function is shown in Fig.~\ref{fig:covloc}c.  Hence, choosing $c=L$, we define
$[\brho]_{ij} = G(|i-j|/L)$. The regularized covariance matrix, $\brho \circ \bP^\rme$ is shown in Fig.~\ref{fig:covloc}d.
As expected, the spurious correlations are tapered.

One of the drawbacks of covariance localization is that it still uses huge regularized covariance matrices (a priori of size $m \times m$).  Possible remedies are (i) avoiding the brute-force computation and storage of $\brho \circ
\bP^\rme$ exploiting for instance the possible sparsity of $\brho$, (ii) combining CL with domain localization (see below), (iii) or transferring the Schur product from state space
to observation space for sparse in situ observations. Solution (iii) can for instance be used to compute the Kalman gain
(cf.\ Eq.~\eqref{eq:KF-anl1}) in a more efficient manner \citep{houtekamer2001,ott2004,greybush2011}.

In a multivariate setting, variable localization was also proposed and tested by
\citet{kang2011}, where cross-correlations between carbon dioxide concentration and meteorological variables other than
wind variables are set to zero.

\paragraph{Domain localization}

An alternative to covariance localization is \emph{domain localization} (DL).  In DL, one performs the
global EnKF analysis as a collection of independent local EnKF analyses, each of them assimilating a subset of the
observations.  Typically, one seeks to update a state at location/grid-cell $x$. The EnKF analysis is locally carried
out by assimilating observations contained in a disk of radius $L$ centered on $x$. The number of observations in this
local domain should be small enough so that the ensemble can accommodate the information content of these observations,
i.e., that $\bP^\rme$ is locally full rank. This is equivalent to setting the observation error covariance matrix
$\bR$ to $\bR_x$ where all the entries associated to sites outside the disk are large or infinite, which rules out these
outer observations.  In DL, both the state vector and anomalies are locally updated. A global analysis is then recovered
from this collection of local analyses.  Because the subdomains overlap, one assumes that the transition between
analyses is smooth enough so that the global updated ensemble remains physically balanced.  A way to improve on these
transitions, is to obtain $\bR_x$ by tapering (Schur product) $\bR^{-1}$ with a regularizing cut-off function, typically
the GC, that decreases with the distance from $x$.

Because the number of local EnKF analyses to be carried out scales linearly with the size of the system $m$, the method
may seem numerically doomed.  However, on the one hand, all of the analyses can be performed in parallel and, on the
other hand, their numerical complexity is significantly reduced as it is driven by the ensemble size and the number of
local, not global, observations. The numerical feasibility of the approach has been demonstrated by
\citet{ott2004,hunt2007}.

\citet{sakov2011} have shown that CL and DL are not mathematically equivalent in general but that they can become so in
regimes where the analysis is dominated by the prior.  Another important point to mention is the physical imbalance that
may result from the gluing of local analyses and from the true long-distance correlations that have necessarily been
discarded in the localization process \citep{kepert2009,greybush2011}.  For chaotic systems, it has been suggested (see
Sect.~\ref{sec:AUS}) that the ensemble size below which localization is mandatory is equal to the dimension of the
unstable and neutral subspace of the dynamics.

\subsubsection{Inflation \label{sec:infl}}

Using localization addresses the rank-deficiency issue and gets rid of spurious correlations.  Unfortunately, sampling
errors are not entirely removed in the process. Residual errors are still often sufficient to make the EnKF diverge for
long-term runs \citep[see][for a discussion on the need for inflation in reduced order filter]{grudzien2018}.  One very simple {\it ad hoc} means to account for this residual sampling error is to inflate the error
covariance matrix by a multiplicative factor $\lambda^2 \ge 1$, $\bP^\rme \longrightarrow \lambda^2 \bP^\rme$, which can
also be implemented by inflating the anomalies $\x_{n} \longrightarrow \barx + \lambda \(\x_{n}-\barx\) $
\citep{pham1998,anderson1999}.  Note that inflation is not only used to cure sampling errors
\citep{bocquet2011,whitaker2012,bocquet2015}, but is also often used to counteract model error impact.

Inflation can also come in an additive, clearly non-equivalent, form: $\x_{n} \longrightarrow \x_{n} + \epsi_{n}$
with ${\mathbb E} \left[ \epsi_{n} \(\epsi_{n}\)^\T\right] = \bQ $.

As a drawback, inflation often needs to be tuned in order to obtain satisfactory performance of the EnKF, which is
numerically costly. Hence, adaptive schemes have been developed to make the task more automatic
\citep{wang2003,anderson2007,li2009,zheng2009,brankart2010,bocquet2011,miyoshi2011,bocquet2012,liang2012,ying2015,bocquet2015}. For instance, a variant of the deterministic EnKF, the DEnKF from \citet{sak08b}, was introduced as an approximation of the EnKF with an implicit inflation.

\subsection{\sffamily \Large Ensemble Variational methods}
\label{sec:EnVar}

4DVar and the EnKF are now both used in operational meteorological centers
as well as in academic studies \citep[see, e.g.,][]{buehner2013,bauer2015quiet}. 
In spite of apparent similar performances in the context of synoptic scale
meteorology, they are not equivalent. 4DVar searches for a nonlinear estimation of the maximum a posteriori of the
underlying pdf using nonlinear optimization techniques, while the EnKF only performs a linear update \citep[with the notable exception of the MLEF;][]{zupanski2005}. The EnKF propagates the statistics of the
errors, hence estimating the \emph{errors of the day}, whereas the traditional 4DVar only relies on time-independent
statistics of the errors. An important weakness of 4DVar, of technical nature, is the derivation of the adjoint models,
which remains a computer sciences challenge, whereas the EnKF estimates the tangent linear of the observation operator and its adjoint within the ensemble subspace.  These and other pros and cons of both approaches have been discussed by
\citet{lorenc2003,kalnay2007,yang2009comparison,bocquet2013}.

Hence, it is tempting to combine both approaches, for theoretical reasons (nonlinear analysis, errors of the day) and
technical reasons (no adjoint models).  The resulting algorithms are called \emph{ensemble variational} methods
(EnVar). They have been reviewed in chapter 7 of \citet{asch2016data} on theoretical grounds and with a focus on
operational implementation by \citet{bannister2017}, although the iterative ensemble Kalman smoother 
(see Sect.~\ref{sec:IEnKS}) is lacking in this latter
review. A classification of these methods and the terminology has been proposed by \citet{lorenc2013}.

Here, we distinguish between (i) the hybrid methods, (ii) the methods based on an ensemble of variational data
assimilation (EDA), (iii) the so-called 4DEnVar, and (iv) full nonlinear four-dimensional EnVar methods, of which the 
IEnKS is a deterministic exemplar.

\subsubsection{Hybrid methods}

Hybrid can refer to any combination of a variational method and of an EnKF method. Yet, it quite often specifically refers
to the hybridizing of a static error covariance matrix with a dynamical one sampled from an ensemble.  The idea was
first introduced by \citet{hamill2000}. The objective was to use the static covariance of a 3DVar and perform a
variational analysis to be used for the analysis step of an EnKF. The effective covariance matrix used for the prior is:
\be
\label{eq:hybrid}
\bB = \alpha \bC + (1-\alpha)\bX^\mathrm{f}\(\bX^\mathrm{f}\)^\T,
\ee
where $\bC$ is the static error covariance matrix \citep[as used in 3DVar or optimal interpolation;][]{Kalnay2002}, $\bX^\mathrm{f}$ is the
matrix of the forecast ensemble anomalies, and $\alpha \in [0,1]$ is a scalar that weights the static and dynamical
contributions.  The updated ensemble can be obtained within the framework of a stochastic EnKF (cf.\ Sect.~\ref{sec:EnKF}) 
 using several
stochastically perturbed variational problems.  On the other hand, if the mean is estimated in a deterministic
EnKF framework (cf.\ Sect.~\ref{sec:EnKFSQRT}), the anomalies update has to rely on an approximation due to the necessary reduction of the rank from the
hybrid covariance to the new dynamical covariances. There are ways to improve this update, such as using a
left-transform update \citep{sakov2011,bocquet2015}, or Lanczos vectors \citep{auligne2016}.  Note that, as with the
EnKF, it may be necessary to enforce localization of the sample covariances, which would be based on $\bB = \alpha
\bC + (1-\alpha)\brho \circ \left[ \bX^\mathrm{f}\(\bX^\mathrm{f}\)^\T \right]$, using CL.

\subsubsection{Ensemble of data assimilation \label{sec:EDA}}

Methods known as \emph{ensemble of data assimilations} (EDA) stem from meteorological prediction centers that operate a
4DVar, and in particular M\'et\'eo-France and the European Centre for Medium-Range Weather Forecasts (ECMWF). The idea
is to introduce dynamical errors that are absent in the traditional 4DVar. In order to build on the existing 4DVar
systems, one considers an ensemble of $N$ 4DVar analyses.  Each analysis, indexed by $i$, uses a different first guess
$\x_0^i$, and observations perturbed with $\epsi_k^i \sim {\mathcal N}(\bzero,\bR_k)$ to maintain statistical
consistency. Hence, each analysis $i$ carries out the minimization of (cf.\ Eq.~\eqref{eq:J-strong})
\be
   {\mathcal J}^\mathrm{EDA}_i(\x_0) = \onehalf \sum_{k=0}^K \left\| \y_k + \epsi^i_k -{\mathcal H}_k \circ {\mathcal
     M}_{k:0}(\x_0) \right\|^2_{\bR^{-1}_k} + \onehalf \left\| \x_0-\x^i_0\right\|^2_{\bB^{-1}}   .
   \ee
The background covariance $\bB$ is typically a hybrid one as in Eq.~\eqref{eq:hybrid} because it still uses the static
covariances of the traditional 4DVar and incorporates the sampled covariances from the dynamical perturbations.  The
procedure yields an updated ensemble, from which it is possible to assess a dynamical part of the error covariance
matrix. This is rather close to the idea of the hybrid EnKF-3DVar, but with a 4DVar scheme. It is worth to note that, in the linear case and Gaussian case, EDA is exactly Bayesian, in the sense that it produces an ensemble of independent realizations of the conditional PDF, though not necessarily in the
nonlinear or non-Gaussian case \cite[][and references therein]{liu2017,npg-2018-5}.
The method has been
implemented at M\'et\'eo-France \citep{raynaud2009,raynaud2011,berre2015} and at the ECMWF
\citep{bonavita2011,bonavita2012}.  Like for the hybrid method, it may be necessary to localize the sampled part of the
hybrid covariances, using for instance covariance localization (CL).  In this context, it may be convenient to enforce CL via
techniques such as the \emph{${\boldsymbol \alpha}$ control variable} trick \citep{lorenc2003,buehner2005,wang2007} -- which is
mathematically equivalent to CL -- or using wavelet truncations \citep{berre2015}.

\subsubsection{4DEnVar \label{sec:4DEnVar}}

NWP centers which have been using 4DVar were also the ones to offer the most reliable forecasts.  But the future of
4DVar as an operational tool is uncertain because of its poor scalability and of the cost of the adjoint model
maintenance, although weak-constraint 4DVar has scalable implementations \citep{fisher2017}.  The 4DEnVar
method emerged in these centers as a way to circumvent the development of the adjoint of the dynamical model.  The key idea is
based on the ability of the EnKF to estimate the sensitivities of the observation to the state variables using the full
observation model in place of the tangent linear one and of its adjoint in the computation of the Kalman
gain.

\citet{liu2008} essentially proposed to generalize this idea to the computation of the sensitivities of the
observations within a data assimilation window (DAW) to the variables of the initial state. Hence, these sensitivities
are associated to the composition ${\mathcal H}\circ {\mathcal M}$ defined over the DAW, rather than to observation
operator ${\mathcal H}$ as in the EnKF.  As a consequence, the sensitivities of the observations with respect to the degrees of freedom of
each anomaly can be approximated with
\be
\varepsilon^{-1}{\mathcal H}_k \circ {\mathcal M}_{k : 0} \(\barx_0\b1^\T +  \varepsilon\bX^\mathrm{f}\) \(\bI_m-\frac{\b1\b1^\T}{m}\)   .
\ee
This formula first builds an ensemble of mean $\barx_0$ with anomalies $\bX^\mathrm{f}$ scaled by $\varepsilon$; then
propagates this ensemble through ${\mathcal H}_k \circ {\mathcal M}_{k : 0}$; rescales the output ensemble by
$\varepsilon^{-1}$; and finally generates an ensemble of centered perturbations. With $0 < \varepsilon \ll 1$, these
perturbations can be seen as generated by the tangent linear of ${\mathcal H}_k \circ {\mathcal M}_{k : 0}$
(using finite-differences), whereas with $\varepsilon = \sqrt{m-1}$ these perturbations account for the nonlinearity of
${\mathcal H}_k \circ {\mathcal M}_{k : 0}$ as in the EnKF.  The set of these sensitivity vectors form a basis for the
analysis, an idea quite close to that of the reduced-rank 4DVar put forward in oceanography by \citet{robert2005}.

In 4DEnVar, the perturbations are usually generated stochastically, for instance resorting to a stochastic EnKF
\citep{liu2009,buehner2010a}.  Hence, in addition to avoiding the need for a model adjoint, flow-dependent error
estimation is introduced. Given that 4DEnVar has been developed in NWP centers using 4DVar, which relies on building a
sophisticated static background error covariance matrix, the background is usually of hybrid nature.  If the
perturbations are generated by a deterministic scheme, then the method could be seen as similar to the 4D-ETKF
\citep{hunt2007}, where the linear analysis is carried out in a basis made of trajectories (the perturbations) over the
DAW obtained from the previous deterministic analysis.

Because of the introduction of the limited set of perturbations, localization is again needed to mitigate the limited
rank of the sample error covariance matrix. However, in a 4D ensemble context and in contrast with the EnKF whose update uses observations
of given time, localization must be applied to covariances that are not only distant in space but also possibly in time within the
DAW. Unfortunately localization and the dynamics do not commute in general \citep{fairbairn2014,bocquet2014}, yielding
inconsistencies in the implementation of a static localization within the flow. A solution is to implement a
\emph{covariant} localization that changes in time with the dynamical flow \citep{bocquet2016,desroziers2016}.  For
instance, if localization is enforced via domain localization (DL), one would pull-back the influential observations
within the DAW using a surrogate, hyperbolic-like, model for the full evolution model and check in which local domains
their antecedent would fall in \citep{bocquet2016}. This type of solutions has also been put forward in a sensitivity
analysis context \citep{kalnay2012}.  Note, however, that this issue is circumvented if one possesses the adjoint of the
dynamical model.

Many variants of the 4DEnVar are possible depending on the way the perturbations are generated, or if the adjoint
model is available or not \citep{buehner2010a,buehner2010b,zhang2012b,clayton2013,poterjoy2015,bocquet2016}.  Full
4DEnVar operational systems are now implemented or are in the course of being so \citep{buehner2013, gustafsson2014,
desroziers2014, lorenc2015comparison, kleist2015osse, buehner2015, bowler2017a}.

\subsubsection{The iterative ensemble Kalman smoother \label{sec:IEnKS}}

Most of these EnVar methods, with the noticeable exception of the more firmly grounded EDA ones (Sect.~\ref{sec:EDA}), have been designed
heuristically blending theoretical insights and operational constraints. This led to many variants of the schemes,
even when this is not mathematically justified. Most of these ideas stemmed from the variational DA community, with the remarkable exception
of the Running in Place (RIP), an EnKF with an outer loop \citep{kalnay2010}.
By contrast, the iterative ensemble Kalman smoother \citep[IEnKS,][]{bocquet2014}, is a four-dimensional
EnVar method that is derived from Bayes' rule and where all approximations are understood at a theoretical level.  It
comes from ensemble-based DA and, specifically, extends the iterative ensemble Kalman filter
\citep{sakov2012} to a full DAW as in 4DVar.  The name \emph{smoother} reminds us that the method smooths
trajectories like 4DVar. However, it can equally be used for smoothing and filtering.  The name also pays an homage to
the iterative Kalman smoother of \citet{bell1994} which corresponds to the non-ensemble precursor of the method.
Basically, the IEnKS can be seen as an EnKF, for which each analysis corresponds to a nonlinear 4DVar analysis but
within the reduced subspace defined by the ensemble. Hence, the associated cost function is of the form
\be
\label{eq:IEnKS}
{\mathcal J}(\w) = \onehalf \left\| \w \right\|^2 + \sum_{k=L-S+1}^{L} \onehalf  \left\| \y_k-{\mathcal H}_k \circ {\mathcal M}_{k : 0}\(\barx_0+\bX_0\w\)
\right\|_{\bR^{-1}_k}^2  ,
\ee
where $L$ is the length of the DAW, and $S$ is the length of the forecast in between cycles (in units of $t_{k+1}-t_k$). Because the IEnKS catches
the best of 4DVar (nonlinear analysis) and EnKF (flow-dependence of the error statistics), both parameters could be
critical. In Eq.~\eqref{eq:IEnKS}, $\bX_0$ is the matrix containing along its columns the normalized (by $\sqrt{N-1}$) anomalies of the ensemble members at the initial time, while $\w$ is the vector of coefficients used to parametrize the state vector in the span of the ensemble, $\x_0=\barx_0+\bX_0\w$.  

The minimization of $\mathcal J$ can be performed in the ensemble subspace using any nonlinear optimization method,
such as Gauss-Newton (cf.\ Appendix {\color{red} B}), Levenberg-Marquardt or trust-region methods \citep{bocquet2012,mandel2016,ninoruiz2016}.  The
required sensitivities (derivative of the observation with respect to the initial state) are computed using an ensemble
forecast within the DAW, as in 4DEnVar. Hence, it does not require the tangent and adjoint models but emulates them,
similarly to 4DEnVar.  The anomalies update is computed using an approximate Hessian of this cost function, using
again these sensitivities.  The ensemble forecast step is similar to the EnKF but over $S$ time units.

The parameters $L$ and $S$ are constrained to satisfy $1 \le S \le L+1$ if all observations are to be assimilated. In
the case $S=L+1$, with a single iteration of the minimization and further less important restrictions, the IEnKS identifies
with the 4D-ETKF \citep{hunt2004}. In the case $S=1$ and $L=0$, it coincides with the MLEF \citep{zupanski2005}, which
can be seen as a precursor to the IEnKF and IEnKS.  The IEnKS represents an ideal tool to study deterministic four-dimensional EnVar
methods and test new strategies.

\begin{figure}[htbp]
  \begin{tabular}{cc}
    \includegraphics[width=0.48\textwidth]{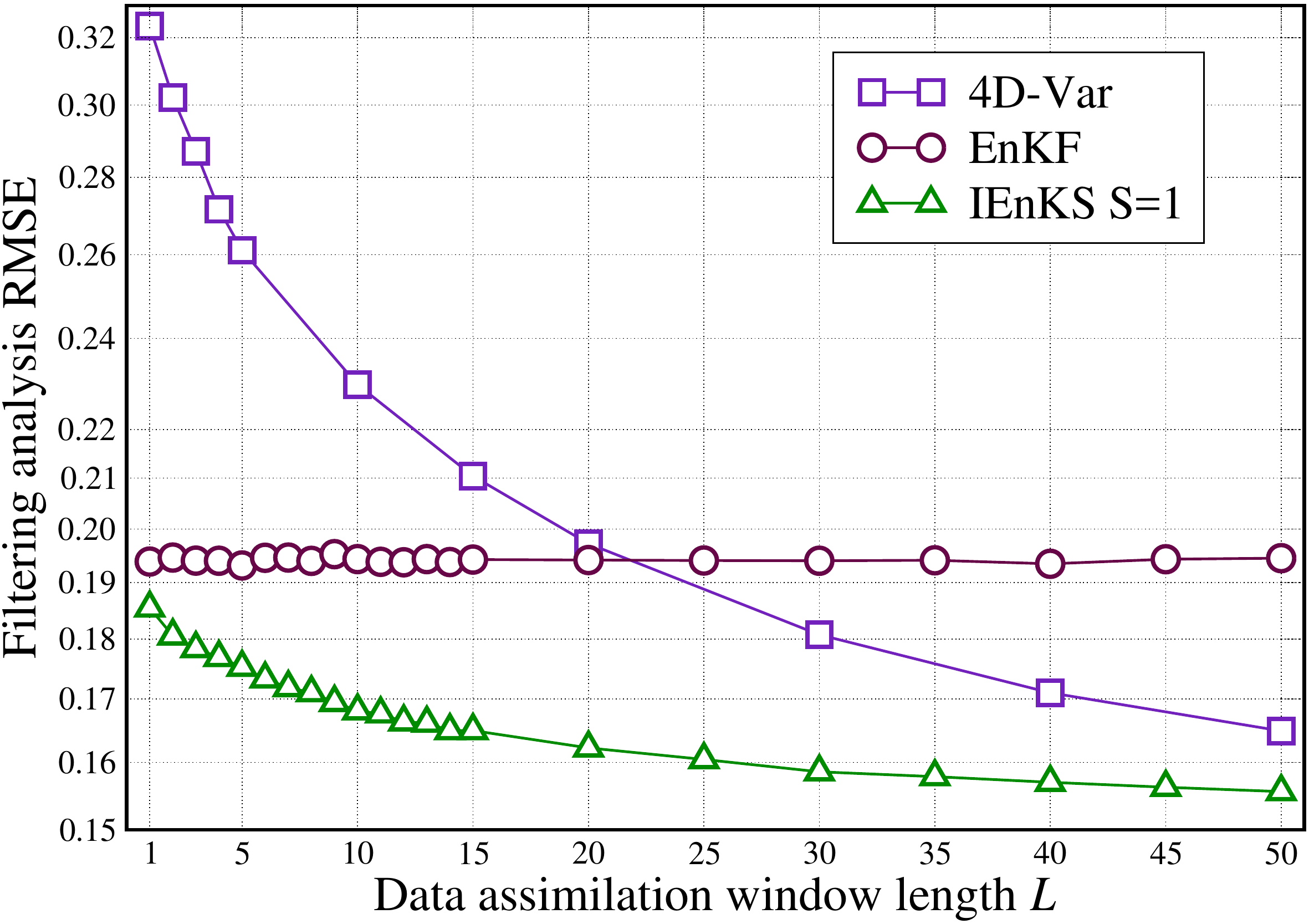} &
    \includegraphics[width=0.48\textwidth]{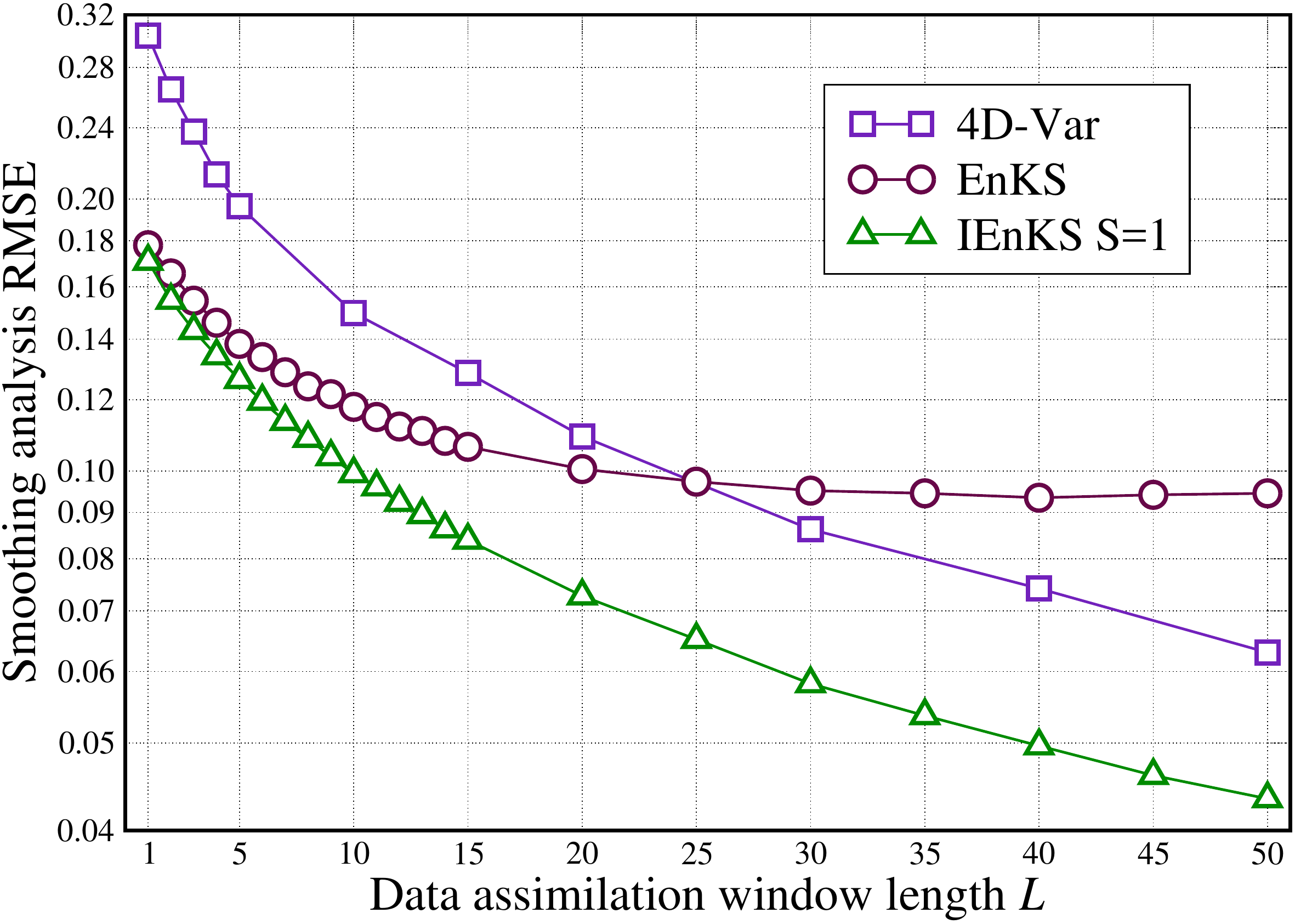}
  \end{tabular}
  \caption{\label{fig:rmse-ienks} Average root mean square error of several DA methods computed from synthetic experiments with the Lorenz-96 model.
    The left panel shows the filtering analysis root mean square error of optimally
    tuned EnKF, 4DVar, IEnKS assimilation experiments, as a function of the length of the DAW. The right panel shows the smoothing analysis root mean square error of optimally tuned EnKS, 4DVar and IEnKS as a
    function of the length of their data assimilation window. The optimal RMSE is chosen within the window for 4DVar and
    it is taken at the beginning of the window for the IEnKS.  The EnKF, EnKS and IEnKS use an ensemble of $N=20$, which
    avoids the need for localization but requires inflation.  The length of the DAW is $L \times \Delta t$,
    where $\Delta t = 0.05.$  }
\end{figure}

In chaotic systems, the IEnKS outperforms any reasonably scalable DA method in terms of accuracy. By construction, it
outperforms the EnKF, the EnKS and 4DVar for smoothing but also filtering.  This has been checked numerically on
several low-order models \citep[Lorenz models;][]{Lorenz1963,lorenz1998,lorenz2005}, 2D turbulence and
quasi-geostrophic models \citep{bocquet2014,bocquet2016}. Figure~\ref{fig:rmse-ienks} compares the performance of the IEnKS with that of 4DVar, the EnKF and the EnKS
with optimally tuned multiplicative inflation of the background covariance matrix so as to minimize the RMSE, for filtering and smoothing on the 40-variable Lorenz-96 model \citep{Lorenz96}. 
As for the other EnVar methods, localization needs to be used
in high-dimensional systems.  It can be implemented using covariant DL or CL \citep{bocquet2016}.

Note that, as a variational smoother, the IEnKS can theoretically and within its DAW account for errors that are correlated in time.
For instance it can account for a systematic constant in time bias \citep{bocquet2013}.

To conclude this brief review of the EnVar set of techniques, the simplified table \ref{tab:envar} summarizes some key properties of a selection of the methods we described.

\newcommand{\stag}[1]{\multicolumn{1}{l}{\rlap{\rotatebox{60}{#1}~}}}
\newcommand{\stagc}[2]{\multicolumn{1}{l}{\rlap{\rotatebox{60}{\color{#1} #2}~}}}
\newcolumntype{a}{>{\columncolor{pink}}c}
\newcolumntype{b}{>{\columncolor{SkyBlue}}c}
\newcolumntype{d}{>{\columncolor{LimeGreen}}c}

\begin{table}
  \caption{ \label{tab:envar} Comparison of EnVar data assimilation techniques. This table answers the following questions: (i) Is the analysis based on a linear or nonlinear scheme? (ii) Is the adjoint of the evolution model required? (iii) Is the adjoint of the observation operator required? (iv) Is the background flow-dependent? (v) Are the updated perturbations stochastic or deterministic? (vi) Are the updated perturbations fully consistent with the analysis, i.e., are they a faithful representation of the analysis uncertainty? (vii) Is localization of the ensemble analysis required? (viii) Is a static background used? To some extent, all algorithms can accommodate a static background; the answer tells whether the published algorithm
has a static background. Blank answers correspond to irrelevant questions.
}
\begin{tabular}[htbp]{laaabbbbd}
algorithm & \stagc{red}{analysis type} & \stagc{red}{evol. model adjoint required?} & \stagc{red}{obs. operator adjoint required?}  &  \stagc{blue}{background flow-dependence?} &  \stagc{blue}{sto. or det. perturbations?} &  \stagc{blue}{consistent perturbations?} &  \stagc{blue}{localization required?} & \stagc{Green}{static background} \\
\hline
EnKF & linear & & no & yes & both & yes & yes & no\textsuperscript{4} \\
3DVar & nonlinear & & yes & no & & & & yes \\
4DVar & nonlinear & yes & yes & no  & & & & yes \\
\hline
EDA with 4DVar & nonlinear & yes\textsuperscript{1} & yes\textsuperscript{1} & yes & sto. & yes & part. & yes\textsuperscript{3} \\
4DEnVar & linear & no & no & yes & sto. & no\textsuperscript{2} & yes & yes\textsuperscript{3} \\
IEnKS & nonlinear & no & no & yes & det. & yes & yes & no\textsuperscript{4} \\ 
MLEF & nonlinear &  & no & yes & det. & yes & yes & no\textsuperscript{4} \\ 
4D-ETKF & linear & no & no & yes & det. & yes & yes &  no\textsuperscript{4} \\
\hline
\end{tabular}
\\
  \small\textsuperscript{1} The adjoint models could be avoided considering an EDA of 4DEnVar. \\
  \small\textsuperscript{2} It depends on the implementation of 4DEnVar; the perturbation are often generated by a concomitant EnKF.  \\
  \small\textsuperscript{3} With an hybridization of the covariances.  \\
  \small\textsuperscript{4} But possible with an hybridization of the covariances.  \\
\end{table}
  
\section{\sffamily \Large Special topics \label{sec:SpecTop}}

This section discusses four selected topics related to the application of DA to the geosciences.
The issues for the DA and the countermeasures connected to the chaotic nature of the atmosphere and ocean are the content of Sect.~\ref{sec:AUS}. 
Section~\ref{sec:anamo} discusses the techniques that allow to deal with non-Gaussian variables in the framework of DA methods originally devised to tackle with Gaussian quantities. 
The DA for chemical species is the subject of Sect.~\ref{sec:DA-chem}: the topic is of great relevance (pollutant dispersion, accidental nuclear release etc.) and has received much attention in recent years. Finally Sect.~\ref{sec:TOPAZ} describes one example of operational DA, the TOPAZ system for ocean prediction.

\subsection{\sffamily \large Dealing with chaotic dynamics \label{sec:AUS}}

Many natural systems, including atmosphere and ocean, are examples of chaotic dissipative dynamics \citep{Dijkstra_2013}. Among the characteristic features of this class of system, the one having the largest impact on prediction, is the extreme sensitivity to initial conditions \citep{Lorenz1963}. Infinitesimal errors are bound to grow exponentially in the short time (and later saturate), and the rate and directions of such a growth are themselves highly state-dependent \citep{trevisan2011chaos}. Forecasters have been long aware of this flow-dependent behavior of the forecast uncertainty and ``prediction of predictability'' (i.e., the prediction of the uncertainty) has become an important issue in NWP \citep{kalnay1987forecasting,Lorenz96}.
The difficulties inherent to the chaotic nature of the climate on the one hand, and the social pressure to deliver reliable forecast with associated prediction of the uncertainty, on the other, have motivated a huge bulk of research on predictability \citep[see the review by][and references therein]{vannitsem2017predictability}. In forecasting practice, the problem of predicting uncertainty, thus moving from a "single-realization deterministic forecast" to a "multi-realizations probabilistic forecast",  has been tackled by ensemble prediction schemes \citep{leutbecher2008ensemble}.

Dealing with such a flow-dependent error growth is obviously also a challenge for DA: one must in fact be able to properly track and incorporate this dependency in the DA description of the state estimation error. A situation that is further complicated by the large size of the geophysical models ($m=\mathcal{O}(10^9)$). Nevertheless, the dissipative character (due to friction) of the dynamics, induces an effective dimensional reduction, in the sense that the error dynamics, in deterministic systems, is often confined to a subspace of much smaller dimension, $n_0\ll m$. This subspace, referred to as the {\it unstable-neutral subspace}, is defined as the vector space generated by the backward Lyapunov vectors with non-negative Lyapunov exponents, i.e., the space of the small perturbations that are not growing exponentially under the action of the backward dynamics \citep[see][for a topical review]{legras1996guide}. 
The full phase-space can thus be seen as split in a (usually much smaller) unstable-neutral subspace and a stable one \citep{kuptsov2012}. For instance, \citet{carrassi2007} have shown how a quasi-geostrophic atmospheric model of $~O(10^5)$ degrees of freedom possesses an unstable-neutral subspace of dimension as small as $n_0=24$.

The existence of this underlying splitting of the phase-space has enormous impact on the skill of DA with chaotic models.
In EnKF-like methods, \citet{carrassi2009,ng2011,bocquet2014} have studied the relation between 
the properties of the model dynamics and the performance of DA, and revealed how the knowledge about the former can be used in the design of the square-root EnKF and EnKS: for purely deterministic models, the minimum ensemble size required to achieve good performance must be at least as big as the number, $n_0$, of non-negative Lyapunov exponents \citep{carrassi2009,bocquet2014}. Along similar lines, but for variational DA, \citet{pires1996} showed that the relative projection of the state estimation error on the stable and unstable subspaces depend on the length of the DAW (see Sect.~\ref{sec:4DEnVar}): for long DAW the stable components are all dampened and the error is essentially confined on the unstable subspace.

\begin{figure}[H]
\begin{center}
\includegraphics[scale=0.37]{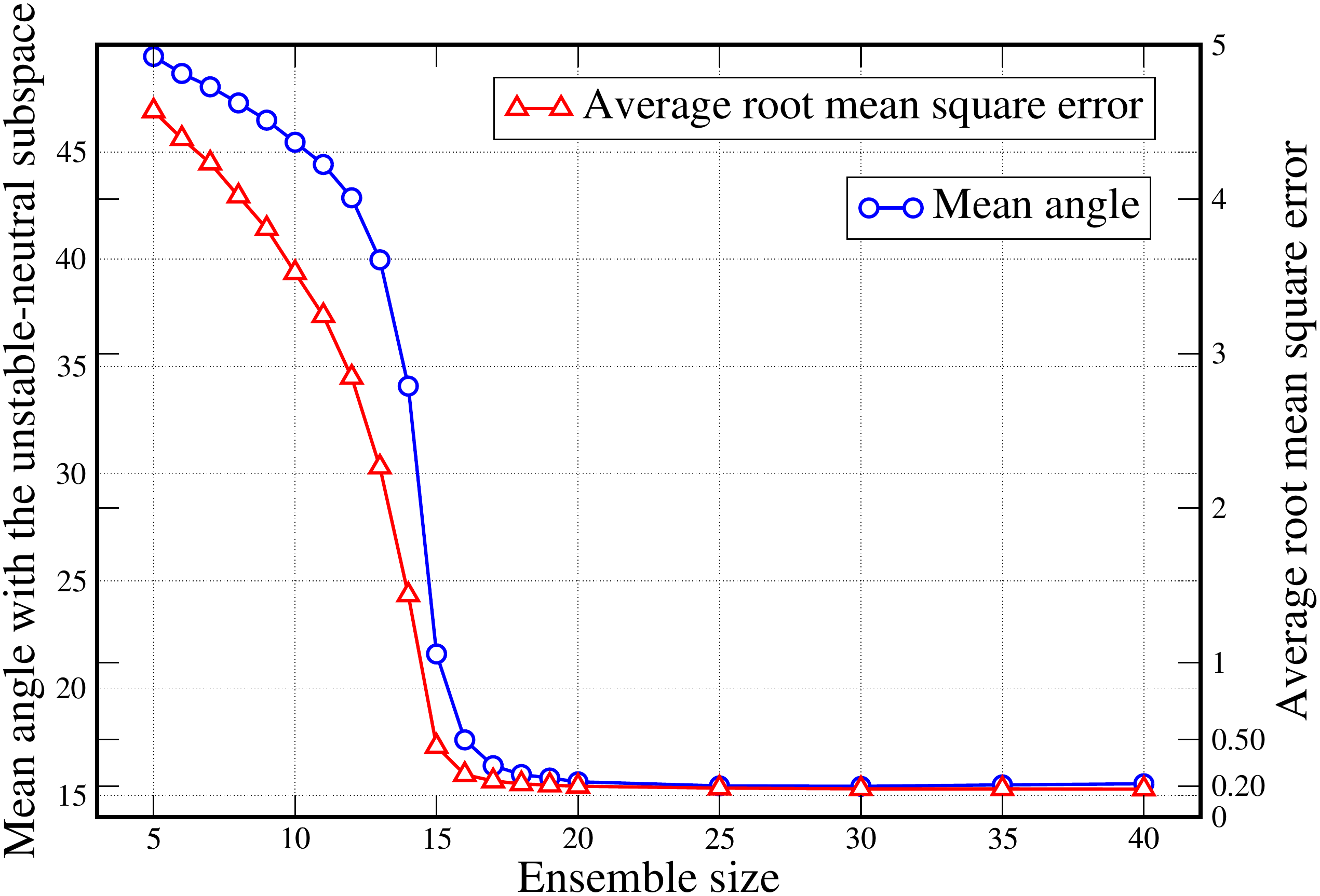}
\caption{\label{fig:angle_rmse}
Time- and ensemble- averaged angle (in degree) between an anomaly from the EnKF ensemble and the unstable-neutral subspace
  as functions of the ensemble size $N$ (left y-axis) and corresponding time-averaged root mean square error of the
  EnKF (right y-axis). The numerical experiments are performed on the Lorenz-96 model with $m=40$ 
 variables \citep{lorenz1998} and the DA setup is $\bH=\bI_d$, $\bR=\bI_d$, observational frequency $\Delta t = 0.05$ and 
 error variance $\sigma=1$.
}
\end{center}
\end{figure}

These results have recently been corroborated by a proper mathematical understanding of the behavior of the KF and KS (cf.\ Sect.~\ref{sec:KF} and \ref{sec:KS}) for linear unstable systems in the absence of model noise. 
It has been proven analytically that the span of the error covariance matrices of the KF and KS tends asymptotically to the unstable-neutral subspace \citep{gurumoorthy2017rank}, independently from the initial condition \citep{bocquet2017degenerate}, and that the rate of such convergence is faster for the KS \citep{bocquet2017four}. 
Remarkably, \citet{bocquet2017four} have shown that, even in the case of a four-dimensional EnVar method (the IEnKS, cf.\ Sect.~\ref{sec:IEnKS}) applied to a noiseless nonlinear system, the full alignment of the ensemble anomaly subspace with the unstable-neutral subspace is accompanied by the maximum reduction of the state estimation error. Figure \ref{fig:angle_rmse} illustrates this mechanism for the EnKF using the Lorenz-96 model \citep{lorenz1998}.

These results indicate that, in deterministic chaotic systems, the unstable-neutral subspace fully describes (in an average, asymptotic, sense) the uncertainty in the state estimate. This fact is at the basis of a class of DA algorithms known as {\it assimilation in the unstable subspace} (AUS) pioneered by Anna Trevisan and collaborators \citep{trevisan2004,palatella2013a}, in which the unstable-neutral subspace (or a suitable numerical approximation of it) is explicitly used in the DA scheme to parametrize the description (both temporally and spatially) of the uncertainty in the state estimate. 
Given that usually $n_0\ll m$, AUS represents an interesting alternative to operate DA efficiently on high-dimensional system (large $m$) by monitoring only $n_0$ unstable-neutral directions. Applications to large atmospheric \citep{carrassi2008b} and oceanic models \citep{uboldi2006} have confirmed its capabilities. 
The AUS paradigm has been successfully incorporated in both the EKF \citep{trevisan2011} and the 4DVar \citep{trevisan2010} and its extension to cope with the deviations from the linear error dynamics and perfect model assumptions on which it relies, has been studied recently by \citet{palatella2015} and \citet{grudzien2018JUQ} respectively.

\subsection{\sffamily \large Dealing with non Gaussian variables}
\label{sec:anamo}

There are cases for which the linear analysis is sub-optimal. 
These arise when either the state variables, observations or model parameters are not Gaussian distributed. 
We consider first the case in which the distribution of at least one variable is not Gaussian, for example it exhibits a clear 
skewness or kurtosis, but the pdf remains continuous. 

Ignoring the non-Gaussian nature of a variable will normally make the least-squares estimator sub-optimal (cf.\ Appendix {\color{red} A}): 
either by systematic biases, by under-estimating the correlation length scales \citep{brankart12},  
or more visibly by returning nonphysical values (for example negative concentrations of tracer variables).
In a DA framework, the state variables (called the "control variable" in this context) need not be exactly defined as in the forward model: 
nonlinear transformations are allowed and will change the outcomes of the minimization problem \citep{ber03a}. 

Transformations of the variables have been practiced extensively in the geostatistical literature under the term {\it Gaussian
anamorphosis} (also known as "normal-score transform"): a nonlinear function is applied to the cumulative pdf in order to make it Gaussian \citep{chilesdelfiner}. In DA, the Gaussian anamorphosis function can either be
applied to the state variables $\tildx=\phi(\x)$ or to the observations $\tildy=\varphi(\y)$ or the model parameters or to all of them at the
same time. The prerequisite is that the transformed variables $\tildx$ and $\tildy$ should be "more Gaussian" and better suited to a 
linear analysis update than the original ones. 
The anamorphosis function must be strictly increasing so that backward transformations from the Gaussian values to the ``natural''
pdf can be given by $\phi^{-1}$ and $\varphi^{-1}$. Since $\bE(\phi^{-1}(\tildx))\neq
\phi^{-1}(\bE(\tildx))$, Monte Carlo techniques are often employed to obtain unbiased statistics. 
This makes the anamorphosis particularly convenient to apply in conjunction with ensemble techniques \citep{ber03a}. 

In practice, the stochastic models (\ref{eq:modellin}) and (\ref{eq:obslin}) are valid for the transformed $\tildx$ and $\tildy$, 
so that the EnKF and EnKS equations can be re-used with $\tildx$ and $\tildy$ in replacement
of $\x$ and $\y$. 
Equations~(\ref{eq:Efa}) and (\ref{eq:Eps}) become
\begin{align}
    \bE^{{\rm f/a}}&=(\tildx^{{\rm f/a}}_{[1]}, \tildx^{{\rm f/a}}_{[2]}, \ldots , \tildx^{{\rm f/a}}_{[N]}) \in \Re^{m\times N}, \label{eq:Aanam} \\
    \bY^\rmo&=(\tildy_{[1]}, \tildy_{[2]}, \ldots , \tildy_{[N]}) \in \Re^{d\times N} \label{eq:Epsanam} .
\end{align}

The observations are preferably transformed before perturbing the observations, since the perturbations are more conveniently applied 
in the Gaussian space (even when using a deterministic version of the EnKF, the $\bR$ matrix characterizes more adequately Gaussian
observation errors). 
After the analysis, the inverse transformation $\phi^{-1}$ is applied to each member of the analyzed ensemble $\x^a_i = \phi^{-1}(\tildx^a_i)$ 
before the next propagation step. 
If one variable is both an observation and a model state variable, then the same transformation should be applied to both so that the
observation operator $\bH$  keeps a simple form \citep{amezcua14}. Alternatively, using independent transformations can be a practical 
way to correct for biases \citep{Lien2016a,Lien2016b}. 

There are infinitely many of strictly increasing functions: choosing a good anamorphosis function is therefore an open question. 
When a set of observations is available or a free model run, it is possible to approximate the anamorphosis function based on the
histogram of available data, using for example a piece-wise linear fit \citep{Simon2009a}. 
Adaptive types of anamorphosis functions have also then been explored \citep{Simon2012a}, in which the anamorphosis function  is
fitted to the values present in the ensemble forecast at the assimilation time. These have proven advantageous in twin experiments
\citep{Simon2012a,Li2012}, but their stability remains to be tested in real cases, where the true density is unknown and the results can
become excessively sensitive to the choice of the tails of the distribution \citep{Simon2012a}. 
In practical cases, simple analytical functions are
sometimes preferred like the exponential, gamma or logit functions, which can be supported theoretically knowing the nature of the
problem at hand \citep{Simon2015,Gharamti2017}. 

One immediate benefit of the Gaussian anamorphosis is the ability to impose positivity constraints in the EnKF which is useful 
for example in ecosystem models \citep{Simon2009a}. 
The use of an anamorphosis function has been demonstrated with the EnKF in hydrological models \citep{Li2012,Zhou2012}, 
NWP models \citep{Lien2013,Lien2016a,Lien2016b,Kotsuki2017}, 
in joint parameter-state estimation in ocean ecosystem models \citep{Simon2012a,Gharamti2017,Gharamti2017a}, 
for the evaluation of covariance length scales in various ocean modeling applications \citep{brankart12} and with moderate success in a coupled
ice-ocean model \citep{Barth2014}. It has been as well introduced in a snow model with an ensemble smoother \citep{aalstad18}. 

One limitation of the Gaussian anamorphosis as described above is that it does not consider the multi-Gaussian case: only the marginal
distributions of the variables are transformed independently from each other. Although this does not guarantee that the joint distributions
become multi-Gaussian, the experience still shows that the multi-Gaussian properties are generally improved after transforming the
marginals \citep{hans,amezcua14}. The anamorphosis framework can however be further extended by the use of copulas \citep{Scholzel2008}. 

A more serious limitation in many practical cases is that the Gaussian anamorphosis does not lend itself to discontinuous pdf. These are better addressed in the framework of truncated Gaussians: a threshold is defined on the Gaussian
distribution, and all values exceeding the threshold are reset to its precise value, generating an atom of probability distribution.  
Truncated Gaussians were first applied in optimal interpolation, by introducing a Lagrange parameter in the minimization \citep{tha07}, 
only in the cases when assimilated values exceeded the threshold, leading to a two-step interpolation process (detect and constrain). 
The issue of sampling a multivariate
truncated Gaussian in the framework of an EnKF was then explored applying a Gibbs sampler \citep{Lauvernet2009}, 
under the simplifying assumption of small truncations. 
Another approach is the use of Quadratic Programming \citep{Janjic2014}, more
general in the sense that it can accommodate nonlinear constraints. 
An alternative and inexpensive approach is to "moderate" the strength of DA in order to avoid reaching the discontinuities of 
the distribution.
This has been proposed for assimilation into an ocean model in isopycnic coordinates, for which the values of layer thickness
must be positive \citep{wang2016}. Combinations of Gaussian anamorphosis and truncations have been successfully tested by \citep{Lien2013,Lien2016b} 
by setting the Gaussian value of the threshold to the median of the out-of-range Gaussian pdf (zero precipitations in their case). 

The above examples are extending the Gaussian framework introduced in Sect.~\ref{sec:KFS} and extend as well the use of the EnKF and EnKS
methods. The Gaussian anamorphosis can be included at almost no additional costs, but the costs of the truncated Gaussian methods are
potentially a limitation in realistic cases \citep[a 500\% increase of the analysis step is reported with the Gibbs sampler;][]{Lauvernet2009}. 
Still, these methods benefit from the advantages of the Gaussian framework in high dimensions and represent attractive alternatives to
more general Bayesian methods that suffer from the curse of dimensionality (see Sect.~\ref{sec:PF}). 

\subsection{\sffamily \large Data assimilation for chemicals}
\label{sec:DA-chem}

There is a huge literature on the use of DA for chemical constituents of the atmosphere.  There are several quasi
exhaustive reviews about the advances in this field, such as \citet{carmichael2008, zhang2012a, bocquet2015b}.  We here briefly
explain what chemical DA is, what the motivations for using DA in this field are, and, finally, which methods are used.

The specific set of chemical reactions that prevails in the atmosphere depends on the species but also on the considered
spatial and temporal scales.  For instance, one typically distinguishes between global atmospheric chemistry and air
quality. The former is concerned with species transported and reacting over long distances and within the whole column
of the atmosphere, whereas the latter is more concerned with the complex chemistry within the boundary layer (about 1 km
from the ground) at regional scales.  As an example of species, greenhouse gases are global pollutants with long life
spans considered in the first type of study, even though regional studies of their transport are also of interest.
Another example is ozone, which, in the lower troposphere, is studied and forecast in air quality as a harmful
pollutant whereas, in the stratospheric layer, or when transported in between continents, is the focus of global
studies. Besides the global and regional scales, DA is also nowadays used for chemical studies at urban scale.

Chemistry and transport models are fundamentally multivariate with many species to be accounted for (about a hundred for
gaseous species, hundreds when considering aerosols).  This also significantly increases the dimensionality of the models
compared to atmospheric and ocean models independently.  That is why the choice of the control variables (cf.\ Sect.~\ref{sec:Var}) is critical in the
feasibility of chemical DA.

One also distinguishes between online and offline models. Offline models (also called CTM for \emph{chemical transport
  model}) only consider emissions and uptakes, chemistry, loss processes and transport driven by meteorological fields
that are supposed to be provided.  Online models (also called CCMM for \emph{coupled chemistry meteorology model}) couple
meteorology with chemistry and transport. CCMMs are therefore much more costly but allow for two-way coupling between
meteorology and chemistry. Most of chemical DA studies are performed offline.

One major difference with atmospheric and ocean models is that the dynamics of CTMs are mostly stable. In practice two
model trajectories with distinct initial conditions will coalesce, quite differently from chaotic geofluid
dynamics \citep{haussaire2016,vannitsem2017predictability}. Hence, the initial conditions are rarely the most important control variables in chemical DA. Quite often, the
species emissions are the most influential input parameters, if not control variables. More generally, model error is
more influential for chemical DA than for geofluids DA. One consequence of this dynamical stability is that air quality
forecast are rarely useful beyond 48 hours, even with the help of DA.  Another consequence is the importance of all
other parameters and input of these models and their uncertainty and hence of model error.

For these many reasons, but also to serve the enforcement of regulations, chemical DA is used for (i) chemical forecast,
nowcasting and possibly reanalysis, (ii) inverse modeling of species emission and uptake, and (iii) parameter
estimation, such as the boundary conditions, constants of the physical and chemical parametrizations.

The methods used in chemical DA are largely inspired by meteorological DA.  Optimal interpolation \citep{Kalnay2002} has been largely used
from the very beginning.  Since the beginning of the 2000s, 4DVar has also been used, which can advantageously be
adopted for the simultaneous estimation of inputs and model parameters \citep{elbern2007}. Methodological advances in chemical DA also
had some influence, with the early development of reduced Kalman and ensemble methods in the 90s' \citep{hanea2007,
  constantinescu2007a, constantinescu2007b, wu2008}.  Four-dimensional ensemble variational methods have also been
tested in this field \citep{haussaire2016, emili2016}.

\subsection{\sffamily \large An example of operational data assimilation and reanalysis: the TOPAZ integrated ice-ocean system}
\label{sec:TOPAZ}
\begin{figure}[H]
\begin{center}
\includegraphics[scale=0.45]{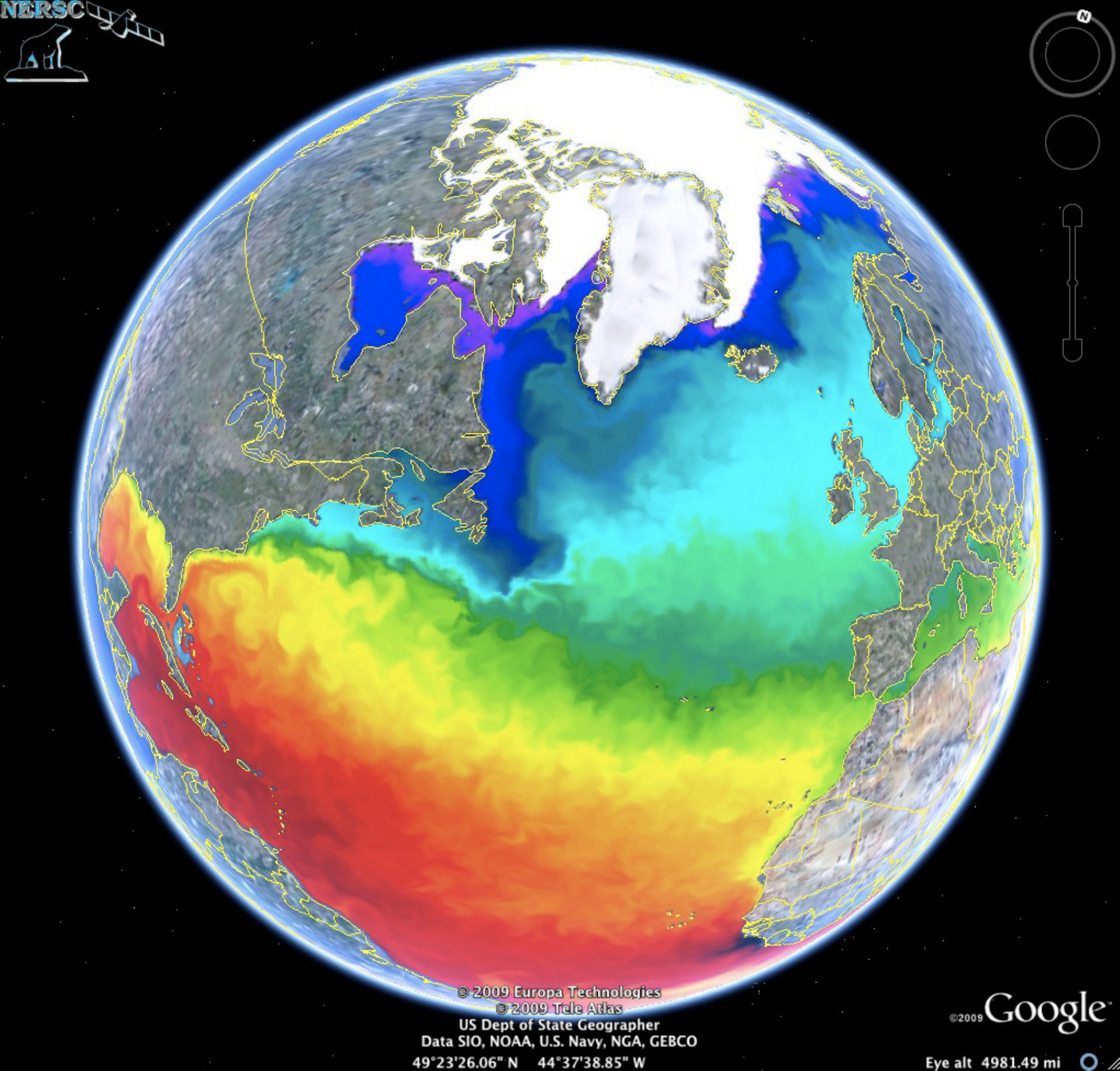}
\caption{
\label{fig:topaz}
Example of sea surface temperature (in color) and sea ice concentration (in white) real-time analysis by the TOPAZ system on the $28^{{\rm th}}$ November 2009.}
\end{center}
\end{figure}

The concept of operational oceanography has gradually arisen in the 1990's under the joint pressure from its users in the navy, the authorities in
charge of emergency response (oil spills, search and rescue operations) and the offshore oil and gas industry. Ocean monitoring
was becoming possible from remote sensing and increasing in-situ observations to fulfill the needs of both the operational and the climate research
communities and the development of ocean models was accelerated. The conditions for ocean DA were then ripe and a community was
formed around the international Global Ocean Data Assimilation Experiment (GODAE, now GODAE-OceanView, \href{www.godae-oceanview.org}{www.godae-oceanview.org}). 
The EnKF, originally introduced in a QG model, was then extended to a multi-layer isopycnal model (Miami Isopycnal Coordinate Model, MICOM) and then
its offspring in generalized vertical coordinates HYCOM \citep[HYbrid Coordinate Ocean model][]{ble02}. The EnKF \citep{evensen2009}, SEEK filter \citep{pham1998} as well as the EnKS \citep{evensen2009}
were applied successfully to this model to assimilate ocean remote sensing observations from satellite altimeters and sea surface
temperatures in a North Atlantic configuration called DIADEM \citep{bru03a}. In parallel, the assimilation of sea ice concentrations in a coupled
ice-ocean model was demonstrated \citep{lis03a} as well as the assimilation of remotely sensed ocean color in a coupled physical-ecosystem
model \citep{nat03a}, both with the EnKF. These experiments were all using similar settings of the EnKF: same number of members, same sources and
statistics of model errors and were therefore encouraging signs that a single integrated system could monitor the ocean, the sea ice and the
ocean ecosystem consistently, with the uncertainties of the three components represented by the same ensemble. Such an integrated system for
the North Atlantic and the Arctic Oceans was then built and named TOPAZ, including downscaling capabilities to forecast the coastal zones at
higher resolution. An illustration showing a typical output from the TOPAZ system is given in Fig.~\ref{fig:topaz}, that shows the sea surface temperature and sea-ice concentration over the Atlantic Ocean. The TOPAZ system is run at a horizontal resolution of $10$km that makes it eddy-permitting and of very large dimension. The scales resolved by the model are visible on the figure. The model resolution is gradually increasing from one configuration to the next one, although without much incidence on the data assimilation source code.   

The TOPAZ system has been set in near real-time forecasting mode in January 2003, initially using the perturbed-observations EnKF \citep{eve03a} (cf.\ Sect.~\ref{sec:EnKF}), and then moved on to the DEnKF \citep{sak08b} (cf.\  Sect.~\ref{sec:infl}) in January 2010. 

The HYCOM ocean model uses isopycnic coordinates in the deep stratified ocean, the following state vector is then the closest possible to the
original model prognostic variables: 
\begin{itemize} 
\item 3-dimensional variables, defined for each hybrid vertical layer: layer thickness, u- and v-components of the current velocity, temperature and
salinity 
\item 2-dimensional variables: barotropic pressure, u- and v-components of the barotropic velocity, sea ice characteristics (ice concentration,
ice thickness, snow depth). 
\item Static parameters, 2-dimensional bias fields for mean sea level amplitude and mean sea surface temperature. 
\item If the ecosystem module is coupled: 
   \begin{itemize} 
      \item 3-dimensional nutrients concentrations (nitrate, phosphate and silicates in North Atlantic and Arctic regions)
      \item 3-dimensional phytoplankton (diatoms and flagellates) and zooplankton species (meso- and microzooplankton). 
      \item 3-dimensional detritus variable, as well as oxygen and biogenic silicate. 
      \item 2-dimensional maps of model parameters (different plankton loss rates). 
   \end{itemize}
\end{itemize} 
Recalling that the EnKF analysis is conserving linear equilibrium, the above definition of the state vector in generalized (isopycnic) vertical
coordinates is advantageous for preserving the geostrophic balance, which relates linearly the velocity vector averaged over one ocean layer to the
gradient of the layer thickness \citep{eve03a}. The use of isopycnic coordinates in the state vector has also the advantage of reducing
diapycnal mixing during the linear update in Eq.~\eqref{eq:X5ana} or \eqref{eq:sqrtanaD} since the analyzed temperature and salinity
tracers are the results of combinations of forecast temperature and salinity from different members but at the same density. The
application of the EnKF in the MICOM isopycnic model has also proven to make the assimilation of surface temperature properties more
efficient in the context of a climate reanalysis \citep{Counillon2016}. 

On the downside, not all the above state variables are Gaussian distributed, so according to the discussion in Sect.~\ref{sec:anamo}, biological variables are
log-transformed \citep{Simon2015}. However variables with discontinuous distributions (sea ice variables, ocean layer thickness) are corrected in case nonphysical values occur after the analysis \citep{sakov2012topaz4}, and TOPAZ is thus exposed to assimilation biases. Yet, the experience has not revealed strong
biases. 

TOPAZ has used a local analysis (cf.\ Sect.~\ref{sec:local}) ever since its inception, a global analysis having proven utterly incapable of controlling millions of state
variables with only 100 members. The inflation is implicit in the formulation of the DEnKF \citep{sak08b} and additionally in the form
of a factor applied to the $\bR$ matrix, used only in the update of the anomalies \citep{sakov2012topaz4}. The use of a constant multiplicative
inflation (cf.\ Sect.~\ref{sec:infl}) of the forecast anomalies has proven problematic in the case of the Arctic due to the inhomogeneous observation coverage, even with
a small inflation of $1\%$. 

The present TOPAZ4 physical system with its horizontal resolution of about 12 km and 28 hybrid vertical layers counts about 80 million
unknowns. The data assimilated in a weekly cycle count the along-track sea level anomalies from satellite altimeters, the interpolated sea surface
temperatures and sea ice concentrations from satellites as well, the Lagrangian sea ice drift from continuous maximum cross-correlation of
scatterometer data, and the temperature and salinity profiles from various in-situ platforms, including the Argo autonomous buoys and
Ice-Tethered Profilers. This amounts to about $400$ thousands observations after averaging high-resolution observations onto the model grid. The altimeter and the sea ice drift data are
assimilated asynchronously \citep{sak10a}, accounting for time-delayed covariances. 

The computer costs of the TOPAZ4 system are mainly related to the propagation of the 100 members ($1200$ CPU hours per week, but embarrassingly
parallel as the 100 members are submitted independently and the HYCOM model is itself parallel, running on $144$ CPUs), the analysis step takes
20 hours only, and is also parallelized because all local analysis updates are independent from each other. 

\begin{figure}[H]
\begin{center}
\includegraphics[clip=true, trim={2cm, 2cm, 1cm, 2cm}, width=\textwidth]{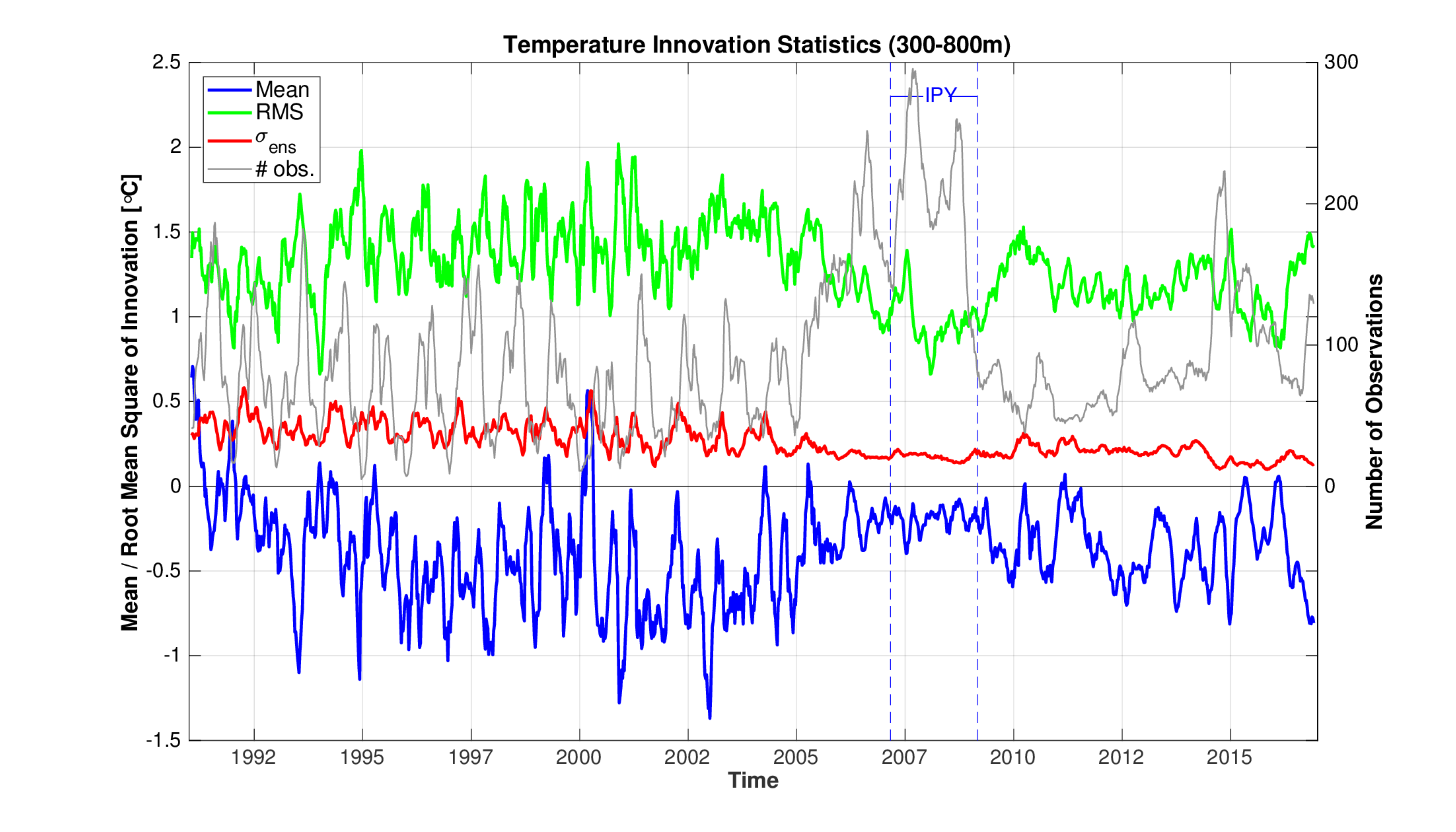}
\caption{
\label{fig:Fig8}
Time series of data assimilation diagnostics across the 24-years reanalysis for all temperature profiles in the depths 300 m to 800 m in the whole Arctic. The blue line is the average of all innovations, the green line is the related standard deviation (Root Mean Square Error, RMS), the red line is the ensemble spread and the grey line is the number of temperature observations. The IPY was officially taking place between the two vertical lines, but the observations were increasing progressively. 
} 
\end{center}
\end{figure}

The TOPAZ4 reanalysis \citep{xie2017quality} has been conducted over 24 years (covering the altimeter era 1991-2015) and thus 1250 assimilation steps.
During this time, the Arctic observing system has undergone large changes with for example the International Polar Year (IPY, 2007-2009)
during which several Ice-Tethered Profilers have been deployed in the Central Arctic, measuring temperature and salinity profiles in areas
previously unobserved. The large increase of observations during the IPY is visible in the time series of Fig.~\ref{fig:Fig8} and 
does, as expected, reduce the bias and root mean square error, but
the latter increases again during the 6 months that follow the end of the IPY. This implies that the special observation efforts done on the
occasion of the IPY should be the rule rather than the exception for the sake of monitoring the Arctic Ocean. 
The diagnostics shown for other variables indicate that the reanalysis is stable all through the 1250 assimilation cycles \citep{xie2017quality} and
how much each observation type contributes to the minimization. 

The coupled physical-biological reanalysis with assimilation of surface chlorophyll estimates from remotely-sensed ocean color data is a
much more difficult endeavor than the physical reanalysis, both on technical and scientific levels \citep{Simon2015}. 

\section{\sffamily \Large Where we are and where we go: a look at the perspectives \label{sec:Persp}}

\begin{figure}[H]
\begin{center}
\includegraphics[width=\textwidth]{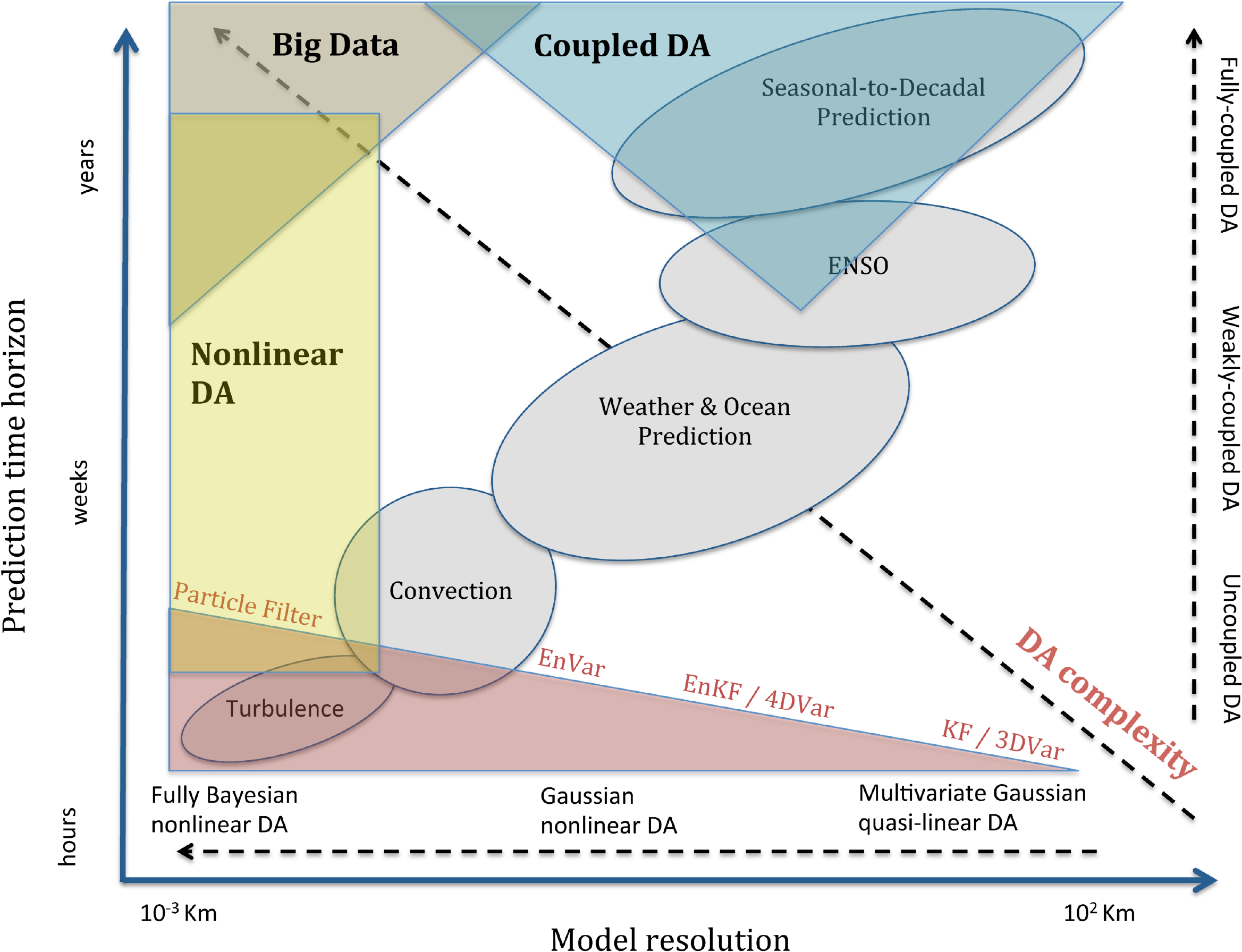}
\caption{\label{fig:DACompl}
Required DA method versus model resolution and prediction time horizon; examples of corresponding natural phenomena are also shown for illustrative purposes. The degree of sophistication of the DA grows commensurately with the increase in prediction time horizon and the decrease of the model grid size. 
}
\end{center}
\end{figure}

Data assimilation is nowadays a key ingredient of the atmospheric and oceanic prediction machinery and substantial computational power and economical resources are allocated to its maintenance and updates. What are the challenges DA is facing at this present time? Which solutions are being proposed to cope with the requirements of present days climate research?
Figure~\ref{fig:DACompl} illustrates this quest by displaying the required DA approach as a function of the model resolution (x-axis) and of the prediction horizon (y-axis). 
The constant increase of the numerical model resolution, i.e., reducing the model grid size, implies resolving more and more small scale processes (e.g., convection or turbulence) that are often inherently nonlinear and non-Gaussian. The transition toward high-resolution models must thus be accompanied by corresponding DA developments where the Gaussian and linear assumptions are relaxed \citep[see, e.g.,][]{yano2017scientific}. This identifies one of the more challenging and active lines of research in DA at the current time: fully Bayesian methods. 
The horizontal dashed line in Fig.~\ref{fig:DACompl} illustrates the DA method required features as a function of the model grid size; particle filters, the subject of Sect.~\ref{sec:PF}, are fully Bayesian and nonlinear methods and are placed at the top of the algorithms hierarchy in Fig.~\ref{fig:DACompl}.

Together with the increase in model resolution, the current era is also characterized by a growing interest in long-term forecasts. Indeed, beyond the meteorological time horizon of two weeks, predictions over seasonal-to-decadal (s2d) time scales potentially bear higher societal relevance: they can guide adaptation to near-term climate change and related risks \citep{doblas2013seasonal}. Such long-term predictability arises from the interactions between the atmosphere and the more slowly varying components of the climate system, like the ocean, land surface and cryosphere, so that predictions are issued using fully coupled models (Earth System Models, ESMs). 
The fruitful use of DA with coupled models necessitates the development of adequate coupled DA (CDA) methods that allow for a consistent and balanced propagation of the informational content of the observations across all model components (see vertical dashed line in Fig.~\ref{fig:DACompl}). As will be discussed in Sect.~\ref{sec:CDA}, this is not straightforward with standard DA methods and substantial efforts are currently being undergone to develop CDA strategies.

In addition to the increase in numerical model resolution and of the prediction time horizon, we are also experiencing a dramatic growth and refinement of the observations suppliers. The Earth is now observed over a wide range of spatial and temporal scales, thanks to an increasingly wider variety of sustained observing systems, which includes satellites, but also new, not stationary, ocean measurements such as floats, drifters and more recently gliders \citep[see, e.g.,][]{kuznetsov2003method}. The assimilation of data derived from instruments that follow the flow has come to be known as {\it Lagrangian DA} \citep[see, e.g.,][]{ide2002lagrangian,nodet2006variational}. 
These methods have gained popularity as they offer a suitable way to consistently incorporate modern observation of the ocean, and have experienced a flourishing stream of improvements in the last decade, driven by addressing two main challenges: the use of indirect measurements of the state variables \citep[see, e.g.,][]{salman2006method} and the inherent nonlinear character of the underlying dynamics \citep{apte2013impact}.
More recently, the Lagrangian dimension of the DA problem has also involved the model component, and not just the data, with the appearance of numerical models discretised on a spatio temporal varying mesh \citep[see, e.g.,][for an example of Lagrangian sea-ice model]{rampal2016nextsim}. 
This feature represents another methodological challenge for DA, which is no longer only demanded to update the value of the physical variables on the grid points, but also the physical location of grid itself \citep[see, e.g.,][]{bonan2017data}.

\subsection{\sffamily \large Bayesian data assimilation: particle filters}
\label{sec:PF}

From the Bayesian standpoint, the state estimation problem is best formulated as Eq.~\eqref{eq:Bayes-Th}:
\be
\label{eq:Bayes-Th-repeat}
p(\x\vert\y)=\frac{p(\y\vert\x)p(\x)}{p(\y)}\, .
\ee
We have seen in Sect.~\ref{sec:form2} how to exploit and formulate the sequential estimation problem through the conjunction of Bayes' rule and
the Chapman-Kolmogorov equations.

Yet, these formulae were claimed to be impractical, at least for high-dimensional models, and we moved on to solutions
based on Gaussian approximations. The algebra involved with these approximations is numerically demanding with
inversions of covariance matrices in the analysis step, which contrasts with the apparent simplicity of
Eq.~\eqref{eq:Bayes-Th-repeat}.

A direct and brute force approach to the DA problem would be to use Monte Carlo methods and draw $N$ samples from
Eq.~\eqref{eq:Bayes-Th}. The hope is that in the asymptotic limit, i.e., $N \rightarrow \infty$, one would properly
estimate the conditional density $p(\x\vert\y)$.  In a sequential context, this approach is called the particle filter
(PF), or sequential Monte Carlo. In the following we show how to justify and implement such simple algorithm but that it unfortunately
comes with its own drawbacks.

\subsubsection{Bootstrap particle filter and resampling}

In the following, most pdfs -- mainly designated by $p$ -- will be identified by their arguments. This is not mathematically rigorous but the notation has the merit of offering sleek expressions.

As a first step, let us see how to solve the filtering problem sequentially with the PF.  The forecast pdf $p(\x_k | \y_{k-1:})$
from $t_{k-1}$ to $t_k$ is assumed -- and this is the main approximation at finite $N$, i.e. with a limited ensemble --
to be of the form
\be
\label{eq:forecastpdf}
p(\x_k | \y_{k-1:}) = \sum_{n=1}^N \omega^n_{k-1} \delta(\x_k-\x_k^n) ,
\ee
which is an empirical distribution with delta-Dirac masses $\delta$, each one centered on an $\x_k^n$. 
In Eq.~\eqref{eq:forecastpdf} the notation $\y_{k:}$ stands for the, ideal, infinite sequence of observations from the far past until time $t_k$, $\y_{k:}=\{\y_k,\y_{k-1},\dots,\y_{-\infty}\}$.
The $\left\{
\x_k^n \right\}_{n=1,\ldots,N}$ are the particles (i.e., the ensemble members or the samples) and the $\left\{
\omega_{k-1}^n \right\}_{n=1,\ldots,N}$ are positive weights attached to the particles; $\x_k^n$ is a shortcut for the
particle $\x_{n}$, as denoted in Sect.~\ref{sec:EnsMeth}, at time $t_k$. Since Eq.~\eqref{eq:forecastpdf} represents a pdf, the weights need
to be normalized to one, $\sum_{n=1}^N \omega^n_{k-1} = 1$. Hence, those weights tell how probable a particle is.

The analysis at $t_k$ consists in the assimilation of $\y_k$ using Bayes' rule:
\begin{align}
    \label{eq:pf-analysis}
  p(\x_k | \y_{k:}) & =  \frac{p(\y_k | \x_k)}{p(\y_{k:})}  p(\x_k | \y_{k-1:})
  \nn
  & =  \sum_{n=1}^N \omega^n_{k-1} \frac{p(\y_k | \x_k)}{p(\y_{k:})} \delta(\x_k-\x_k^n) \nn
  & \propto  \sum_{n=1}^N \omega^n_{k-1} p(\y_k | \x^n_k) \delta(\x_k-\x_k^n) .
\end{align}
Equation~\eqref{eq:pf-analysis} suggests to define the updated weights as
\be
\label{eq:wu1}
\omega^n_{k} \propto \omega^n_{k-1} p(\y_k | \x^n_k) ,
\ee
where the proportionality factor can be determined afterwards by the condition that the normalized updated weights
should sum up to $1$. Hence, the analysis elegantly sums up to a simple multiplication of the weights by the likelihood
of each particle.

The forecast step amounts to applying Chapman-Kolmogorov Eq.~\eqref{eq:Chap-Kol}:
\begin{align}
  \label{eq:pf-forecast}
  p(\x_{k+1} | \y_{k:}) &= \int \! \mathrm{d}\x_{k} \, p(\x_{k+1} | \x_k) p(\x_k | \y_{k:}) \nn
  & = \sum_{n=1}^N \int \! \mathrm{d}\x_{k} \, p(\x_{k+1} | \x_k) \omega^n_{k} \delta(\x_k-\x_k^n) \nn
  & = \sum_{n=1}^N \omega^n_{k} p(\x_{k+1} | \x^n_k) .
\end{align}
If the model, defined by its transition density $p(\x_{k+1} | \x_k)$, as in Eq.~\eqref{eq:trans}, is deterministic, the
forecast pdf $p(\x_{k+1} | \y_{k:})$ is, as the update pdf $p(\x_k | \y_{k:})$, in the form of a delta-Dirac density; otherwise it is not so.
To obtain $p_{k+1|k}$ as a delta-Dirac pdf, which would be necessary in order to cycle the algorithm, we need to sample
from Eq.~\eqref{eq:pf-forecast}.

One solution to obtain a delta-Dirac pdf from Eq.~\eqref{eq:pf-forecast} consists, for each particle $n$, in sampling
$\x_{k+1}^n$ from the density $p(\x_{k+1} | \x^n_k)$ by simply forecasting $\x_{k}^n$ from $t_k$ to $t_{k+1}$ using the
(possibly stochastic) model associated with the transition density $p(\x_{k+1} | \x_k)$, which yields:
\be
p(\x_{k+1} | \y_{k:}) \approx \sum_{n=1}^N \omega^n_{k} \delta(\x_{k+1}-\x_{k+1}^{n}) .
\ee
Alternatively, one can sample particle $n$ from Eq.~\eqref{eq:pf-forecast} by randomly selecting one of the particles,
say $n'$, with a probability proportional to its importance weight. Once $\x_k^{n'}$ is selected, one can forecast it
and define $\x_{k+1}^{n}$ using, as in the previous case, the model associated with the transition density $p(\x_{k+1} |
\x^{n'}_k)$, which yields
\be
p(\x_{k+1} | \y_{k:}) \approx \frac{1}{N}\sum_{n=1}^N \delta(\x_{k+1}-\x_{k+1}^{n}) .
\ee

\begin{figure}[H]
  \centering
  \includegraphics[scale=0.7]{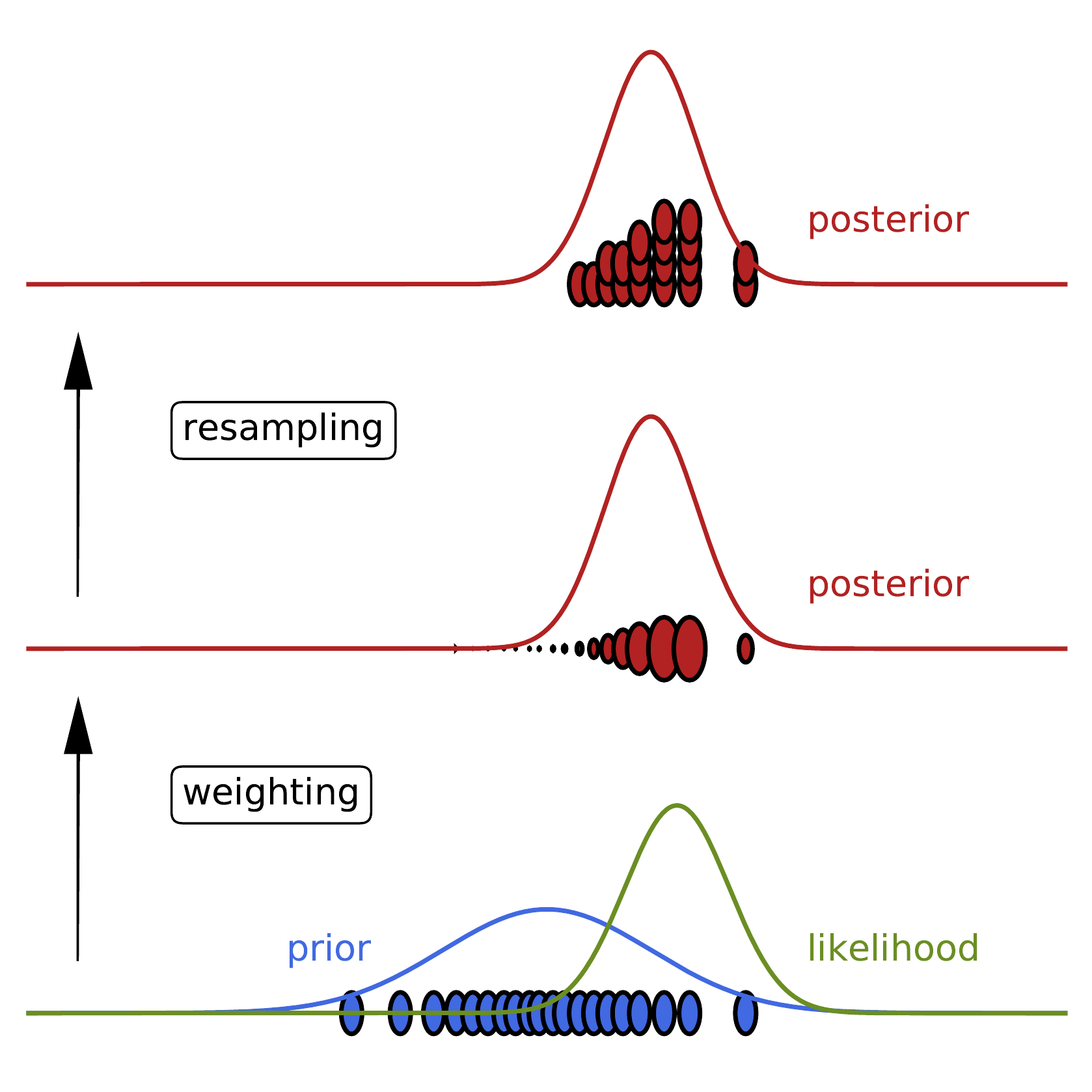}
  \caption{\label{fig:pf} Principle of the SIR particle filter (here $N=19$). The lower panel curves are the pdfs of the
    prior and the observation. The initial equal-weight particles are also displayed.  The middle panel shows the
    updated unequal weights of the particles as computed by the likelihood. The upper panel shows the outcome of
    resampling with multiple copies of several of the initial particles.}
\end{figure}

The first option corresponds to the \emph{bootstrap} PF \citep{gordon1993} or sequential importance sampling
(SIS) PF.  The second option adds, before forecasting, a \emph{resampling} step that uniformly resets the weights to $N^{-1}$;
it is called sequential importance resampling (SIR) PF and is depicted in Fig.~\ref{fig:pf}.  There are several ways to resample the particles given their
weights. Popular resampling schemes include multinomial resampling, residual resampling, and stochastic universal (or systematic)
resampling, the latter minimizing the sampling noise introduced in the procedure \citep{douc2005}.

\subsubsection{Importance sampling with a proposal}
\label{sec:PF-proposal}

In statistics, most smart sampling strategies usually come with the possibility to draw samples from an ancillary, more accessible,
distribution, naturally called the \emph{proposal} distribution. However, the weights of these drawn particles must be
corrected so that their empirical distribution is an unbiased estimator of the targeted distribution.  Particle filters can
also crucially benefit from this approach. Since this subject is more technical, the reader not interested in the details of
its implementation is invited to move on to the next subsection Sec.~\ref{sec:PF-degeneracy}.

A richer class of particle filters can be formalized if we consider a density of trajectories
$p(\x_{k:} | \y_{k:})$ in place of model states (i.e., the density $p(\x_k|\y_{k:})$), a pdf which is usually associated to a smoothing problem (cf.\ Sect.~\ref{sec:form2}).
Now, we assume the existence of a smoothing density $q(\x_{k:}|\y_{k:})$ from which it is easier to sample, instead of the "desired" $p(\x_{k:}|\y_{k:})$.
It has the following delta-Dirac representation:
\be
q(\x_{k:} | \y_{k:}) = \sum_{n=1}^N \omega^n_{k} \delta(\x_{k:}-\x_{k:}^n) .
\ee
We can exploit this \emph{proposal} density, which is auxiliary to the conditional pdf under study, to compute any
statistical moment of the conditional probability $p(\x_{k:} | \y_{k:})$.  If $\phi$ is a generic test function,
we have
\begin{align}
{\mathbb E} \left[ \phi(\x_{k:}) \right] &= \int \! \mathrm{d}\x_{k:} \, \phi(\x_{k:})p(\x_{k:} | \y_{k:}) \\
& = \int \! \mathrm{d}\x_{k:} \, \phi(\x_{k:}) \frac{p(\x_{k:} | \y_{k:})}{q(\x_{k:} | \y_{k:})} q(\x_{k:} | \y_{k:}) \\
& = \sum_{n=1}^N \omega^n_{k} \phi(\x^n_{k:})\frac{p(\x^n_{k:} | \y_{k:})}{q(\x^n_{k:} | \y_{k:})} .
\end{align}
Hence, the conditional pdf can be represented using weighted samples:
\be
\label{eq:aux-rep}
p(\x_{k:} | \y_{k:}) \approx \sum_{n=1}^N \frac{p(\x^n_{k:} | \y_{k:})}{q(\x^n_{k:} | \y_{k:})} \omega^n_{k} \delta(\x_{k:}-\x_{k:}^n) .
\ee
We further assume that the proposal density factorizes according to
\be
\label{eq:prop-fact}
q(\x_{k:}|\y_{k:}) = q(\x_{k} | \x_{k-1:},\y_{k:}) q(\x_{k-1:}|\y_{k-1:})
\ee
such that it is easy to sample $\x_{k}^n$ from $q(\x_{k} | \x^n_{k-1:},\y_{k:})$.

The particle trajectories $\left\{\x^n_{k-1:}, \omega^n_{k-1}\right\}_{n=1,\ldots,N}$ can then be extended to $t_{k}$ by
sampling $\x_{k}^n$ from $q(\x_{k} | \x^n_{k-1:},\y_{k:})$ so as to obtain $\left\{\x^n_{k:},
\omega^n_{k-1}\right\}_{n=1,\ldots,N}$.
Hence, the smoothing conditional pdf at $t_{k}$ is
\be
\label{eq:aux-rep2}
p_{k|k}(\x_{k:} | \y_{k:}) = \sum_{n=1}^N \frac{p(\x^n_{k:} | \y_{k:})}{q(\x^n_{k:} | \y_{k:})} \omega^n_{k-1} \delta(\x_{k:}-\x_{k:}^n) .
\ee
Assuming Markovian dynamics, the sequential evolution of the smoothing pdf decomposes as
\be
p(\x_{k:} | \y_{k:}) \propto p(\y_{k}|\x_{k}) p(\x_{k}|\x_{k-1}) p(\x_{k-1:} | \y_{k-1:}),
\ee
which, together with Eq.~\eqref{eq:prop-fact}, yields
\be
p(\x_{k:} | \y_{k:}) \propto \sum_{n=1}^N \frac{p(\x^n_{k}|\x^n_{k-1})p(\y_{k}|\x^n_{k})}{q(\x^n_{k} | \x_{k-1:}, \y_{k:})} \omega^n_{k-1} \delta(\x_{k:}-\x_{k:}^n) . 
\ee
By comparison with the generic Eq.~\eqref{eq:aux-rep}, the weights should be updated at $t_{k}$ according to
\be
\label{eq:wu2}
\omega^{n}_{k} \propto \frac{p(\x^n_{k}|\x^n_{k-1})p(\y_{k}|\x^n_{k})}{q(\x^n_{k} | \x_{k-1:}, \y_{k:})} \omega^n_{k-1} ,
\ee
up to a normalization to $1$ of the updated weights.
The filtering solution of the estimation problem is simply obtained without further computation from
$\left\{\x^n_{k:}, \omega^n_k\right\}_{n=1,\ldots,N}$ by marginalization, i.e., keeping the states at $t_k$
with the same weights: $\left\{\x^n_{k}, \omega^n_k\right\}_{n=1,\ldots,N}$.

Importantly, if we choose $q(\x_{k} | \x_{k-1:}, \y_{k:}) \equiv p(\x_{k}|\x_{k-1})$, then we obtain $\omega^{n}_{k}
\propto p(\y_{k}|\x^n_{k}) \omega^n_{k-1}$ and recover the bootstrap PF (cf.\ Eq.~\eqref{eq:wu1}).  Furthermore, if we choose $q(\x^n_{k} |
\x_{k-1:}, \y_{k:}) \equiv p(\x_{k}| \y_{k}, \x_{k-1}) $, then we obtain $\omega^{n}_{k} \propto
p(\y_{k}|\x^n_{k-1}) \omega^n_{k-1}$. This corresponds to the \emph{optimal importance proposal} PF
\citep{doucet2000}. It is optimal only in the sense that it minimizes the theoretical variance of each weight $\omega_{k}^n$
conditional on $\x^n_{k-1:}$ and $\y_{k:}$; over the realizations of $\x_{k}^n$. This variance is actually $0$.

\subsubsection{Degeneracy of the particle filter}
\label{sec:PF-degeneracy}

The algebra required by these PFs is very simple and elegant, and in principle offers a nice and asymptotically exact
alternative to the Gaussian approximations to DA. Unfortunately, the PF is plagued by the \emph{curse of
  dimensionality} as the dimensionality of the model is increased.  In a sequential scheme, this curse manifests itself
by the degeneracy of the weights: one weight will be close to $1$, while the others essentially vanish \citep{kong1994}.
That is to say, the ensemble of particles collapses onto one single particle, while updating the weights via
Eq.~\eqref{eq:wu1} or Eq.~\eqref{eq:wu2}.  This collapse is very inefficient as far as state estimation is concerned.
Resampling does help by resetting to uniform weights, but it is, quite often, not enough to counteract this curse.
Moreover, this trend essentially grows exponentially with the dimension of the system. More precisely, it has been shown
in very simple but instructive Gaussian models that the particle numbers required to avoid degeneracy should scale like
the variance of the likelihood \citep{snyder2008}:
\be
\label{eq:varlik}
\ln(N) \propto \mathrm{Var}\left[ \ln(p(\y|\x) \right] .
\ee
Equation~(\ref{eq:varlik}) has the merit to show that $N$ could scale exponentially with the size of simple systems,
but the derivation of such scaling is not straightforward in general.  A carefully designed proposal, such as the
optimal proposal mentioned in Sec.~\ref{sec:PF-proposal}, and which can be shown to minimize the variance of the weights, does not change this exponential
trend. Yet, it does reduce the constant in the exponent \citep{mackay2003,snyder2015}. Numerical investigations confirm
this trend \citep{Bocquet2010,slivinski2016}.

\subsubsection{Smarter particle filters for the geosciences}

Particle filtering is a well developed field of statistics and engineering \citep{doucet2001,arulampalam2002,chen2003}.
It is very successful with low-dimensional models (such as object tracking). The number of contributions to the topic
has substantially grown in geophysical DA in recent years \citep{vanleeuwen2009,Bocquet2010}. Yet, the curse of
dimensionality remains a major obstacle.  That is why appealing schemes have been proposed to reduce its impact.

The implicit particle filter \citep{chorin2009,morzfeld2012} combines smoothing and particle filtering over a DAW, similarly to
4DVar or the IEnKS.  It stands as a possible extension of the optimal proposal particle filter but over a several-step
long DAW.

In order to circumvent the curse of dimensionality, one can restrict the full Bayesian analysis to a limited number of
degrees of freedom, while the rest of the control variables are estimated via a Gaussian-based method, typically the
EnKF. This strategy has been developed in Lagrangian DA \citep{slivinski2015}.

Another strategy relies on mitigating the degeneracy of the PF by hybridizing with the EnKF
\citep{santitissadeekorn2015two}. A family of algorithm can be created and parametrized by a mixing coefficient, which
can be tuned or adaptively selected \citep{stordal2011,frei2013,chustagulprom2016}.

The equal weight PF and variants thereof \citep{ades2015,zhu2016} build a proposal such that the particles effectively
get the same numerical weight out of the analysis, a procedure which is obviously meant to avoid the degeneracy.

\subsubsection{Localization \label{sec:local}}

Similarly to the EnKF, a strategy to mitigate the curse of dimensionality that plagues the PF is to reduce the number of
degrees of freedom by making local analyses. However, its implementation is trickier than with the EnKF.  Indeed,
considering a local domain localization, a particle will be given distinct local weights, even though they might vary
smoothly in space.  Hence, there is no natural updated particle that could emerge from the former if the local weights
are unequal, as opposed to the local EnKF.  One should devise a gluing of local parts of particles through resampling
and try to avoid nonphysical discontinuities at the intersections of the local domains.  Such strategies have recently
been proposed and implemented \citep{reich2013, poterjoy2016, penny2016, robert2017, beskos2017stable, farchi2018}.  Their success
is mitigated by the still large number of particles required in each local domain. As a consequence, the size of such
domain (localization radius) is generally diagnosed to be significantly smaller than with the local EnKF.  Yet,
localization is thought to be a necessary ingredient of a successful particle filter in high-dimensional geophysical systems.

\subsection{\sffamily \large Coupled data assimilation \label{sec:CDA}}

Data assimilation algorithms have been conceived mainly for NWP applications and have been usually designed for state estimation in systems with a single dominant dynamical scale and/or for an observational network having a dominant spatio-temporal density. 
The sustained increase of model resolutions, the deployment of more and more observation platforms and the use of coupled ESMs for seasonal to decadal (s2d) predictions, altogether bind to a deep re-thinking of the DA procedures. 
The design of efficient CDA methods, able to keep simultaneously control of all resolved scales and propagate adequately information across the climate system components, has been recently recognized to have primary importance.

Several research groups and institutions, including weather and climate services, are currently studying and developing CDA \citep[see][for an updated report on these efforts]{penny2017coupled}. 
 Early attempts include the case of sparsely observed flow possessing a wide range of scales with a KF-like procedure \citep{harlim2010filtering}, or a study of the performance of the EnKF in a prototypical nonlinear dynamics possessing two scales of motion \citep{ballabrera2009data}. On the side of variational methods, \citet{lorenc20074d} have nicely illustrated a modification of the 4DVar which might be successfully applied to global high-resolution coupled models. 

Seasonal-to-decadal prediction spans time horizons of up to approximately $10$ years, falling between NWP and centennial projections \citep{doblas2013seasonal}.
Correct initialization of the model is known to improve forecast quality on horizons of several years \citep{carrassi2016sources}, and for a long time predictions have been initialized with observations of the present climatic state using either the full field (FFI) or the anomaly initialization (AI) \citep[see, e.g.,][]{carrassi2014full}. 
FFI makes use of the best possible available estimate of the real state: it reduces the initial error, but the unavoidable presence of model deficiencies causes the model trajectory to drift away from the observations \citep[see, e.g.,][]{stockdale1997coupled}. Anomaly initialization assimilates the observed climate anomalies on top of an estimate of the model mean climate. This initial state, at the expense of an initial error of the size of the model bias, is expected to be closer to the model’ attractor \citep[see, e.g.,][]{smith2007objective}, so that drift is reduced. Comparisons between FFI and AI have revealed respective advantages and drawbacks, the strong regional and model-dependency of the results 
\citep[see, e.g.,][]{magnusson2013evaluation,smith2013comparison,hazeleger2013multiyear,carrassi2014full}), and the fact that AI is a viable option only when the model and the observed statistics differ largely on their first moments alone (i.e., the bias) \citep{weber2015linking}.

However it was made clear that such a “decoupled” initialization approach induced problems, particularly imbalances at the boundary between the ocean and the atmosphere. 
To cope with this issue, a solution has shown some success: the {\it weakly coupled data assimilation (wCDA)}. In the wCDA, a coupled model is used to run the predictions but the observations of the different model compartments (atmosphere, ocean, land and sea-ice) are used independently, so that each component is subject to a separate analysis. 
A first attempt to create a weakly coupled reanalysis has been done in the USA at the National Center for Environmental Prediction (NCEP) \citep{saha2010ncep} and at the Japanese Agency for Marine-Earth Science and Technology (JAMSTEC) \citep{sugiura2008development}, based on global ESMs and using 3DVar and 4DVar respectively. The wCDA reanalysis showed a marked improvement over the standard uncoupled formulation. In the JAMSTEC implementation the control variable includes the ocean initial conditions plus a set of parameters related to the air-sea fluxes. The approach acted as a proof-of-concept for successfully producing balanced initial conditions for the coupled system and optimal coupling parameters, and enhancing the skills of the s2d prediction. The UK MetOffice has designed a weakly coupled atmosphere-ocean assimilation using the incremental 4DVar \citep{laloyaux2016coupled} (cf.\ Appendix {\color{red} D}) and the global coupled model, but the corrections for atmosphere and ocean are calculated independently. Similarly, the ECMWF has produced a 20-th century reanalysis based on wCDA \citep{poli2016era}. 
The EnKF in a wCDA setting has been recently used to assimilate ocean observations to initialize s2d predictions with the Norwegian Earth System Model (NorESM) \citep{counillon2014seasonal}. 

Atmosphere and ocean are constrained independently using the ensemble-based approach at the Geophysical Fluid Dynamics Laboratory (GFDL) using the EAKF \citep{zhang2007system}. Using the same framework, \citet{lu2015strongly} have achieved some success in a controlled simulated scenario using {\it strongly CDA (sCDA)}, in which the different model compartments are coupled together also at the analysis times, so that observations on one compartment, say the atmosphere, impact on another, say the ocean. The assimilation reconstructed successfully relevant climate fields over the period of interest and provides automatically the initial conditions to run an ensemble of forecasts. 
One of the first attempts of sCDA for a coupled ocean and sea-ice model has been used operationally in TOPAZ (cf.\ Sect.~\ref{sec:TOPAZ}), demonstrating that successful assimilation of sea ice concentrations requires a coupled, multivariate and time-dependent assimilation method \citep{sakov2012topaz4}. A recent interesting result using sCDA is due to \citet{sluka2016assimilating} that shows improvements over wCDA in using only atmospheric observations in a coupled atmosphere-ocean model. 
Coupled data assimilation with the EnKF to recover the Atlantic meridional overturning circulation (AMOC) with simulated observations in a low-order coupled atmosphere-ocean model has been studied by \citet{tardif2014coupled}, and subsequently with data from a millennial-scale simulation of a comprehensive coupled atmosphere-ocean climate model in \citet{tardif2015coupled}.
These studies suggest that atmospheric observations alone, albeit frequent, do not suffice to properly recover the slowly evolving AMOC. Interestingly, it was shown that, in the lack of enough observations in the ocean, CDA of time-averaged atmospheric measurements can successfully track the AMOC.   

A comparison of different CDA approaches in the context of incremental 4DVar (cf.\ Appendix {\color{red} D}) using an idealized single-column atmosphere ocean model is given in \citet{smith2015exploring}, and revealed the benefit of CDA as being able to produce more balanced analysis fields than its uncoupled counter-part, thus reducing initialization shock and the subsequent predictions. The same idealistic model setup has been used to investigate the impact of the model error and of the window length of the 4DVar showing that while uncoupled DA may reduce the analysis error more than the CDA, the latter better reduces the imbalances and thus reduces the forecasts error. 

Coupled data assimilation is one of the main areas of research at present time and more advancements can be expected in the coming years; a review on the current status of the field can be found in \citet{penny2017coupled}.


\section{\sffamily \Large Conclusion}

The goal of this work is to provide an up-to-date review of data assimilation (DA) methods for the geosciences. We hope that it may serve as a first guide for scientists who are confronting with the use of DA methods, by providing a complete outlook of the approaches and of their foundations. This work offers a detailed introduction to the topic aimed at being a starting point from where interested readers may later expand their knowledge.

\subsection{Summary of content}
We have introduced the estimation problem, along with the definition of the dynamical and observational models, in Sect.~\ref{sec:form}. A statistical, Bayesian, point of view has been adopted to derive the filter and smoother equations. The statistical approach offers notable conceptual and mathematical advantages: it genuinely accommodates the treatment of the uncertainty in terms of probability and the model/observation outputs as realizations of random variables. The assumption of time-uncorrelated model and observational errors has been employed and led to formulate the problem as an hidden Markov model. Nevertheless, such an assumption does not generally hold in geosciences applications and we have thus mentioned methods to overcome it even though their extensive description is beyond our scope here. 
 
The huge dimension of typical DA problems in the geosciences makes the full Bayesian approach computationally unaffordable in many realistic cases, and a parametric description of the probability density function is required. The Gaussian hypothesis is thus employed in most DA methods and this has been the content of Sect.~\ref{sec:GaussMeth}, where in particular we have described the popular Kalman filter (KF), smoother (KS) and the variational approach. The latter class of methods in particular does not rely much on the hypothesis of uncorrelated model error. Section~\ref{sec:GaussMeth} has four complementary Appendices in which more details, properties and features of these methods are explained. The Monte Carlo formulations of the KF and KS, known as ensemble Kalman filter (EnKF) and smoother (EnKS), have made possible the successful extension of the classical KF and KS to high-dimensional non-linear situations. The EnKF and EnKS, in their stochastic and deterministic (square-root) formulations, are the main subject of Sect.~\ref{sec:EnsMeth}, that also discusses the two popular {\it ad hoc} fixes that are functional to the success of the EnKF and EnKF in high-dimension: localization and inflation. Section~\ref{sec:EnsMeth} includes as well an updated survey of the very recent class of hybrid methods known as ensemble-variational that are being increasingly applied in operational weather services. 

To gain more insight and getting the flavor of the scientific challenges encountered by DA in its application to the geosciences Sect.~\ref{sec:SpecTop} exposes four selected topics: (i) DA for chaotic dynamics such as the atmosphere or the ocean, (ii) DA for non Gaussian variables, (iii) DA for chemical constituents of the atmosphere, and, (iv) an example of operational DA for the ocean prediction. 

We have finally presented a prospect of the challenges that DA is facing nowadays in Sect.~\ref{sec:Persp}, with special attention to coupled data assimilation, needed to perform DA with coupled climate systems, and to particle filters, that are experiencing a tremendous trend of development aimed at making computationally viable the use of a fully nonlinear, Bayesian, DA for high-dimensional systems.

\subsection{Forward looking}
From its origin in the context of numerical weather prediction, DA has later expanded to the broad areas of environmental prediction, including seasonal, interannual and decadal time-scales. The current efforts toward the implementation of seamless predictions, where the same high-resolution coupled models are used from short to long term forecasts, are also accompanied by advancements in DA. In particular, DA has to be conceived to tackle the nonlinearities emerging from the increase in resolution and the coupling mechanisms giving rise to long term predictability. 

Nevertheless, the range of applications of DA have not remained confined to the state (and/or parameters) estimation to initialize prediction, but are progressively involving other problems. 
A notable example is the detection and attribution of climate change, or climate related events \citep{Stott13}, which is the issue of providing evidence for either the existence or the non-existence of a causal relationship between a hypothetical external forcing (e.g., anthropogenic emission) to a system (e.g., the climate) and an observed response (e.g., increase of temperature), for which novel methods based on DA have proven to be very efficient \citep{Hannart-et-al-2016}. More generally, DA can be efficiently used to estimate the marginal likelihood of the data, the so called model evidence \citep{carrassi2017estimating}, which is a key statistical metric to perform model selection \citep[see, e.g.,][for an application in the context of glacial-interglacial cycle]{carson2017bayesian} and calibration or parameter estimation \citep[see, e.g.,][for the estimation of a radiological plume]{Winiarek-et-al-2011}, or \citet{Tand15} for the optimization of a subgrid-scale parametrization. 
DA has been used to reconstruct the climate of the past based on observations proxies \citep[see, e.g.,][]{dubinkina2013assessment}, and, in the solid Earth science, for seismology applications \citep[see, e.g.,][]{fichtner2006adjoint}.

In general the use of DA has proven that the consistent data-to-model fusion provides a more insightful view on the phenomena of interest, than any of the two components, the model or the data, independently. 
Future applications of DA within the geosciences and beyond, are expected to be numerous, and to naturally arise by the improvement of our modelling capabilities, as a result of the increased computational power and physical understanding, on the one hand, and by the progresses of the observing facilities (such as, but not only, satellite) on the other. 
Data assimilation is nowadays spreading to many emerging disciplines such as neurosciences, genetics, biology, medicine or even in sociology-demography and traffic managements just to mention a few \citep[see, e.g.,][for an example of applications for traffic flow and biology respectively]{palatella2013nonlinear,kadakia2016nonlinear}. This expansion exposes DA to the need of new theoretical principles and novel methodological solutions and provide new contexts for challenging its effectiveness and robustness.
Data assimilation is thus expected to continue playing a central role to bridge model with data, to maximally exploit their respective informational content.  

We hope that this overview may be a first guide for scientists who are confronting with the use of DA methods, and provide them with a complete first outlook of the approaches and of their foundations. The present work is thus to be intended as a detailed introduction to the topic from where interested readers and researchers may later expand their knowledge.

\subsection*{Appendix A: Some properties of the Kalman filter and smoother \label{sec:KFKSprop}}

Although the straightforward use of the Kalman filter (KF) and Kalman smoother (KS) in geosciences is obviously hampered by the computational limitations and by the inconsistency of their statistical/dynamical hypotheses (Gaussianity and linearity), yet they represent the backbone of many practical DA algorithms. 
The history of the use of KF-like methods in geosciences is the one of a never-ending search for suitable approximations that, even if sub-optimal, can still work satisfactorily in a nonlinear, non-Gaussian, and high dimensional setting. We have seen in Sect.~\ref{sec:EnsMeth} how the KF has served as a key conceptual and factual framework upon which several successful operational DA methods have been built.  
This appendix reviews some of the key properties and issues of the KF and KS.
Our discussion here mainly pertain to the KF, but most of the conclusions apply to the KS too.\\ 

{\bf Time dependent prior}

The KF analysis will be statistically closer to either the observations or the prior depending on their respective accuracy, i.e., on our belief about them as estimated via the covariances $\bR_k$ and $\bP^\mathrm{f}_k$ respectively. 
In the geosciences, the number of observations, albeit large, is usually insufficient to fully cover the state space ($d\ll m$) so that much of how the information is spread from observed to unobserved areas is controlled by the prior. 
Having an informative, accurate, and reliable prior is thus of great importance. As mentioned at the end of Sect.~ \ref{sec:form1}, the situation $d\ll m$ is endemic in NWP, and the use of a short-range numerical forecast in a cyclic DA procedure has been key to the success of DA in that context \citep[see, e.g.,][]{daley1993atmospheric, Kalnay2002}. 
The KF recursion provides a time-dependent estimate of the prior (its mean, $\x^f_k$, and covariance, $\bP^\mathrm{f}_k$) that is highly desirable in environmental systems that are usually chaotic, so that the actual error associated with $\x^f_k$ is itself strongly time-dependent. We have seen in Sect.~\ref{sec:AUS} that this property of the chaotic dynamics, while representing a challenge to the state estimation process, can also be exploited explicitly in the design of DA algorithms for this class of systems. \\ 

{\bf Filter divergence} 

Filter divergence is the name used to refer to the situation in which the solution of the KF deviates dramatically from the true signal that it was supposed to track, and the KF is not longer able to pull back its solution close to the truth \citep{fitzgerald1971divergence, harlim2010catastrophic}. Filter divergence is often the result of progressive and repeated under-estimation of the actual error, $\bP^\mathrm{a}_k<\bP^\mathrm{truth}_k$ \citep[the matrices order relationship is that of the cone of the positive semi-definite matrices,][]{bocquet2017degenerate}.
Under the action of the dynamics, Eq.~\eqref{eq:KF-fcst2}, the analysis error covariance, $\bP^\mathrm{a}_{k-1}$, is transformed into the forecast one at the next time, $\bP^\mathrm{f}_{k}$. If the dynamical model is chaotic (or just unstable) then at least one of the eigenvalues of $\bM_{k:k-1}$ is larger than one and $\bP^\mathrm{f}_{k}\ge\bP^\mathrm{a}_{k-1}$: the estimated error grows during the forecast phase. However, for generic stable dynamics such an error growth is not guaranteed.  
At the analysis times, the term $(\bI_k-\bK_k\bH_k)$ in Eqs.~\eqref{eq:KF-anl2} and \eqref{eq:KF-anl3} represents the forcing due to the observations, and it has a stabilizing effect since its eigenvalues are bounded to be lower or equal to one \citep{carrassi2008a}. This implies that the estimated analysis error covariance is always smaller or equal to the forecast (prior) error covariance, $\bP^\mathrm{a}_k\le \bP^\mathrm{f}_k$. 
The overall fate of the KF error covariance comes by the balances between the (possible) growth during the forecast phases and the (certain) decrease at analysis times. If the dynamics is not able to counteract the covariance decrease occurring at analysis times, the KF error covariance will progressively decrease, and once $\bH_k\bP^\mathrm{f}_k\bH_k^{\mathrm{T}}<<\bR_k$ the filter solution may start to ignore the observations.   
This is not an issue in itself, as long as the actual error is also decreasing and the KF solution is properly tracking the desired signal. Nevertheless, when this is not the case, i.e., when the KF error estimates decrease but the actual error does not, the KF solution starts to deviate from the observations, eventually diverging completely from the true signal. 

Several factors may be at the origin of filter divergence, notably in the misspecification in the DA setup, such as a too strong influence
from the measurements (from wrongly specified error statistics, neglected measurement error covariances, etc.),
or wrongly specified or neglected model errors.
Filter divergence also occurs in ensemble-based DA (see Sect.~\ref{sec:EnsMeth}) and we have seen in Sect.~\ref{sec:Infl-Loc} which countermeasures, inflation and localization, have been placed in order to deal with this issue in real applications.      \\ 

{\bf A diagnostic tool}

A remarkable property of the KF, originating from the linear and Gaussian assumptions, is that the error covariances, $\bP^\mathrm{f}_k$ and $\bP^\mathrm{a}_k$ do not depend on the observation values: they are thus unconditional covariances. This is a direct consequence of the first and second moments of the system's state pdf being independent from each other (and uncoupled with higher order moments); a behavior that no longer holds in nonlinear, non-Gaussian, scenarios.   
Another peculiar feature, which serves to monitor the goodness of the hypotheses, is that the innovation vector sequence, $\v_k=\y_k-\bH_k\x^\mathrm{f}_k$, is Gaussian and uncorrelated in time \citep{jazwinski1970}: one can thus keep checking the innovations and, possibly, to implement corrections \citep{daley1993atmospheric}. \\

{\bf Bias and covariance estimation} 

The optimality of the KF relies upon the veracity of its assumptions: the linearity of the model and observational operator, and the Gaussianity of the true error pdfs. Any mismatch between the real conditions on which the KF operates and its working hypotheses will negate its optimality. Nevertheless, even when the hypotheses are correct, the KF will still depend on the correct specification of its statistical inputs: the model and observational error covariances. 
The initial conditions, $\x^\rma_0$ and $\bP^\rma_0$, are also input but their impact on the filter performance is discussed separately in the following paragraph.
	
Recall from Eqs.~(\ref{eq:modellin}--\ref{eq:obslin}) that the model and observational error are assumed unbiased and Gaussian,
$\eeta_k\sim\mathcal{N}(\bzero,\bQ_k)$ and $\epsi_k\sim\mathcal{N}(\bzero,\bR_k)$. 
If either the actual model or observational errors are biased (or both), the KF analysis will be biased too, unless those biases are removed from the forecast before the analysis update, Eqs.~(\ref{eq:KF-anl2}--\ref{eq:KF-anl3}). These biases can be estimated recursively in time, along with the system's state, using an approach known as {\it state augmentation} in which the state is formally augmented with the bias term \citep[e.g.][]{dee2005bias}. 
The state augmentation strategy is also the classical choice to deal with the simultaneous model state and parameter estimation \citep{jazwinski1970}. 

Likewise, discrepancies can also be present between the actual model and observation error covariances and those stipulated in the filter setup. 
In contrast to the bias, the covariances cannot be corrected using the state augmentation approach, and an additional procedure is required. A possibility is again on the use of the innovations: when all error covariances entering the KF are correct, the innovations are distributed according to $\v_k\sim\mathcal{N}(\bzero,\bSigma_k) $, with $\bSigma_k=\bH_k(\bM_{k:k-1}\bP^a_{k-1}\bM_{k:k-1}^{{\rm T}}+\bQ_k)\bH_k^{{\rm T}}+\bR_k$ \citep{cohn1997introduction}. It is then possible, in principle, to estimate the ``best'' $\bQ_k$ and/or $\bR_k$ as those maximizing the conditional probability, $p(\v_k\vert\bQ_k, \bR_k)$, where the innovation is treated as a random variable \citep{dee1995line}. Given the large dimension of $\bQ_k$ and $\bR_k$, such a maximum likelihood approach can only be feasible if $\bQ_k$ and $\bR_k$ are parametrized based on a very small number of parameters. 

In any case, suitable parametrizations of the covariance matrices are necessary, particularly for model error, given the huge size of the geophysical models and the wide range of possible error sources. The former problem implies the need to estimate large matrices based on a limited number of available observations. The second is related to the multiple sources of model error, such as incorrect parametrization, numerical discretization, and the lack of description of some relevant scale of motion, which makes it difficult to set a unified parametrization.
Recent works have proposed efficient combinations of Bayesian estimation procedures with Monte Carlo approximation to estimate both the observational and model error covariances \citep{ueno2014,ueno2016bayesian,dreano2017estimating,liu2017,pulido_et_al_2018}.   

The computation of the Kalman gain, Eq.~\eqref{eq:KF-anl1}, requires the inversion of the matrix $(\bH_k\bP_k^\mathrm{f}\bH_k^{\rm T}+\bR_k)^{-1}\in{\mathbb R}^{d\times d}$.
To make it computationally tractable, $\bR_k$ is often assumed to be diagonal and full rank, i.e., observations are assumed to be spatially uncorrelated. The estimation of $\bR_k$ is reduced to the task of specifying only its diagonal. It can also negatively affect the filter's performance 
when observations are spatially correlated, which is typically for remotely sensed data. The impact of neglecting observational error correlations, as well as approaches to include them efficiently in the DA setup, have been studied in several works \citep[see, e.g.,][]{stewart2008correlated,miyoshi2013estimating}. \\

{\bf Dependence on the initial condition}

The criticality of the choice of the initial error covariance, $\bP_0$, is related to the filter's stability, 
intended as the convergence of its solutions to an asymptotic sequence, independently of the initial conditions \citep{gelb1974applied}. Stability is a very desirable practical property: a stable filter will always tend to a steady solution and all unwanted errors in the specification of the initial conditions, $\x_0$ and $\bP_0$, do not alter the its final output.
Nevertheless, optimality of the filter alone does not guarantee stability but, for a stochastically-driven dynamical system as in Eq.~\eqref{eq:modellin}, it also requires the filter to be (uniformly) {\it observable} and {\it controllable} \citep[see, e.g., ][]{Kalman1960, jazwinski1970, cohn1988observability}. Roughly, observability is the condition that, given sufficiently many observations, the initial state of the system can be reconstructed by using a finite number of observations \citep{quinn2010state}. 
To see this, let consider the case of a discrete, autonomous (i.e., constant, $\bM_k=\bM$), and deterministic dynamical model $\x_k=\bM\x_{k-1} $ of dimension $n$, that is observed $n$-times, without error, with scalar measurements and a linear operator, so that $y_k=\bH\x_k$ (the operator $\bH$ is in this case a $n$-dimensional row vector). Starting from the initial condition at $t_0$, we have $y_0=\bH\x_0$, $y_1=\bH\x_1=\bH\bM\x_0$, and so on until, $y_{n-1}=\bH\x_{n-1}=\bH\bM^{n-1}\x_0$, that can be written compactly as
\begin{equation*}
  \begin{bmatrix}
    y_0 \\
    y_1 \\
     .  \\
     .  \\
    y_{n-1} 
  \end{bmatrix}
=
  \begin{bmatrix}
    \bH \\
    \bH\bM \\
     .  \\
     .  \\
    \bH\bM^{n-1} 
  \end{bmatrix}
\x_0
=
\bPsi^{\rm T}\x_0.
\end{equation*}
We see therefore that, if one wants to determine uniquely the initial state, $\x_0$, based on the observations, the $n\times n$ matrix $\bPsi$ must be invertible, that is to say its rank must be equal to $n$, or equivalently its determinant must be nonzero. In this case the system is said to be observable by the sequence of observations $y_0...y_{n-1}$. As an example consider the simple $2\times 2$ system 
\begin{equation*}
\bM = 
  \begin{pmatrix}
    2 & 1 \\
    0 & 1 
  \end{pmatrix} \,
\end{equation*} 
such that the dynamics of the first component depends on both the first and second components, while the second component depends only on itself. It is easy to show that observing the first component alone (i.e., $\bH = [1~0]$, a $2$-dimensional row vector), the corresponding $2\times 2$ matrix $\bPsi$ has determinant equal to $1$, is therefore invertible, and the system is observable. On the other hand, if it is second component to be observed (i.e. $\bH = [0~1]$), the determinant of $\bPsi$ is zero and the system is not observable. This result is physically interpretable such that, given that the first component carries also information about the second, but not vice-versa, its observation is more effective in informing about the full $2$-dimensional system.    
Similarly to observability, controllability can be described as the ability to move the system from any initial state to a desired one over a finite time interval, and is related to the properties of the system noise, $\bQ_k$ \citep[see][for a complete discussion on observability and controllability with several examples]{gelb1974applied}.

The KF stability and convergence for purely deterministic systems (i.e., like in Eq.~\eqref{eq:modellin} but with $\bQ_k={\bf 0}$), under the sole condition of uniform observability has been recently proved by \citet{ni2016stability} and further characterized in terms of the stability properties of the dynamics by \citet{carrassi2008a,gurumoorthy2017rank}. The generalization to the case of degenerate (rank deficient) initial condition error covariance is given in \citet{bocquet2017degenerate}, thus corroborating reduced-rank formulations of the KF based on the system's unstable modes \citep{trevisan2011}.

\subsection*{Appendix B: Minimization process in variational methods \label{sec:minim}}

With the gradient, Eq.~\eqref{eq:grad-strong}, in hand, the minimization is iteratively solved searching for the state vector, $\x_0^i$, at the i-th iteration that satisfies ${\mathcal J}(\x_{0}^i) < {\mathcal J}(\x_{0}^{i-1})$ (i.e. the amplitude of the cost function decreases from iteration $i-1$ to $i$), and the process is repeated until a prespecified convergence criterion (a threshold on the amplitude of the gradient, or on the difference of the cost function at two successive iterations, ${\mathcal J}(\x_{0}^{i+1}) - {\mathcal J}(\x_{0}^i)$) is verified. The new state at each iteration is updated as $\x_0^i = \x_0^{i-1} + \gamma^{i-1}\v^{i-1}$, with $\v^i$ being the searching direction and $\gamma^{i-1}$ the step size. 
The various minimization algorithms differ on how $\v^i$ and the step size along it are chosen. 
When the searching direction is chosen to have an angle greater than $90$ degrees with respect to the gradient (i.e., $(\v^i)^{\rm T}\nabla_{\x}{\mathcal J}(\x^i)<0 $), the minimization procedures are referred to as {\it descent methods}.
The most common and straightforward descent methods are the {\it steepest} and the {\it Newton} method. 

In the former, $\v^i$ is taken as opposite to the gradient. 
This strategy works very well when the cost function is uniformly strictly convex (i.e., the Hessian of the cost function is positive definite and it has at most one global minimum), in which case the gradient at any arbitrary point always heads to the absolute (and unique) minimum of the cost function, and the rate of convergence is linear. The computational cost of each iteration is relatively low, but the linear convergence can be so slow that the difference, $\x_{0}^{i}-\x_0^{i-1}$, becomes smaller than computer precision. Furthermore, the assumption of a globally convex cost function is critical in geosciences applications \citep{Milleretal94,pires1996}.

To cope with this, Newton's method assumes that the cost function can be locally approximated by a quadratic expansion around the state point, $\x^i$, ${\mathcal J}(\x) \approx {\mathcal J}^{{\mathrm Newt}}(\x) = {\mathcal J}(\x^i) + \nabla_{\x}{\mathcal J}(\x^i)(\x-\x^i) + \dfrac{1}{2}(\x-\x^i)^{{\mathrm T}}\nabla^2_{\x}{\mathcal J}(\x^i)(\x-\x^i) $. The state at i-th iteration is found by setting the gradient of this approximation to zero, which gives $\v^i = -\nabla^{-2}_{\x}{\mathcal J}^{{\mathrm Newt}}(\x^i)\nabla_{\x}{\mathcal J}^{{\mathrm Newt}}(\x^i)$; the search direction is equal to the opposite of the Hessian matrix of the cost function multiplied by its gradient. 
At the minimum the Hessian of the cost function is positive definite so that the search direction verifies the condition of being oriented with an angle greater than $90$ degrees from the gradient. In practice, and in contrast to steepest descent, the Newton method uses also the local information about the curvature of the cost-function in order to better point toward its minimum. 
Although the convergence of the Newton's method is quite rapid, its operational use in geophysical DA is rendered difficult by the need to invert the Hessian matrix, which is usually huge size and ill conditioned. 

Minimization algorithms used operationally are a trade-off between efficiency and computational limitation and have features that mimic those of the two main algorithms just described. A throughout description of the state-of-art minimization methods goes beyond the scope of this article but interested readers can find more details in, e.g., \citet{fisher2001developments} or \citet{asch2016data}.

\subsection*{Appendix C: Comments on the variational methods \label{sec:VarProp}}

The Gaussian hypothesis has not just allowed to get an analytic expression for the cost function, Eq.~\eqref{eq:Jw4DVar} or \eqref{eq:J-strong}, but it also offered a statistical, and physically plausible, interpretation of the analyzed trajectory. Given the unimodality of the Gaussian pdf, the most likely state is also the mean of the pdf, that is to say the minimum variance estimate. Without unimodality the mean state, while still having minimum variance, may well be of scarce relevance (it may fall in very low probability region) or not have physical plausibility at all. 

In deriving the 4DVar, either in the weak or strong constraint formulations, no assumptions have been made about the characteristics of the dynamical and observational models: they can be assumed nonlinear and so they are in many real applications. 
Nevertheless, whether or not the latter is actually the case it has enormous consequences on the accuracy of the 4DVar analysis, as well as on the complexity of the algorithms used to solve it. 
When both models are linear, all errors are Gaussian and independent, the 4DVar cost-function is quadratic. If furthermore the Hessian of the cost function is spherical the gradient will depend linearly on the control variable and will correctly point to the cost function (unique) global minimum.
In this linear case, the 4DVar solution will match exactly the mean solution of a Kalman smoother (KS), Eq.~\eqref{eq:KS-backrec1}, to which the same input statistics are provided, and it will thus represent an alternative way to get the best mean estimate without the explicit need to compute inverse matrices as in the KS \citep{fisher2005equivalence}. 

In the general nonlinear case however, the exact minimum-variance solution may not be obtained. The approximate analysis will be the outcome of the minimization process and the degree of its accuracy will strongly depend on the degree of nonlineariy in the dynamical and observational models, even if the initial condition and observational error are Gaussian. The cost function will not longer be quadratic and it may possess multiple minima to which the minimization procedure can wrongly be trapped. A number of fixes have been proposed and put in place to overcome this issue, so as to render the cost-function "more quadratic", notably by the use of a precondition under the form of a suitable invertible control variable transformation \citep{zupanski1996preconditioning}. It is beyond our scopes to expand further on this subject, but the readers can find more details in the literature \citep{talagrand2010variational,asch2016data}.    

The variational approach does not automatically solve the complete Gaussian estimation problem: it does not provide the two moments, the mean and the covariance, of the posterior distribution, but only the first one. It is possible to show that the analysis error covariance is indeed given, exactly/approximately for the linear/nonlinear case respectively, by the inverse Hessian matrix of the cost function;  
at its minimum the (inverse) Hessian must be positive definite (see Appendix {\color{red} B}), consistently with a feature of a covariance matrix. Nevertheless, estimating the Hessian matrix for a realistic geophysical applications is extremely difficult, and usually the same (fixed in time) error covariance matrix is used to characterize the background errors at the beginning of each DA cycle. When solving the s4DVar the background error covariance is implicitly evolved within the window so that, effectively, a dynamically evolved estimate of the prior error is used at the observation times \citep{pires1996}, but such an updated covariance is not explicitly accessible to initialize the next cycle.

This inherent limitation of the variational approach marks a key distinction with respect to sequential methods like the Kalman filter or smoother, that provide a time-dependent description of the uncertainty associated to the state estimate. This aspect has largely, but not solely, contributed to the popularity of KF-like approaches for DA with chaotic models \citep{vannitsem2017predictability} where a time dependent description of the estimation error is highly desirable (cf.\ Sect.~\ref{sec:AUS}). 
We have seen in Sect.~\ref{sec:EnVar} that the recent promising efforts toward hybrid variational-ensemble methods are also aiming to cope with this issue, thus endowing the 4DVar with a flow-dependent estimate of the error covariance \citep{lorenc2015comparison,kleist2015osse,buehner2015implementation}.

\subsection*{Appendix D: Some popular approximations \label{sec:approx}}

We describe briefly some of the early successful approximations of the Kalman filter and of the variational approach that have made their implementation possible in the geosciences.

\subsubsection*{Extended Kalman filter \label{sec:EKF}}

The {\it extended Kalman filter} (EKF) represents a first-order expansion of the Kalman filter (KF) and extends its use to nonlinear dynamics \citep{jazwinski1970}. Like KF, it is sequential: the system’s state and associated error covariance are updated at discrete observation times and evolved in between them.
In the EKF, the mean state estimate is propagated by the full nonlinear model, but the error covariance evolution is approximated using the tangent linear one. The linearization is taken around the nonlinear model solution, so that the Jacobian of the model is evaluated upon it and it is thus state dependent.

As with the standard KF for linear dynamics, the EKF also assumes that errors are all Gaussian distributed. Nevertheless, under the action of the nonlinear dynamics, even a possible
initial Gaussian error covariance will not stay Gaussian, and the EKF will only provide an approximate description of the actual estimation error distribution. In general, the accuracy
of the EKF scales with the degree of nonlinearity in the model \citep{Milleretal94}. For instance, \citet{eve92} implemented the EKF with a multilayer ocean model finding that the tangent linear operator led to unbounded error growth since the nonlinear 
saturation that should occur at climatological level is contained in higher order moments
of the error covariance equations, and those are all neglected in the closure used in the EKF.

The EKF has been successful in a number of pioneering applications of DA for meteorology \citep{Ghil1981,Dee_et_al_1985} and oceanography \citep{ghil1991data}. It has also been used in one of the early study of coupled DA (cf.\ Sect.~\ref{sec:CDA}) with an atmosphere-ocean model of intermediate complexity \citep{sun2002data}.
The joint state and parameter estimation is possible with the EKF using the state-augmentation approach and its efficiency for this purpose has been demonstrated in the context of DA for seasonal forecasts \citep{kondrashov2008data} or land surface DA \citep[see, e.g.,][]{de2014initialisation}.  
A formulation of the EKF for parameter estimation in the presence of time correlated model error has been proposed by \citet{carrassi2011state} and later applied to a soil model \citep{carrassi2012short}.

Along with the linear assumption on which it is built, another limitation of the EKF is due to the enormous computational requirements of the error covariance propagation. This involves the storage of full covariance matrices, the derivation of the tangent linear model, and its application a number of times twice the state vector dimension \citep{asch2016data}.

\subsubsection*{Incremental 4DVar and 3DVar \label{sec:incr4DVar}}

The {\it incremental formulation} \citep{courtier1994strategy} employs a linearization of the problem around the background trajectory: both the dynamical and observational models are linearized and the cost function of the incremental (strong-constraint) 4DVar reads   
\be
\label{eq:Jincs4DVar}
{\mathcal J}^{\mathrm{s4DVar-Incr}}(\delta\x_{0}) = \frac{1}{2}\sum_{k=0}^K\left\| \v_k - \bH_k\bM_{k:0}\delta\x_0\right\|_{\bR^{-1}_k}^2 +\frac{1}{2}\left\| \delta\x_0 \right\|_{\bB^{-1}}^2,
\ee
where the increment, $\delta\x_0=\x_0-\x^\rmb $, is now the control variable for the minimization, and $\v_k=\y_k-{\mathcal H}_k(\x_k)=\y_k-{\mathcal H}_k\circ{\mathcal M}_{k:0}(\x_0)$ is the innovation vector (cf.\ Appendix {\color{red} A}). 

The cost function is now quadratic, it possesses a unique absolute minimum, and it can be minimized much more easily.  
The minimization can be carried out by first computing the innovations (outer loop) using the nonlinear models, ${\mathcal H}$ and ${\mathcal M}$. In the inner loop, Eq.~\eqref{eq:Jincs4DVar} is evaluated using the linearized models, $\bH$ and $\bM$, and then the gradient using the adjoint $\bM^{\rm T}$. This procedure returns the analysis increment, $\delta\x_0$, to be used for the next outer loop and so on until convergence. The incremental 4DVar allows thus to deal with small nonlinearities in an incremental way, given that the linearized models are cyclically updated when a new outer loop trajectory is computed. Usually a simplified version of the model (coarser resolutions, simplified physics, etc.) is used in the inner loop \citep{lawless2008using}, and this feature along with the quadratic form of the cost function, have been pivotal for the operational implementation of the incremental 4DVar \citep[see, e.g.,][]{courtier1994strategy}. 

3DVar is a special case of the 4DVar where the time dimension is removed and only the observations at the analysis time are assimilated (see Fig.~\ref{fig:Fig2} and \citet{Kalnay2002}). In this case the control variable is the state at $t_0$, $\x_0$, like for the strong-constraint 4DVar, but in contrast to it only the observations a $t_0$ are used in the update. In operational implementations of the 3DVar all observations within a specific interval, $[t_0-\Delta t, t_0+\Delta t]$ (typically $\Delta t=3~hrs$), are used to update $\x_0$. It is also worth to mention the {\it First Guess at Appropriate Time 3D-Var} (FGAT 3D-Var), in which the unity operator for the resolvent of the tangent linear system and its adjoint is used. In practice FGAT 3D-Var has the form of a 4DVar but it reduces the necessary computations to those of a 3DVar \cite{fisher2001developments}.

\section*{\sffamily \Large ACKNOWLEDGEMENTS}
The authors wish to thank Eugenia Kalnay and another anonymous reviewer for their detailed, deep and critical reviews of the original version of this work. Their comments and suggestions have substantially improved its readability to a wider audience and have also helped in clarifying the discussion in many instances. 
The authors are thankful to P.~N.~Raanes (NERSC), A.~Farchi (ENPC) and C.~Grudzien (NERSC) for their comments, suggestions and insightful discussions and to J. Xie (NERSC) for providing Figure~\ref{fig:Fig8}. Finally, the authors also wish to thank R.~Davy (NERSC) who provided a critical review of the second version of the manuscript that helped to smooth further the mathematics and to make the work more accessible to the geosciences community at large.
A. Carrassi has been funded by the project REDDA (\#250711) of the Norwegian Research Council. 
G. Evensen has been partly funded by the project EmblA of Nordforsk.
CEREA is a member of Institut Pierre-Simon Laplace (IPSL).

\bibliographystyle{./agufull}
\renewcommand{\bibfont}{\small}
\bibsep=1.0pt

\end{document}

%% file: definitions.tex
\def\be{\begin{equation}}
\def\ee{\end{equation}}
\def\bea{\begin{eqnarray}}
\def\eea{\end{eqnarray}}
\def\nn{\nonumber\\}
\def\({\left(}
\def\){\right)}

\def\x{\mathbf{x}}
\def\barx{\overline{\mathbf x}}

\def\tildx{\tilde{\mathbf x}}
\def\tildy{\tilde{\mathbf y}}

\def\y{\mathbf{y}}
\def\w{\mathbf{w}}
\def\v{\mathbf{v}}

\def\bA{\mathbf{A}}
\def\bB{\mathbf{B}}
\def\bC{\mathbf{C}}

\def\bR{\mathbf{R}}
\def\bP{\mathbf{P}}
\def\bM{\mathbf{M}}
\def\bH{\mathbf{H}}
\def\bI{\mathbf{I}}
\def\bQ{\mathbf{Q}}
\def\bK{\mathbf{K}}

\def\bX{\mathbf{X}}
\def\bY{\mathbf{Y}}

\def\bZ{\mathbf{Z}}

\def\bS{\mathbf{S}}
\def\bT{\mathbf{T}}
\def\dY{\mathbf{Y}'}

\def\bGamma{{\boldsymbol \Gamma}}
\def\bPsi{{\boldsymbol \Psi}}
\def\bpsi{{\boldsymbol \psi}}

\def\bSigma{{\boldsymbol \Sigma}}
\def\bPhi{{\boldsymbol \Phi}}
\def\brho{{\boldsymbol \rho}}

\def\eeta{{\bm \eta}}
\def\epsi{{\bm \epsilon}}

\def\blam{{\bm \lambda}}
\def\bLam{{\bm \Lambda}}

\def\Re{{\mathbb R}}

\def\bzero{{\mathbf 0}}
\def\b1{{\mathbf 1}}
\def\T{\mathrm{T}}

\def\onehalf{\frac{1}{2}}

%% file: definitions_ext.tex
\def\be{\begin{equation}}
\def\ee{\end{equation}}
\def\bea{\begin{eqnarray}}
\def\eea{\end{eqnarray}}
\def\nn{\nonumber\\}
\def\({\left(}
\def\){\right)}

\def\x{\mathbf{x}}
\def\barx{\overline{\mathbf x}}

\def\y{\mathbf{y}}
\def\w{\mathbf{w}}
\def\v{\mathbf{v}}

\def\bA{\mathbf{A}}
\def\bB{\mathbf{B}}
\def\bC{\mathbf{C}}

\def\bE{\mathbf{E}}

\def\bR{\mathbf{R}}
\def\bP{\mathbf{P}}
\def\bM{\mathbf{M}}
\def\bH{\mathbf{H}}
\def\bI{\mathbf{I}}
\def\bQ{\mathbf{Q}}
\def\bK{\mathbf{K}}

\def\X{\mathbf{X}}

\def\bZ{\mathbf{Z}}

\def\bS{\mathbf{S}}

\def\bGamma{{\boldsymbol \Gamma}}
\def\bPsi{{\boldsymbol \Psi}}

\def\bSigma{{\boldsymbol \Sigma}}

\def\eeta{{\bm \eta}}
\def\epsi{{\bm \epsilon}}

\def\blam{{\bm \lambda}}

\def\bzero{{\mathbf 0}}

\def\T{\mathrm{T}}

%% file: def.tex
\newcommand {\lp}{\left(}
\newcommand {\rp}{\right)}

\newcommand{\rma}{{\rm a}}
\newcommand{\rmb}{{\rm b}}

\newcommand{\rme}{{\rm e}}
\newcommand{\rmf}{{\rm f}}

\newcommand{\rmo}{{\rm o}}

\newcommand{\rmT}{{\rm T}}

\newcommand{\ol}{\overline}